\documentclass[aps,twocolumn,prd,showpacs,showkeys,preprintnumbers,superscriptaddress,nobibnotes,floatfix,longbibliography,notitlepage,nofootinbib,groupedaddress]{revtex4-2}

\pdfoutput=1
\usepackage{amsmath}
\usepackage{amsfonts}
\usepackage{amssymb}
\usepackage{mathrsfs}
\usepackage{graphicx}
\usepackage{color}
\usepackage{bbding}
\usepackage{tabularx}
\usepackage[usenames,dvipsnames,table]{xcolor}
\usepackage[normalem]{ulem}
\usepackage{makecell}
\usepackage{siunitx}
\usepackage{enumerate}
\usepackage{multirow}
\usepackage{makecell}
\usepackage{upgreek}

\usepackage{marvosym}
\definecolor{darkred}{rgb}{0.5,0,0}
\definecolor{darkblue}{rgb}{0,0,0.5}
\definecolor{firebrick}{rgb}{0.75,0.125,0.125}
\definecolor{darkgreen}{rgb}{0,0.5,0}
\usepackage[colorlinks=true,linkcolor=firebrick,citecolor=darkgreen,urlcolor=darkblue]{hyperref} 
\usepackage[capitalise]{cleveref}

\newcommand{\ie}{{\it i.e.}}

\newcommand{\eg}{{\it e.g.}}

\newcommand{\eq}{Eq.}

\newcommand{\fig}{Fig.}
\newcommand{\Fig}{Fig.}

\newcommand{\Refe}{Ref.}
\newcommand{\Refs}{Refs.}

\newcommand{\equ}[1]{\eq~(\ref{equ:#1})}
\newcommand{\figu}[1]{\fig~\ref{fig:#1}}

\graphicspath{{figures/}{figures_appendix/}}

\begin{document}

\title{The ultra-high-energy neutrino-nucleon cross section:\\measurement forecasts for an era of cosmic EeV-neutrino discovery}

\author{V\'ictor B.~Valera}
\email{vvalera@nbi.ku.dk}
\affiliation{Niels Bohr International Academy, Niels Bohr Institute,\\University of Copenhagen, DK-2100 Copenhagen, Denmark}

\author{Mauricio Bustamante}
\email{mbustamante@nbi.ku.dk}
\affiliation{Niels Bohr International Academy, Niels Bohr Institute,\\University of Copenhagen, DK-2100 Copenhagen, Denmark}

\author{Christian Glaser}
\email{christian.glaser@physics.uu.se}
\affiliation{Department of Physics and Astronomy, Uppsala University, Uppsala, SE-752 37, Sweden}

\date{\today}

\begin{abstract}
Neutrino interactions with protons and neutrons probe their deep structure and may reveal new physics.  The higher the neutrino energy, the sharper the probe.  So far, the neutrino-nucleon ($\nu N$) cross section is known across neutrino energies from a few hundred MeV to a few PeV.  Soon, ultra-high-energy (UHE) cosmic neutrinos, with energies above 100~PeV, could take us farther.  So far, they have evaded discovery, but upcoming UHE neutrino telescopes endeavor to find them.  We present the first detailed measurement forecasts of the UHE $\nu N$ cross section, geared to IceCube-Gen2, one of the leading detectors under planning.  We use state-of-the-art ingredients in every stage of our forecasts: in the UHE neutrino flux predictions, the neutrino propagation inside Earth, the emission of neutrino-induced radio signals in the detector, their propagation and detection, and the treatment of backgrounds.  After 10~years, if at least a few tens of UHE neutrino-induced events are detected, IceCube-Gen2 could measure the $\nu N$ cross section at center-of-mass energies of $\sqrt{s} \approx 10$--100~TeV for the first time, with a precision comparable to that of its theory prediction.
\end{abstract}

\maketitle


\section{Introduction}
\label{section:introduction}

Neutrino interactions with matter are powerful probes of particle physics: they map the deep structure of nuclei and nucleons, and may unearth evidence of new physics.  Broadly stated, the higher the energy of the interacting neutrino, the sharper its probing power.  Today, high-energy neutrino-matter interactions---in the form of the neutrino-nucleon ($\nu N$) cross section---are known experimentally up to PeV neutrino energies, the highest detected so far.  Yet, a trove of further insight likely lies in the measurement of the $\nu N$ cross section at higher energies.  Presently, those energies are practically out of the reach of existing detectors, but this limitation will likely be overcome in the coming years. 

\begin{figure*}[t]
 \centering
 \includegraphics[width=\textwidth]{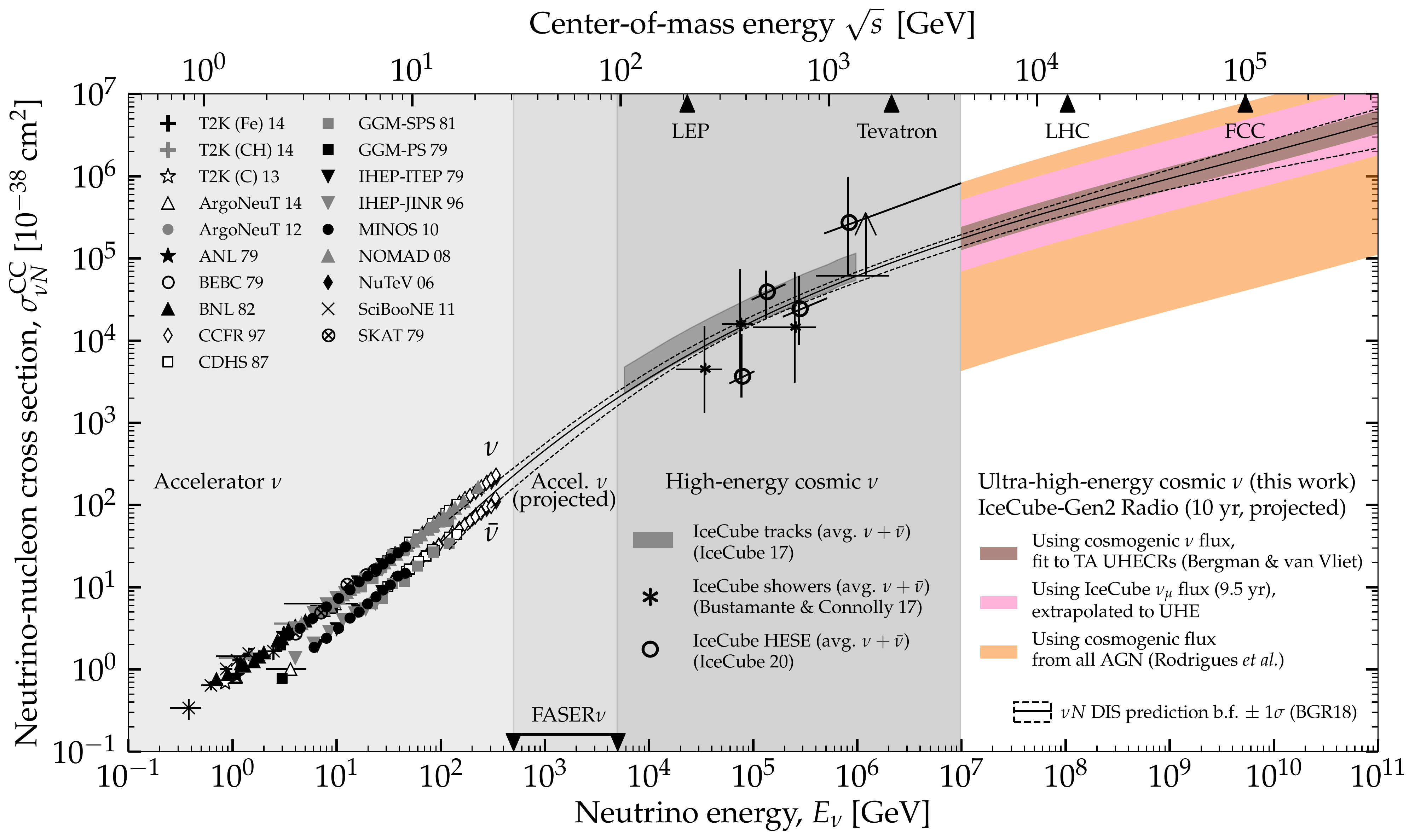}
 \caption{\label{fig:panorama}Neutrino-nucleon ($\nu N$) charged-current (CC) cross section, measurements and predictions.  Sub-TeV measurements are from accelerator-neutrino experiments~\cite{Mukhin:1979bd, Baranov:1978sx, Barish:1978pj, Ciampolillo:1979wp, deGroot:1978feq, Colley:1979rt, Morfin:1981kg, Baker:1982ty, Berge:1987zw, Anikeev:1995dj, Seligman:1997, Tzanov:2005kr, Wu:2007ab, Adamson:2009ju, Nakajima:2010fp, Abe:2013jth, Acciarri:2014isz, Abe:2014nox}.  Few-TeV measurements will be covered by the upcoming FASER$\nu$ accelerator-neutrino experiment~\cite{FASER:2019dxq}.  TeV--PeV measurements use high-energy astrophysical neutrinos detected by the IceCube neutrino telescope~\cite{IceCube:2017roe, Bustamante:2017xuy, IceCube:2020rnc}.  We forecast cross-section measurements above 100~PeV---the ultra-high-energy (UHE) range---where no measurement exists presently.  Our forecasts are for radio-detection of UHE neutrinos in the planned IceCube-Gen2~\cite{IceCube-Gen2:2020qha}, and are based on sophisticated simulations of UHE neutrino propagation inside Earth (Section~\ref{section:neutrino_propagation}) and detector response (Section~\ref{section:radio_detection}).  In this plot, we showcase 10-year forecasts for three attractive flux scenarios that may afford the most precise cross-section measurements~\cite{Rodrigues:2020pli, IceCube:2021uhz} (models 2, 4, and 7 below).  Table~\ref{tab:sensitivities} shows results for many more flux models~\cite{Fang:2013vla, Padovani:2015mba, Fang:2017zjf, Heinze:2019jou, Muzio:2019leu, Rodrigues:2020pli, Anker:2020lre, IceCube:2020wum, Muzio:2021zud, IceCube:2021uhz} (Section~\ref{section:uhe_neutrinos}).  The precision of cross-section measurements that we report accounts for the significant uncertainty on the flux normalization (Section~\ref{section:cs_extracting}), and detector resolution (Section~\ref{section:radio_detection}).  For comparison, we show the state-of-the-art BGR18 calculation of the $\nu N$ deep-inelastic-scattering cross section~\cite{Bertone:2018dse} (Section~\ref{section:cs_basics}) on isoscalar matter, averaged between neutrinos and anti-neutrinos.}
\end{figure*}

Figure~\ref{fig:panorama} shows the current landscape of measurements of the $\nu N$ cross section.  At neutrino energies from about 100~MeV to 350~GeV, the cross section is measured precisely using artificial neutrino beams from particle accelerators~\cite{Mukhin:1979bd, Baranov:1978sx, Barish:1978pj, Ciampolillo:1979wp, deGroot:1978feq, Colley:1979rt, Morfin:1981kg, Baker:1982ty, Berge:1987zw, Anikeev:1995dj, Seligman:1997, Tzanov:2005kr, Wu:2007ab, Adamson:2009ju, Nakajima:2010fp, Abe:2013jth, Acciarri:2014isz, Abe:2014nox}.  Soon, the planned accelerator-neutrino experiment FASER$\nu$~\cite{FASER:2019dxq} will reach TeV-scale energies, but not more.  Beyond TeV neutrino energies, there is no existing or planned artificial neutrino beam.  Thus, TeV--PeV cross-section measurements~\cite{IceCube:2017roe, Bustamante:2017xuy, IceCube:2020rnc} used instead the high-energy cosmic neutrinos discovered by the IceCube neutrino telescope~\cite{IceCube:2013cdw, IceCube:2013low, IceCube:2014stg, IceCube:2015qii, IceCube:2016umi, Ahlers:2018fkn, IceCube:2020wum}.  These are the most energetic neutrinos known so far, though not the most energetic neutrinos predicted.  At even higher energies, of 100~PeV and above, the existence of {\it ultra-high-energy} (UHE) cosmic neutrinos was firmly predicted more than fifty years ago~\cite{Greisen:1966jv, Zatsepin:1966jv}.  They represent the only feasible way to extend $\nu N$ cross section measurements to higher energies.  Yet, because their predicted flux is low, they have so far evaded discovery~\cite{IceCube:2018fhm, PierreAuger:2019ens}.

Fortunately, a host of new neutrino telescopes~\cite{Katz:2013svu, Aguilar:2019jay, GRAND:2018iaj, Romero-Wolf:2020pzh, P-ONE:2020ljt, IceCube-Gen2:2020qha, POEMMA:2020ykm, PUEO:2020bnn, Baikal-GVD:2020irv, Hallmann:2021kqk, Schumacher:2021hhm, Bishop:2021cvr,  Krampah:2021ysn, Brown:2021ane, Abraham:2022jse, Ackermann:2022rqc}, designed to discover UHE neutrinos even if their flux is low in the next 10--20 years, may provide a way forward.  For astrophysics, discovering UHE neutrinos would bring critical progress in understanding the long-standing origin of ultra-high-energy cosmic rays~\cite{Ackermann:2019ows, Ackermann:2022rqc}.  For particle physics, discovering UHE neutrinos, in general, would allow access to tests of fundamental physics in a new energy regime and, in particular, would allow us to further cross-section measurements~\cite{Ackermann:2019cxh, Arguelles:2019rbn, Abraham:2022jse, Ackermann:2022rqc}.  

However, detailed and realistic predictions for the capability of upcoming neutrino telescopes to measure the $\nu N$ cross section, considering their design elements, are still lacking; see, however, \Refs~\cite{Huang:2021mki, Huang:2021mki} for important first estimates.  To address this, and in order to capitalize on this upcoming opportunity, we present state-of-the-art forecasts for the measurement of the UHE $\nu N$ cross section.  We gear our forecasts to the promising case of radio-detection of UHE neutrinos in the planned IceCube-Gen2 neutrino telescope~\cite{IceCube-Gen2:2020qha}, one of the leading next-generation detectors under design.

To make our forecasts realistic, we use state-of-the-art ingredients in every stage of the calculation (see Section~\ref{section:synopsis}).  We model the propagation of UHE neutrinos inside the Earth and their radio-detection in detail.  For the latter, we estimate the radio-detection response of the detector, via dedicated simulations of in-medium shower development and radio emission in IceCube-Gen2, and its energy and directional resolution.  To capture the large uncertainty that exists in the prediction of UHE neutrinos, our cross-section forecasts factor in the uncertainty in the size of their flux and a wide variety of shapes of their energy spectrum from the literature~\cite{Fang:2013vla, Padovani:2015mba, Fang:2017zjf, Heinze:2019jou, Muzio:2019leu, Rodrigues:2020pli, Anker:2020lre, IceCube:2020wum, Muzio:2021zud, IceCube:2021uhz}.  

Figure~\ref{fig:panorama} shows that, in optimistic flux scenarios, IceCube-Gen2 may be able to measure the UHE $\nu N$ cross section to within 50\% of its predicted value.  This would be the first measurement of neutrino interactions at center-of-mass energies of $\sqrt{s} \approx 10$--100~TeV, comparable to those of particle collisions at the Large Hadron Collider and the Future Circular Collider.  Measuring neutrino interactions at these energies has potentially transformative consequences.  First, it will test Standard Model predictions of the $\nu N$ cross section~\cite{Bertone:2018dse}.  Second, it will probe non-linear effects in the distribution of quarks and gluons inside nucleons~\cite{Lipatov:1976zz, Kuraev:1976ge, Kuraev:1977fs, Balitsky:1978ic}, the existence of color-glass condensates~\cite{Gelis:2010nm}, and electroweak sphalerons~\cite{Ellis:2016dgb};  see, \eg, \Refs~\cite{Henley:2005ms, Anchordoqui:2006ta, Albacete:2015zra, Ellis:2016dgb}.  And, third, it will probe a large number of new-physics effects that could modify the cross section, including, \eg, leptoquarks, extra dimensions, and new gauge bosons~\cite{Alvarez-Muniz:2001efi, Connolly:2011vc, Chen:2013dza, Marfatia:2015hva, Mack:2019bps, Huang:2021mki}.  

The goal of our forecasts is double.  On the one hand, they are intended to showcase the reach of IceCube-Gen2 to make particle-physics measurements in a new energy regime, in as realistic a way as it is presently possible.  On the other hand, and more generally, our forecasts are intended to stimulate the development of the particle-physics research programs of upcoming high-energy neutrino telescopes.  The calculation framework that we introduce as part of our analysis can be adapted to other neutrino telescopes, and other measurement goals.  Because the design of telescopes that will run in 10--20~years is being decided upon presently, our forecasts are timely.

This paper is organized as follows.  Section~\ref{section:synopsis} showcases the salient points and strengths of our analysis.  Section~\ref{section:cross_section} presents a brief introduction to neutrino-nucleon deep inelastic scattering and outlines the strategy that we use to measure the cross section.  Section~\ref{section:uhe_neutrinos} introduces the various benchmark models of the cosmic UHE neutrino flux that we adopt.  Section~\ref{section:neutrino_propagation} describes the effect of neutrino propagation through Earth and how we compute it.  Section~\ref{section:radio_detection} gives an overview of radio-detection of UHE neutrinos, introduces the response of IceCube-Gen2, and shows how we estimate event rates in it.  Section~\ref{section:cs_extracting} introduces the statistical analysis that we use to forecast cross-section measurements.  Section~\ref{section:results} shows our resulting forecasts for the different benchmark flux models, and the effect on them of changing detector parameters.  Section~\ref{section:limitations} points out potential future research directions. Section~\ref{section:summary} summarizes and concludes.


\section{Synopsis and context}\label{section:synopsis}

\begin{figure*}[t]
 \centering
 \includegraphics[width=\textwidth]{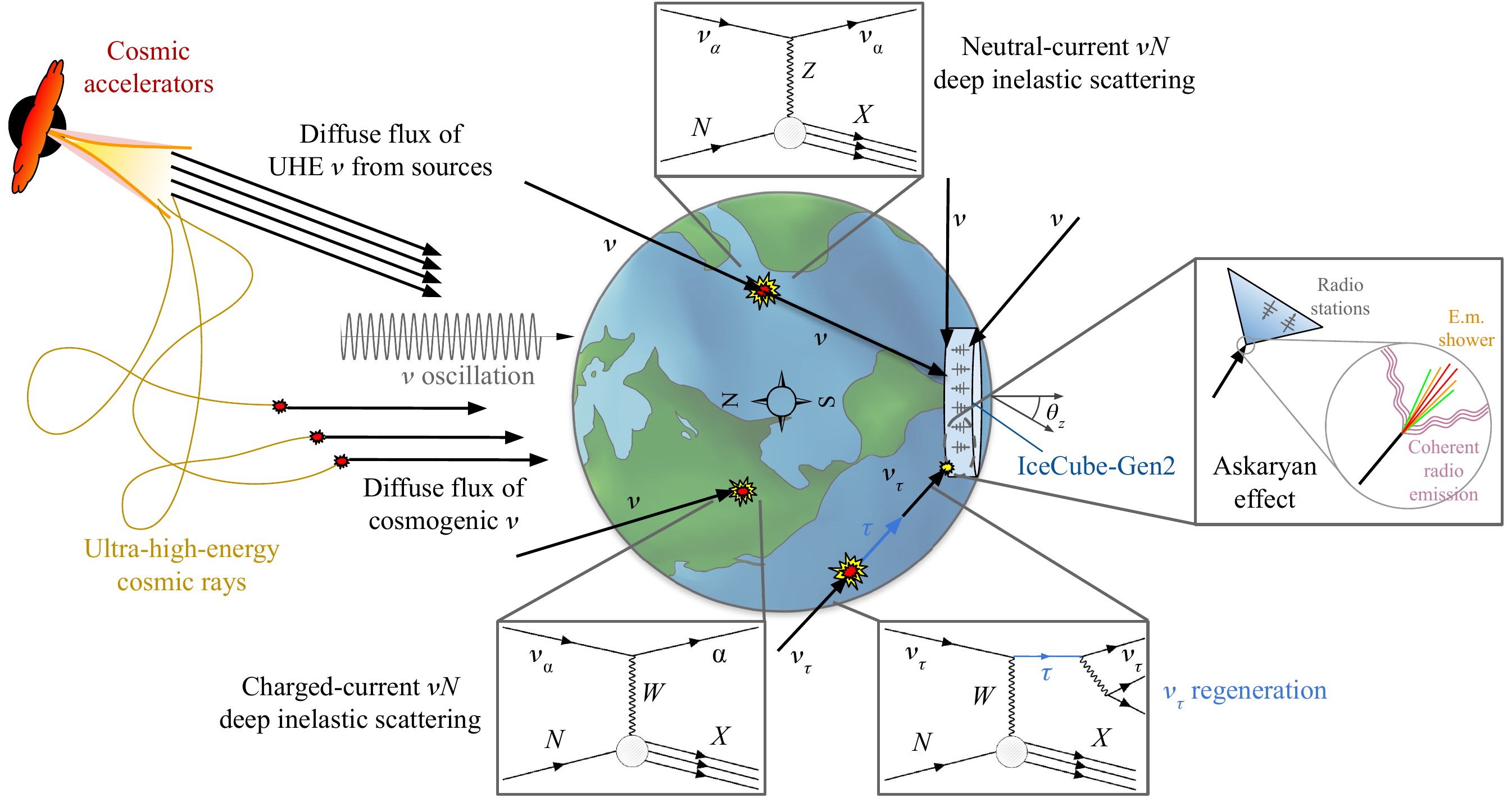}
 \caption{\label{fig:cartoon}Overview of the calculation used to forecast ultra-high-energy (UHE) cross-section measurement capabilities.  A diffuse, isotropic flux of UHE neutrinos, of energies in excess of 100~PeV, cosmogenic or from astrophysical sources, reaches the Earth.  UHE neutrinos propagate through the Earth, across different directions, toward the envisioned IceCube-Gen2 neutrino telescope, located in the South Pole.  Inside Earth, neutrinos interact with matter, mainly via neutrino-nucleon ($\nu N$) deep inelastic scattering.  As a result, their flux is attenuated and shifted to lower energies.  At IceCube-Gen2, neutrinos interact in the ice and induce particle showers that radiate coherent radio emission that may be detected by antennas buried in the ice.  The sensitivity to the $\nu N$ cross section stems from differences in the distribution of neutrino arrival directions, especially of neutrinos coming to the detector horizontally or near-horizontally.  See the main text for details.}
\end{figure*}

Figure~\ref{fig:cartoon} illustrates the flow of calculations involved in producing our forecasts of the measurement of the UHE $\nu N$ cross section.  For a given prediction of the diffuse UHE neutrino flux, we propagate it through the Earth, where neutrinos interact with matter, and model its detection in the radio component of IceCube-Gen2. To make our forecasts realistic, timely, and representative of current unknowns, we use state-of-the-art ingredients at every stage of the calculation.  In brief, these are:
\begin{description}
 \item[UHE $\nu N$ cross section] 
  UHE neutrinos interact with matter while propagating through the Earth, and when detected.  The sensitivity to the $\nu N$ cross section stems from an interplay of both effects; see Section~\ref{section:cs_measurement_overview}.  In our forecasts, we use a state-of-the-art prediction of the UHE cross section~\cite{Bertone:2018dse}, and deviations from it, to compute neutrino propagation and detection.  See Section~\ref{section:cross_section} for details.
 
 \item[Diffuse neutrino flux]
  We have not discovered a flux of UHE neutrinos yet, so we are forced to consider different possibilities.  In our forecasts, we assume a wide variety of UHE neutrino flux predictions that span the breadth available in the literature, from pessimistic to optimistic~\cite{Fang:2013vla, Padovani:2015mba, Fang:2017zjf, Heinze:2019jou, Muzio:2019leu, Rodrigues:2020pli, Anker:2020lre, IceCube:2020wum, Muzio:2021zud, IceCube:2021uhz}.  We model in detail the flavor content of neutrinos and anti-neutrinos in the flux, and keep track of it throughout.  See Section~\ref{section:uhe_neutrinos} for details.
 \item[Neutrino propagation through Earth]
  When propagating UHE neutrinos through the Earth, we compute how neutrino interactions with matter modify the neutrino flux that reaches the detector.  The modifications are energy-, direction-, and flavor-dependent, and are slightly different for neutrinos and anti-neutrinos.  In our forecasts, we account for the dominant neutrino interaction---$\nu N$ deep inelastic scattering---and for other interactions that are collectively non-negligible, via {\sc NuPropEarth}~\cite{Garcia:2020jwr, NuPropEarth}.  See Section~\ref{section:neutrino_propagation} for details.
 \item[Neutrino detection]
  We gear our forecasts to the detection of UHE neutrinos in the radio component of IceCube-Gen2, optimized for neutrino detection above $10^7$~GeV.  In our forecasts, we model the detector geometry, simulate the development of particle showers in the ice, the emission and propagation of radio signals from them, the detector response, including the direction-dependent response of the different antenna types in the array, via {\sc NuRadioMC}~\cite{Glaser:2019cws} and {\sc NuRadioReco}~\cite{Glaser:2019rxw}, the calculation of the trigger condition, and the uncertainties in reconstructing the energy and direction of detected events.  See Section~\ref{section:radio_detection} for details. 
 \item[Non-neutrino backgrounds]
  In the radio-detection of UHE neutrinos, the main backgrounds that may mimic neutrinos are due to showers induced by atmospheric muons in the ice~\cite{Garcia-Fernandez:2020dhb, Glaser:2021hfi} and to showers induced in the atmosphere, mainly by cosmic rays, that penetrate into the ice.  In our forecasts, we model the detection of atmospheric muons and show its effect on our ability to measure the $\nu N$ cross section.  We comment on the tentative effect of showers induced by cosmic rays, whose importance in neutrino radio-detection is still unclear.
\end{description}
Below, we expand on each of the above elements.  

Previous works studied the sensitivity of upcoming neutrino telescopes to the UHE $\nu N$ cross section~\cite{Connolly:2011vc}, or their capability to measure it~\cite{Denton:2020jft, Huang:2021mki}.  They incorporated some of the above elements, often partially or in less detail; none incorporated all of them, or in full detail.  Our analysis is the first detailed, realistic forecast of the capability to measure the UHE $\nu N$ cross section.  It has the flexibility needed to explore how sensitive the measurements are to changes in detector features.  It is geared at IceCube-Gen2, but serves as a template for other neutrino telescopes under planning~\cite{Anker:2020lre, RNO-G:2020rmc, AlvesBatista:2021gzc, Abraham:2022jse, Ackermann:2022rqc}.


\section{Neutrino-nucleon deep inelastic scattering}
\label{section:cross_section}


\subsection{The neutrino-nucleon DIS cross section}
\label{section:cs_basics}

At neutrino energies above 1~TeV, the neutrino-nucleon ($\nu N$) cross section is dominated by deep inelastic scattering (DIS)~\cite{CTEQ:1993hwr, Conrad:1997ne, Giunti:2007ry, Formaggio:2012cpf}, where the interacting neutrino scatters off the partons---\ie, quarks and gluons---that make up the nucleon.  In the process, the nucleon, $N$, is broken up, and the final-state parton promptly hadronizes into hadrons, $X$.  In charged-current (CC) DIS, mediated by a $W$ boson, the final state contains in addition a charged lepton of the same flavor as the incoming neutrino, \ie, $\nu_\alpha + N \to \alpha^- + X$ ($\alpha = e, \mu, \tau$).  In neutral-current (NC) DIS, mediated by a $Z$ boson, the final state contains instead a neutrino, \ie, $\nu_\alpha + N \to \nu_\alpha + X$.  Anti-neutrinos undergo the same DIS interactions, charge-conjugated. 

In a $\nu N$ DIS interaction, a fraction $y$---the inelasticity---of the initial neutrino energy, $E_\nu$, is transferred to $X$; the remainder fraction $(1-y)$ is transferred to the final-state lepton.  The distribution of inelasticity values is important: it affects the passage of neutrinos through the Earth, where they interact with underground matter (see Section~\ref{section:neutrino_propagation}), and their detection at the neutrino telescopes, where they interact with the detector medium (see Section~\ref{section:radio_detection}).  At the energies that are relevant for us, the average inelasticity is about 0.25~\cite{Gandhi:1995tf, Connolly:2011vc}.  However, the distributions of values of $y$, given by the differential $\nu N$ cross sections $d\sigma^{\rm CC}/dy$ and $d\sigma^{\rm NC}/dy$, are broad, peak at $y = 0$, and vary with neutrino energy; see \figu{cs_diff}.  To produce our results below we use these energy-dependent inelasticity distributions.

\begin{figure}[t!]
 \centering
 \includegraphics[width=\columnwidth]{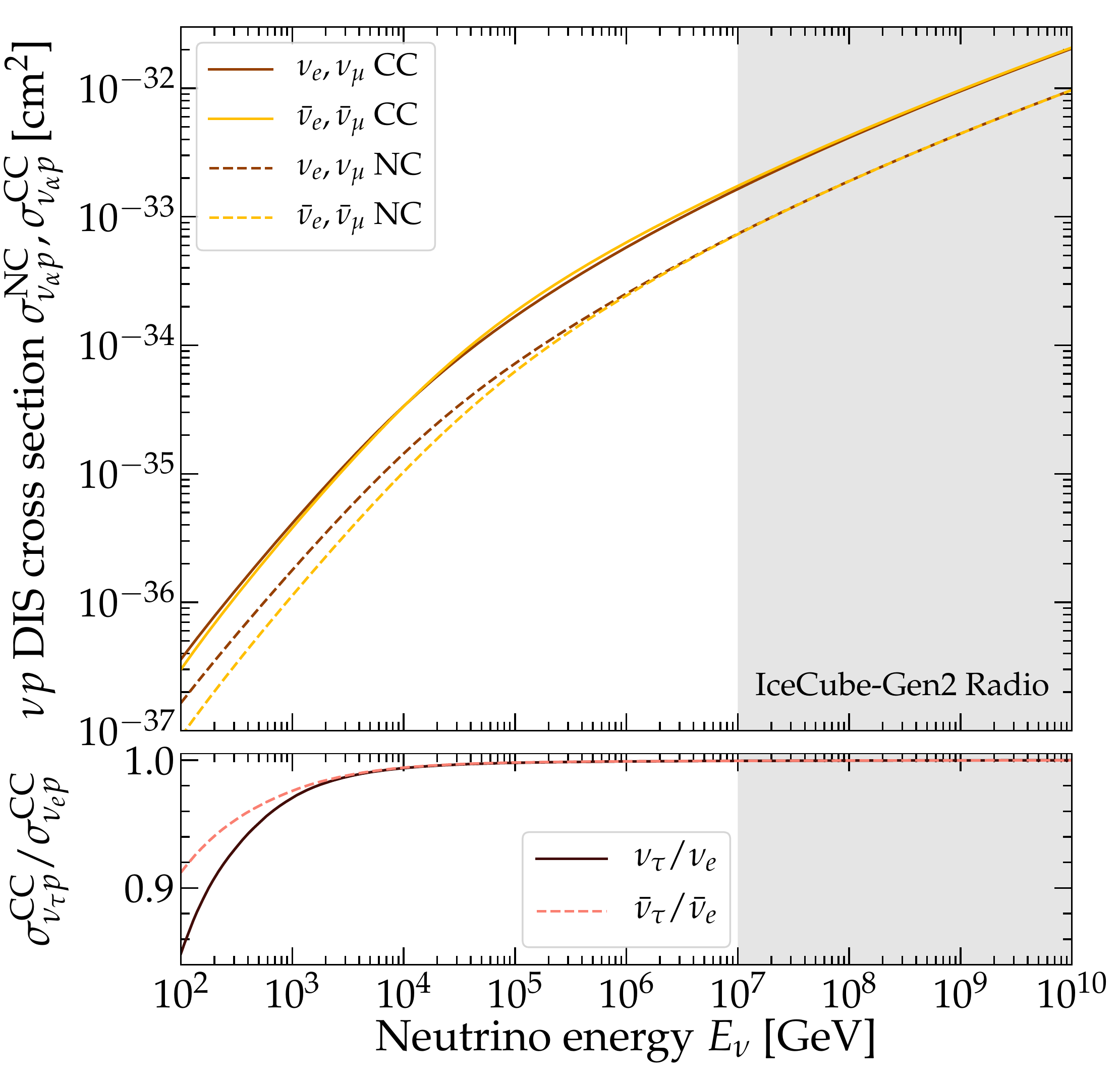}
 \caption{\label{fig:cs}Neutrino-proton ($\nu p$) deep inelastic scattering (DIS) total cross section used in our analysis, during the propagation of neutrinos inside the Earth using {\sc NuPropEarth}~\cite{Garcia:2020jwr, NuPropEarth} and in the computation of event rates at IceCube-Gen2.  See Section~\ref{section:cs_basics} for details.  This is the central value of the BGR18 calculation~\cite{Bertone:2018dse}, extracted from the {\sc HEDIS}~\cite{Garcia:2020jwr} module of {\sc GENIE}~\cite{Andreopoulos:2009rq}.  The shaded region is the approximate energy window where the radio array of IceCube-Gen2 will be sensitive.  {\it Top:} Cross sections for $\nu_e$ and $\nu_\mu$; they are equal.  {\it Bottom:} Ratio of the $\nu_\tau$ to $\nu_e$ CC cross sections showing the low-energy suppression of the former.}
\end{figure}

\begin{figure}[t!]
 \centering
 \includegraphics[width=\columnwidth]{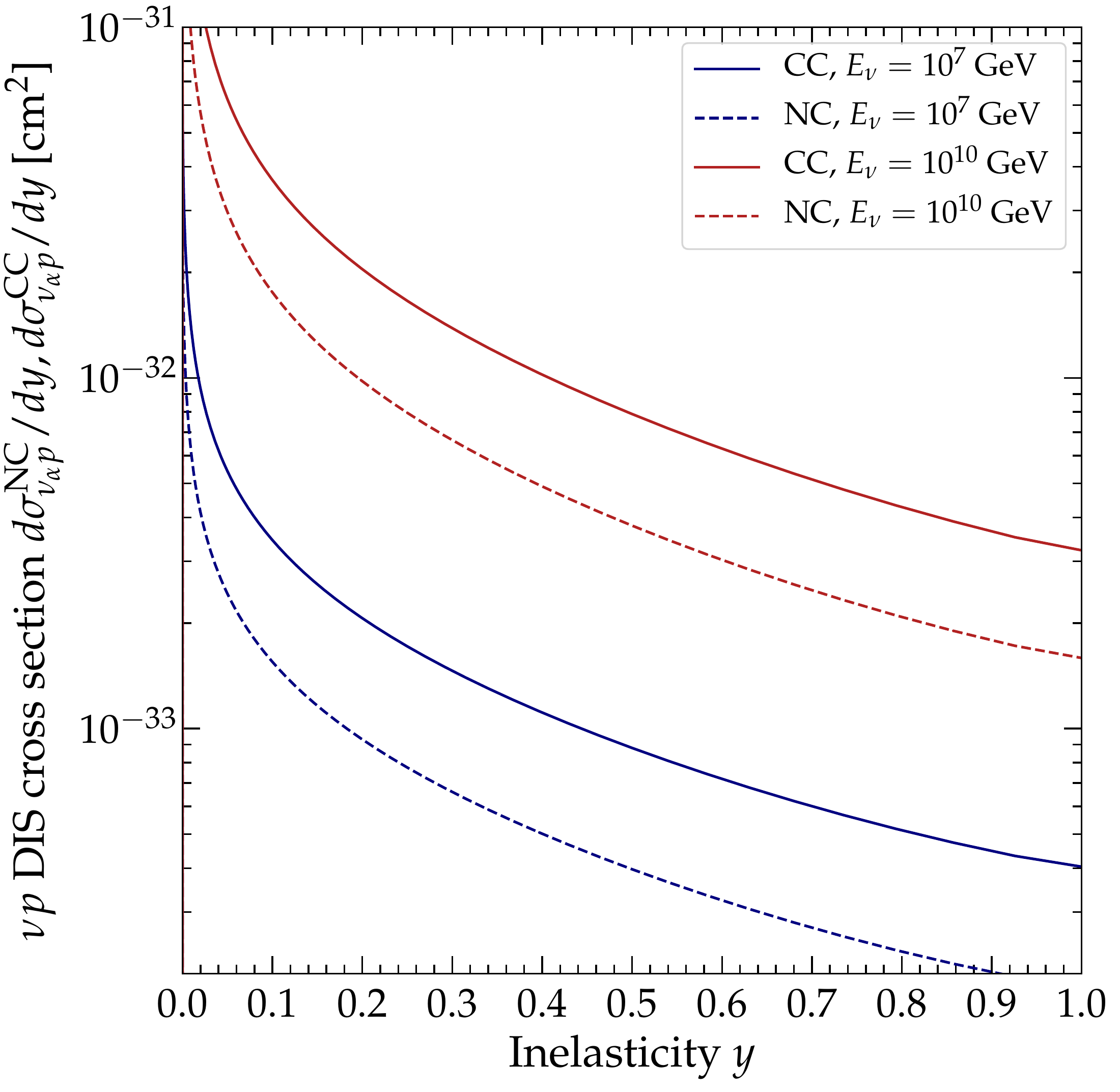}
 \caption{\label{fig:cs_diff}Inelasticity distributions in neutrino-proton ($\nu p$) DIS for the BGR18 cross section~\cite{Bertone:2018dse}, for two illustrative neutrino energies.  We use the inelasticity distributions to compute the expected shower rate at IceCube-Gen2 in Section~\ref{section:radio_detection_rates}.  At the energies relevant to our analysis, the distributions are equal for $\nu_\alpha$ and $\bar{\nu}_\alpha$ of all flavors.  }
\end{figure}

The $\nu N$ DIS cross sections are computed on the basis of the parton distribution functions (PDFs) inside the nucleon, \ie, the probability densities of finding valence and sea quarks of different flavors, and gluons, inside the nucleon~\cite{CTEQ:1993hwr, Conrad:1997ne, Giunti:2007ry, Formaggio:2012cpf}.  They depend on two kinematic variables: the four-momentum transferred to the interacting parton, $Q^2$, and the Bjorken scaling parameter $x$, the fraction of the nucleon moment carried by the interacting parton.  PDFs are measured predominantly in charged lepton-nucleon scattering~\cite{H1:2009pze, H1:2015ubc}; because they describe the nucleon, and not the lepton that probes it, they apply also to neutrino-nucleon scattering.

To compute $\nu N$ DIS cross sections at an energy $E_\nu$, we need to evaluate PDFs at roughly $x \gtrsim m_W/E_\nu$, where $m_W$ is the mass of the $W$ boson.  Presently, PDFs are known down to $x \sim 10^{-5}$.  At TeV--PeV energies, this is sufficient to compute the cross section with low uncertainty.  At EeV energies, relevant to our work, PDFs must be extrapolated to $x \sim 10^{-7}$; this extrapolation is the main source of uncertainty in the calculation of EeV cross sections.  Broadly stated, different calculations of the high-energy $\nu N$ DIS cross section~\cite{Gandhi:1995tf, Gandhi:1998ri, Cooper-Sarkar:2007zsa, Gluck:2010rw, Connolly:2011vc, Block:2014kza, Goncalves:2014woa, Arguelles:2015wba, Albacete:2015zra, Bertone:2018dse} differ in three aspects: the perturbative order to which the cross section is computed, the PDF set that they use, and the procedure they use to extrapolate PDFs to low values of $x$.  Competing calculations are close at TeV--PeV, but diverge at EeV; see, \eg, Fig.~3 in \Refe~\cite{Bustamante:2017xuy}.

In our analysis, we adopt the state-of-the-art BGR18 $\nu N$ DIS cross section calculation from \Refe~\cite{Bertone:2018dse}, computed to next-to-next-to-leading order, as our baseline.  It uses the recent {\sc NNPDF3.1sx} PDFs~\cite{Ball:2017otu}, informed by $D$-meson data from LHCb~\cite{LHCb:2013xam, LHCb:2015swx, LHCb:2016ikn}, including the effect of nuclear corrections and heavy quarks, and treats consistently the behavior of the PDFs at small values of Bjorken-$x$.  The uncertainty in the BGR18 cross section calculation ranges from $\lesssim 3\%$ up to 100~PeV, to $\lesssim 10\%$ at 10~EeV; see \figu{panorama}.  The uncertainty stems from uncertainties in the PDFs~\cite{Garcia:2020jwr}.  Later, in Section~\ref{section:results}, we find that for the most optimistic UHE neutrino flux predictions, we might be able to measure the cross section to within comparable uncertainty; see \figu{panorama}.  (The theory uncertainty in the cross section can be larger in the presence of nuclear corrections~\cite{Bertone:2018dse}, which we ignore for \figu{panorama}, but comment on later.) 

Figure~\ref{fig:cs}, and also \figu{panorama}, shows the central value of the BGR18 CC $\nu N$ DIS cross section (in the notation introduced later, in Section~\ref{section:cs_extracting_overview}, this is $\sigma_{\rm std}$).  The cross section grows with $E_\nu$, and the softer-than-linear dependence on $E_\nu$ due to the $W$ mass is apparent from a few TeV on.  At EeV energies, the cross section grows roughly as $\propto E_\nu^{0.3}$~\cite{Gandhi:1995tf}.  Within the energy window where IceCube-Gen2 will be sensitive, the cross sections are flavor-universal, yet, at low energies, \figu{cs} shows that the $\nu_\tau$ CC cross section is kinematically suppressed due to the large mass of the tauon. 

Figure~\ref{fig:cs_diff} shows the corresponding inelasticity distributions, for NC and CC interactions.  Later, in Section~\ref{section:radio_detection_rates}, we use them to compute the expected event rate at IceCube-Gen2.  The distributions peak at $y = 0$, but they are broad, \ie, there is a large spread in how the incoming neutrino energy in a $\nu N$ DIS is split into the final-state particles.  This, in turn, generates a large spread in the energies of the neutrino-initiated showers at the detector, and on their detectability.

To make our forecasts self-consistent, we use the same BGR18 cross sections to compute the propagation of neutrinos inside Earth and their detection.  At the neutrino energies relevant to us, $E_\nu \gtrsim 10^7$~GeV, the CC cross section is roughly twice the NC cross section, and differences between the cross sections of $\nu_\alpha$ and $\bar{\nu}_\alpha$ and between different flavors are small.  Below 100~TeV, differences are larger, but those energies do not come into play in our analysis.  We always use the CC and NC cross section of each flavor of $\nu_\alpha$ and $\bar{\nu}_\alpha$ individually.  Below, as part of our forecasts, we use the central value of the BGR18 cross section as a baseline ($\sigma_{\rm std}$), and also versions of it shifted up ($> \sigma_{\rm std}$) and down ($< \sigma_{\rm std}$) by constant factors.


\subsection{Other neutrino interactions}
\label{section:nu_other_int}

At ultra-high energies, $\nu N$ DIS is the dominant neutrino interaction.  However, other sub-dominant interactions also affect the propagation of neutrinos inside the Earth.  During propagation, we account for the following interactions, as implemented in {\sc NuPropEarth}~\cite{Garcia:2020jwr, NuPropEarth}; we defer to \Refe~\cite{Garcia:2020jwr} for details:
\begin{itemize}
 \item 
  {\bf $\nu N$ DIS on the partons of the nucleon:} The dominant interaction channel at ultra-high energies; see Section~\ref{section:cs_basics}.
 \item
  {\bf Neutrino DIS on the photon field of the nucleon:} The neutrino interacts with a lepton generated by the photon field of the nucleon.  This is negligible except when the neutrino can produce an on-shell $W$ boson that enhances the cross section resonantly, \ie, when $\sqrt{2 m_N E_\nu} \gtrsim m_W$, or $E_\nu \gtrsim 3~{\rm TeV}$.  It can account for a correction of up to 3\% of the total DIS cross section~\cite{Seckel:1997kk, Alikhanov:2015kla, Gauld:2019pgt, Zhou:2019vxt, Zhou:2019frk}.
 \item
  {\bf Coherent neutrino-nucleus scattering:}  The neutrino interacts coherently with the photon field of the target nucleus~\cite{Czyz:1964zz, Lovseth:1971vv, Ballett:2018uuc}.  The cross section is $\propto Z^2$, where $Z$ is the atomic number of the nucleus; thus, it matters mostly for heavy nuclei. This process is important only in scatterings with small transferred momentum, $Q \lesssim 1$~GeV, where it can contribute up to 10\% of the total cross section.
 \item 
  {\bf Elastic and diffractive neutrino scattering on nucleons:}  The neutrino interacts with the photon field of individual nucleons.  This process is important in scattering with small $Q \sim {\rm GeV}$, where the elastic component of the form factors of the proton become relevant.  When $E_\nu \gtrsim 3~{\rm TeV}$, the process may create an on-shell $W$ boson resonantly, contributing to the resonant cross section of the neutrino DIS on the photon field of the nucleon.
 \item
  {\bf Neutrino scattering on atomic electrons:}  Neutrino scattering on atomic electrons is negligible except for high-energy $\bar{\nu}_e$.  When the center-of-mass energy is $\sqrt{2 m_e E_\nu} \approx m_W$, \ie, when $E_\nu \approx 6.3~{\rm PeV}$, the $\bar{\nu}_e e$ scattering is resonance and produces an on-shell $W$; this is known as the Glashow resonance~\cite{Glashow:1960zz, IceCube:2021rpz}.  
  Around the resonance energy, the $\bar{\nu}_e e$ cross section dominates; it is roughly 200 larger than the $\nu N$ DIS cross section.  We adopt the Glashow-resonance cross section from \Refe~\cite{Gauld:2019pgt}, computed to next-to-leading order.
\end{itemize}
The sub-dominant interactions increase the attenuation of the UHE neutrino fluxes by up to 10\%, when compared to $\nu N$ DIS only, especially for neutrinos coming into the detector from around the horizon~\cite{Garcia:2020jwr}.  Below, when computing the propagation of neutrinos through the Earth and their resulting fluxes at the detector, we always account for all of the above interactions; see Section~\ref{section:neutrino_propagation}.  In all of the interactions, because of the high energies, final-state neutrinos are nearly co-linear with initial-state neutrinos; any transverse momentum is negligible compared to the forward momentum, and we ignore it.   


\subsection{High-energy $\nu N$ DIS using cosmic neutrinos}
\label{section:cs_measurement}


\subsubsection{Motivation}
\label{section:cs_measurement_motivation}

Measuring the high-energy $\nu N$ DIS cross section offers the possibility to probe fundamental physics on two fronts.  First, the higher the energy of the neutrino, the smaller the value of $x$ that it probes; see Section~\ref{section:cs_basics}.  This allows us to probe the structure of nucleons deeper, testing predictions of potentially non-linear behavior of the PDFs, such as from BFKL theory~\cite{Lipatov:1976zz, Kuraev:1976ge, Kuraev:1977fs, Balitsky:1978ic} and color-glass condensates~\cite{Gelis:2010nm}, and to look for electroweak sphalerons~\cite{Ellis:2016dgb}; see, \eg, \Refs~\cite{Henley:2005ms, Anchordoqui:2006ta, Albacete:2015zra, Ellis:2016dgb}.  Second, the higher the neutrino energy, the higher the energy scale probed where new physics could affect the cross section.  Possibilities include, \eg, leptoquarks, extra dimensions, and new gauge bosons~\cite{Alvarez-Muniz:2001efi, Connolly:2011vc, Chen:2013dza, Marfatia:2015hva, Mack:2019bps, Huang:2021mki}.  

Existing measurements of the $\nu N$ DIS cross sections using accelerator neutrinos reach 350~GeV~\cite{Tzanov:2005kr}; see \figu{panorama}.  Upcoming accelerator-neutrino experiments, like FASER$\nu$~\cite{FASER:2019dxq}, should measure the cross section up to a few TeV.  To measure it at higher energies, we must use neutrinos from natural sources: atmospheric neutrinos, from tens of TeV to roughly 100~TeV; high-energy cosmic neutrinos from 10~TeV to 10~PeV; and ultra-high-energy neutrinos, from 100~PeV on.  Measurements using the former two exist~\cite{IceCube:2017roe, Bustamante:2017xuy, IceCube:2020rnc}; we forecast measurements using the latter, where no measurement exists yet.


\subsubsection{Overview}
\label{section:cs_measurement_overview}

We base our forecasts of cross-section measurements on estimates of the expected number of detected neutrino-induced events in IceCube-Gen2.  To understand where the sensitivity to the cross section comes from, we estimate the number of  detected neutrinos of energy $E_\nu$ coming into the detector from zenith angle $\theta_z$ as
\begin{equation}
 \label{equ:event_rate_simple}
 N_\nu(E_\nu, \theta_z) 
 \propto 
 \Phi_\nu(E_\nu) 
 \sigma(E_\nu)
 e^{-L(\theta_z)/L_{\nu N}(E_\nu, \theta_z)} \;.
\end{equation}
Here, $\Phi_\nu$ is the diffuse, isotropic flux of UHE cosmic neutrinos that arrives at the surface of the Earth, $\sigma$ is the $\nu N$ cross section, $L$ is distance from the surface of the Earth to the detector, $L_{\nu N} \equiv (\sigma n_N)^{-1}$ is the mean free path, and $n_N$ is the average number density of nucleons encountered by the neutrino along its way inside Earth.  The latter depends on $\theta_z$ and on the internal matter density of Earth. Equation~(\ref{equ:event_rate_simple}) is merely a simplified expression for the purpose of providing insight, and is not used to produce our results.  The full treatment of neutrino propagation and detection that we use to produce results is in Sections~\ref{section:neutrino_propagation} and \ref{section:radio_detection}, respectively.

Equation~(\ref{equ:event_rate_simple}) shows that the number of events depends on the $\nu N$ cross section doubly.  During propagation, the cross section acts via the exponential, attenuating the flux of neutrinos as they go through Earth (in the full calculation, there is also regeneration of lower-energy neutrinos, which we ignore momentarily).  Higher neutrino energies---and, therefore, larger cross section---and longer distances traveled inside Earth lead to stronger attenuation.  At detection, the cross section acts proportionally; the larger it is, the higher the chances of detecting a neutrino that arrives at the detector.  The sensitivity of neutrino telescopes to the high-energy $\nu N$ DIS cross section stems from the interplay of these two effects.

We extract the cross section by examining the angular distribution of detected events~\cite{Hooper:2002yq, Hussain:2006wg, Borriello:2007cs, Hussain:2007ba, Connolly:2011vc, IceCube:2017roe, Bustamante:2017xuy, IceCube:2020rnc}.  For events induced by neutrinos arriving from above the detector, \ie, {\it downgoing events}, where $L$ is small, the attenuation is negligible.  In this case, the right-hand side of \equ{event_rate_simple} becomes $\propto \Phi_\nu \sigma$ and, therefore, the sensitivity to the cross section is mild.  This is due to the degeneracy between $\Phi_\nu$ and $\sigma$: for a fixed event rate $N_\nu$, a higher flux can be traded off for a lower cross section, and vice versa.  For events induced by neutrinos arriving from well below the horizon, \ie, {\it upgoing events}, where $L$ is large, the attenuation is strong.  In this case, the event rate is low: the higher the neutrino energy---\ie, the higher the cross section---the stronger the attenuation.

For events induced by neutrinos arriving horizontally and nearly horizontally into the detector, the two effects above balance out.  Neutrinos from this direction travel tens to hundreds of kilometers underground, enough for the flux to be attenuated, but not eliminated.  The higher the neutrino energy, the narrower the solid angle around the horizon from where neutrinos arrive to the detector.  At ultra-high energies, neutrinos can only arrive at the detector from a few degrees around the horizon, where the distance traveled underground is not too long (see, \eg, Fig.~A2 in \Refe~\cite{Bustamante:2017xuy}); these are called {\it Earth-skimming neutrinos}~\cite{Fargion:2000iz}.

Combining events from all directions breaks the degeneracy between $\Phi_\nu$ and $\sigma$ in downgoing events, and, via the attenuation of upgoing and near-horizontal events, grants us sensitivity to $\sigma$~\cite{Hooper:2002yq, Hussain:2006wg, Borriello:2007cs, Hussain:2007ba, Connolly:2011vc}.  Yet, to make use of this, the detector requires a good angular resolution around the horizon.  This is necessary to infer the direction of the incoming neutrino precisely, and, therefore, the column density of matter that it traversed on its way to the detector.  In IceCube, at TeV--PeV energies, the angular resolution varies from sub-degree for $\nu_\mu$-induced track events, to tens of degrees for shower events induced by all neutrino flavors.  In our forecasts, geared at radio-detection of EeV-scale neutrinos in IceCube-Gen2, we adopt a baseline resolution of $2^\circ$ to produce our main results, and also explore the effect of alternative choices; see Section~\ref{section:radio_detection} and \figu{angular_resolution} for details. 


\subsubsection{Existing measurements: TeV--PeV}

Figure~\ref{fig:panorama} shows the three existing measurements of the TeV--PeV $\nu N$ DIS cross section, all based on IceCube data~\cite{IceCube:2017roe, Bustamante:2017xuy, IceCube:2020rnc}.  They used different analysis strategies and data sets; they agree with Standard Model (SM) predictions, though the measurement uncertainties are large.  In addition to the TeV--PeV measurements, there are complementary studies to measure the cross section in IceCube from 100~GeV to a few TeV~\cite{IceCube:2021jhz}.

Reference~\cite{IceCube:2017roe} used roughly 10800 through-going neutrino-induced muon tracks, predominantly of atmospheric origin, to measure the cross section in a single neutrino energy bin spanning 6.30--980~TeV.  In a through-going track event, a $\nu_\mu$ interacts at an unknown position outside the detector and creates a high-energy muon that crosses part of it.  This analysis let the normalization of the CC and NC cross sections float, fit it to the data, and found the cross section to be $1.30_{-0.26}^{+0.30}$ times the SM prediction from \Refe~\cite{Cooper-Sarkar:2011jtt}.  The error is approximately equal parts statistical and systematic; the latter is mainly due to the difficulty in reconstructing the neutrino energy from the measured muon energy.  Improvements on this measurement strategy are ongoing~\cite{IceCube:2021keu}.

References~\cite{Bustamante:2017xuy, IceCube:2020rnc} used instead High Energy Starting Events (HESE), predominantly of cosmic origin, to measure the cross section from 20~TeV to a few PeV.  Unlike a through-going track, in a HESE event the neutrino interacts inside the detector.  The ensuing shower deposits a large fraction of its energy in the instrumented volume.  As a result, the neutrino energy is reconstructed accurately, which facilitates measuring the cross section in multiple energy bins.  However, because HESE events are relatively rare, these analyses are limited by low event rates.  The CC interaction of $\nu_e$ or $\nu_\tau$, or the NC interaction of a neutrino of any flavor, induces a shower that is typically contained in the detector.  The NC interaction of $\nu_\mu$ induces in addition a muon track that starts inside the detector, but ranges out of it.  Reference~\cite{Bustamante:2017xuy} used 58 HESE showers collected in 6 years to measure the cross section in bins in the range 18~TeV--2~PeV, with an accuracy of roughly half an order of magnitude in each bin.  Reference~\cite{IceCube:2020rnc} used 60 HESE showers and tracks collected in 7.5 years to measure the cross section in bins in the range 60~TeV--10~PeV with increased accuracy, thanks to reduced detector systematics.  A related analysis~\cite{IceCube:2018pgc} used contained events to make the first measurement of the multi-TeV inelasticity distribution.


\subsubsection{This work: forecasts at EeV}
\label{section:sensitivity}

To produce forecasts of UHE cross-section measurements in IceCube-Gen2, we adopt the BGR18 cross section as the baseline, and explore measurement prospects for a variety of diffuse UHE flux models.  Our procedure is reminiscent of analyses that use HESE events~\cite{IceCube:2017roe, Bustamante:2017xuy}.  Section~\ref{section:cs_extracting} describes it in detail.  In our forecasts, variations in the cross section affect the neutrino propagation inside Earth and their detection.  To account for the degeneracy between the flux and the cross section, we always forecast measuring both simultaneously.  We adopt a Bayesian approach in our statistical analysis, and account for statistical fluctuations in the number of neutrino-induced events and non-neutrino backgrounds.  


\section{Ultra-high-energy neutrinos}
\label{section:uhe_neutrinos}

\begin{figure}[t!]
 \centering
 \includegraphics[width=\columnwidth]{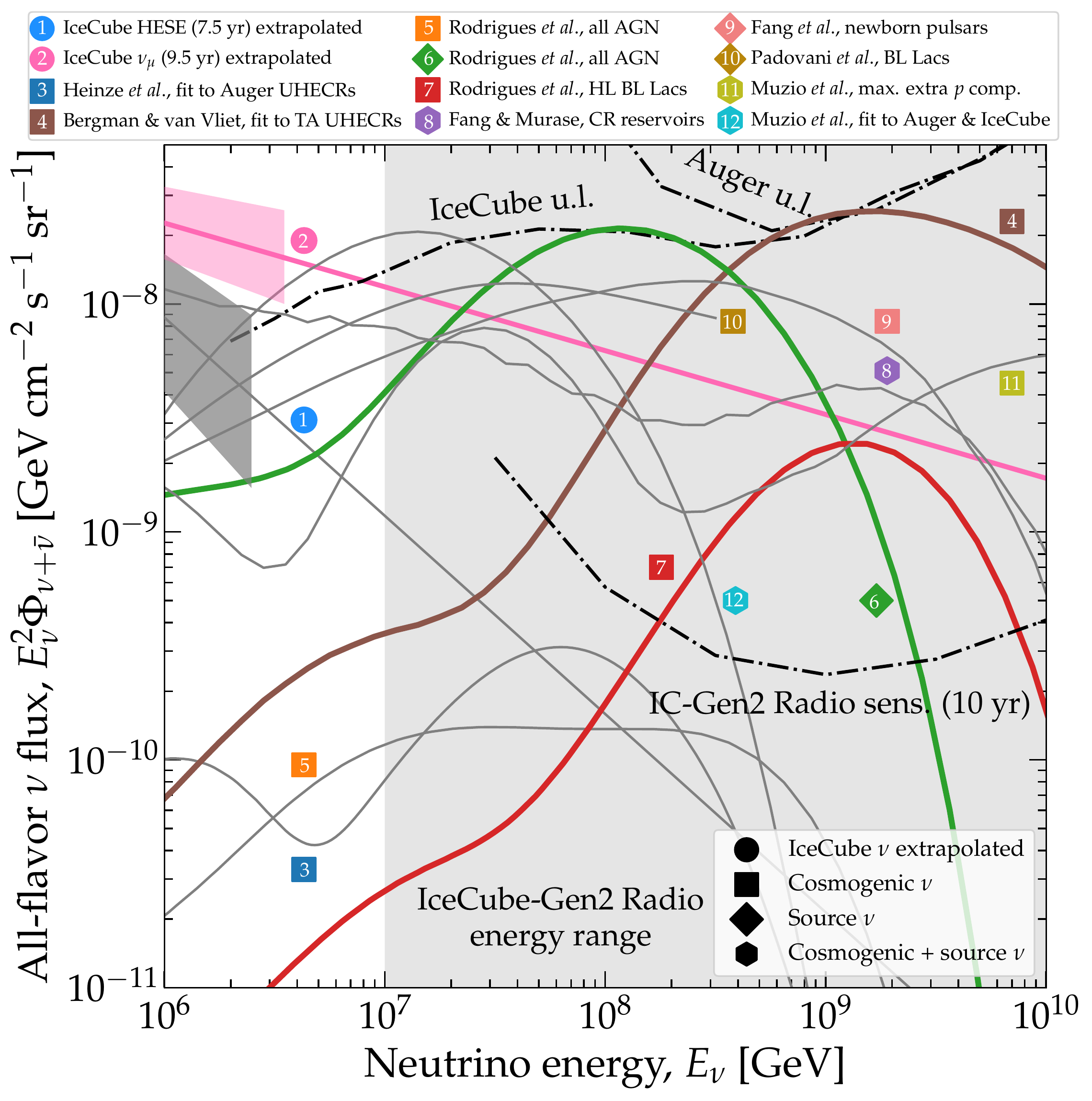}
 \caption{\label{fig:fluxes}Benchmark UHE neutrino flux models used in our analysis~\cite{Fang:2013vla, Padovani:2015mba, Fang:2017zjf, Heinze:2019jou, Muzio:2019leu, Rodrigues:2020pli, Anker:2020lre, IceCube:2020wum, Muzio:2021zud, IceCube:2021uhz}.  Highlighted models 2, 4, 6, and 7 receive special attention in the main text, but results for all models are shown in Tables~\ref{tab:event_rates_total} and \ref{tab:sensitivities}, and in Appendix~\ref{section:appendix_posteriors}.  The upper limits on the UHE neutrino flux are from IceCube~\cite{IceCube:2018fhm} and the Pierre Auger Observatory~\cite{PierreAuger:2019ens}. The projected sensitivity of the radio component of IceCube-Gen2 is from \Refe~\cite{Hallmann:2021kqk}. See Section~\ref{section:uhe_neutrinos} for an overview of the flux models.}
\end{figure}

Ultra-high-energy neutrinos, with energies of 100~PeV and above, are expected to come from interactions of ultra-high-energy cosmic rays (UHECRs), of EeV-scale energies, on radiation or matter.  UHECR interactions may occur inside the cosmic accelerators that are their sources or outside them, en route to Earth.  In the former case, the resulting UHE neutrinos are dubbed {\it source neutrinos} (or {\it astrophysical neutrinos}); in the latter, they are dubbed {\it cosmogenic neutrinos}.  Because of  unknowns in the properties of UHECRs and their sources, there is a large spread in the predicted neutrino flux normalization and the shape of the neutrino energy spectrum.  In our analysis, we consider a wide breadth of benchmark flux models from the literature in order to represent this spread.  Below, we introduce our benchmark flux models, and the choices that we make in building them.


\subsection{Overview}
\label{section:uhe_neutrinos_overview}

Cosmic accelerators are expected to generate a population of non-thermal UHECRs with a power-law energy spectrum~\cite{Bell:2013vxa}.  The interaction of UHECR protons on  matter ($pp$) or radiation ($p\gamma$) often creates a short-lived $\Delta(1232)$ resonance that promptly decays into charged pions.  Upon decaying, they create neutrinos, \ie, $\pi^+ \to \mu^+ + \nu_\mu$, followed by $\mu^+ \to e^+ + \nu_e + \bar{\nu}_\mu$, and their charge-conjugated processes.  Each final-state neutrino receives 5\% of the energy of the parent proton, on average.  En route to Earth, neutrino oscillations change the flavor composition of the flux, \ie, the proportion of $\nu_e$, $\nu_\mu$, and $\nu_\tau$ in it; we account for this in Section~\ref{section:uhe_neutrinos_flavor_composition}.

For UHE neutrinos produced in $pp$ interactions, the resulting neutrino energy spectrum follows the power-law energy spectrum of the parent protons, and may extend unbroken down to low energies~\cite{Fang:2017zjf}. The spectral index of the neutrino spectrum is inherited from the spectral index of the parent proton spectrum.

For UHE neutrinos produced in photohadronic, \ie, $p\gamma$, interactions, the resulting neutrino energy spectrum is determined by the energy spectra of the interacting protons and photons.  Because the proton spectrum is a power law and the photon spectrum is peaked or is a power law with a spectral break, the resulting neutrino spectrum is peaked.  The neutrino spectrum peaks at an energy determined by the $\Delta$ resonance energy; the width of the neutrino peak is determined by the widths of the photon and proton spectra. 

Cosmogenic neutrinos, or {\it GZK (Greisen-Zatsepin-Kuzmin) neutrinos}, were first predicted in the late 1960s~\cite{Berezinsky:1969erk, Greisen:1966jv, Zatsepin:1966jv}.  They are expected to be made during the extragalactic propagation of UHECRs, in photohadronic interactions on the cosmic microwave background (CMB), for neutrinos of energies typically in the EeV-scale, and on the extragalactic background light (EBL), for neutrino of energies typically in the tens of PeV.  (Cosmogenic anti-neutrinos have an additional contribution from the beta-decay of neutrons produced in photohadronic interactions, typically around PeV energies, outside the region of interest for neutrino radio-detection in IceCube-Gen2.)  Because the CMB photon spectrum is well-known, the uncertainty in the prediction of the cosmogenic neutrino flux comes mainly from uncertainties in the energy spectrum, mass composition, and maximum energies of UHECRs, as measured by the Pierre Auger Observatory~\cite{PierreAuger:2017tlx, PierreAuger:2020qqz, PierreAuger:2020kuy} and the Telescope Array (TA)~\cite{TelescopeArray:2015dcv, TelescopeArray:2018bep}, and in the abundance of the UHECR sources at different redshifts.  See, \eg, \Refs~\cite{AlvesBatista:2019tlv, Ackermann:2022rqc} for an overview.  Generally, a harder UHECR energy spectrum, lighter mass composition, higher maximum energy, and a source number density that peaks at intermediate redshifts lead to a higher cosmogenic neutrino flux; see, \eg, \Refs~\cite{Kotera:2010yn, Romero-Wolf:2017xqe, AlvesBatista:2018zui, Heinze:2019jou}. 

UHE source neutrinos are expected to be made in either $pp$ interactions, $p\gamma$ interactions, or both, inside UHECR sources.  When photohadronic interactions are dominant, the spectrum of UHE source neutrinos has a similar shape to that of cosmogenic neutrinos, except that it contains a single $p\gamma$ bump, since there is typically a single relevant target spectrum of photons inside the sources.  (UHE source anti-neutrinos also have an additional contribution from the beta-decay of neutrons produced in photohadronic interactions, typically at energies too low to be relevant for our analysis.)  In some models of UHE neutrino production in cosmic-ray reservoirs~\cite{Fang:2017zjf}, the contribution of neutrinos from $pp$ interactions extends to low energies, and the contribution of $p\gamma$ interactions is dominant at high energies.  

In realistic models of high-energy neutrino production, including some of the ones that we consider in our analysis, different neutrino production channels become accessible at different energies.  In $p\gamma$ interactions~\cite{Mucke:1999yb, Hummer:2010vx, Morejon:2019pfu}, neutrino production occurs dominantly via the $\Delta(1232)$ resonance at the lowest energies, with a sub-leading contribution from direct ($t$-channel) production, via heavier resonances at intermediate energies, and via multi-pion production at the highest energies.  In $pp$ interactions~\cite{Kelner:2006tc}, the neutrino yield evolves with energy as a result of the evolving pion multiplicity.  Moreover, the physical conditions in the production region affect the energy of charged particles---protons, muons, pions, and kaons---whose decay yields neutrinos.  For instance, intense magnetic fields may induce important synchrotron energy losses~\cite{Waxman:1997ti, Waxman:1998yy, Winter:2013cla, Bustamante:2020bxp} that cap the high-energy neutrino yield, while re-acceleration of charged particles might counteract these losses~\cite{Winter:2014tta}.  Further, the presence of nuclei heavier than protons, and the nuclear cascades initiated by their interactions with source environments, introduce additional nuance~\cite{Boncioli:2016lkt, Biehl:2017zlw}.


\subsection{Flavor and $\nu$ {\it vs.}~$\bar{\nu}$ composition in our analysis}
\label{section:uhe_neutrinos_flavor_composition}

Because, at different energies, different neutrino production channels dominate (see Section~\ref{section:uhe_neutrinos_overview}) and the physical conditions at the sources affect charged particles differently, the  flavor composition of the UHE cosmic neutrinos, \ie, the proportion of $\nu_e + \bar{\nu}_e$, $\nu_\mu + \bar{\nu}_\mu$, and $\nu_\tau + \bar{\nu}_\tau$ in the total flux, and the ratio of $\nu_\alpha$ to $\bar{\nu}_\alpha$, are expected to evolve with neutrino energy.  This matters for the purpose of propagating neutrinos through the Earth, on their way to the detector, and of forecasting their detection rates.  Section~\ref{section:neutrino_propagation} shows that neutrinos of different flavor are affected differently by their passage through Earth.  Differences between $\nu_\alpha$ {\it vs.}~$\bar{\nu}_\alpha$ are small, though we keep track of them.  The exception where differences are significant is the case of $\nu_e$ {\it vs.}~$\bar{\nu}_e$, since only $\bar{\nu}_e$ interact via the Glashow resonance~\cite{Glashow:1960zz}.  Section~\ref{section:radio_detection} shows that neutrinos of different flavors have different interaction rates and deposit energy differently at IceCube-Gen2; there, differences between $\nu_\alpha$ {\it vs.}~$\bar{\nu}_\alpha$ are also small, though we keep track of them.

In Section~\ref{section:uhe_neutrinos_benchmarks}, we introduce the benchmark UHE neutrino flux models that we later use to forecast $\nu N$ cross-section measurements in IceCube-Gen2.  In order to make our predictions as informed as possible, we model the flavor composition and $\nu_\alpha$ {\it vs.}~$\bar{\nu}_\alpha$ content of the benchmark fluxes as accurately as possible.  In doing so, neutrino flavor transitions play a key role.  Below, we explain how we compute them.

Because a neutrino of a particular flavor, $\nu_\alpha$ ($\alpha = e, \mu, \tau)$, is a superposition of neutrino mass eigenstates, $\nu_i$ ($i = 1, 2, 3$), it can change flavor as it propagates.  The flavor and mass bases are connected by the Pontecorvo-Maki-Nakagawa-Sakata (PMNS) mixing matrix, $\mathbf{U}$, parametrized~\cite{ParticleDataGroup:2020ssz} via three mixing angles, $\theta_{12}$, $\theta_{23}$, $\theta_{13}$, and one CP-violation phase, $\delta_{\rm CP}$, whose values are measured in neutrino oscillation experiments.

Formally, the probability $P_{\alpha\beta}$ of a flavor transition $\nu_\alpha \to \nu_\beta$ oscillates as a function of the distance traveled by the neutrino.  However, for high-energy cosmic neutrinos, the oscillation length, which is $\propto 1/E_\nu$, is tiny compared to the typical traveled distance of Mpc--Gpc, so the probability oscillates rapidly.  In addition, neutrino telescopes have limited energy resolution~\cite{IceCube:2013dkx}.  As a consequence, in practice, oscillations average out, and we are sensitive only to the average probability~\cite{Pakvasa:2008nx},
\begin{equation}
 P_{\beta\alpha}
 =
 \sum_{i=1,2,3}
 \lvert U_{\beta i} \rvert^2 \lvert U_{\alpha i} \rvert^2 \;.
\end{equation}
Below, we compute the flux of each neutrino flavor at Earth for our benchmark flux models by evaluating $P_{\beta\alpha}$ using values of the mixing parameters forecast for the 2030s~\cite{Song:2020nfh}, anchored on present measurements~\cite{Esteban:2020cvm, NuFit_5.0}.

For this purpose, our benchmark flux models fall into three categories, depending on what information is available to us to build the model with.  For each, we compute the flux of each neutrino species at Earth differently:  
\begin{enumerate}[(a)]
 \item
  {\it Flux models for which we have available the pre-oscillation flux of each neutrino species  separately (models 3--7 below).}  In this case, we compute the flux of $\nu_\alpha$ at Earth, $\Phi_{\nu_\alpha}$, from the pre-oscillation fluxes that we have available, $\Phi_{\nu_\beta, {\rm S}}$, as
  \begin{equation}
   \label{equ:flux_earth_cat_a}
   \Phi_{\nu_\alpha}(E_\nu)
   =
   \sum_{\beta = e, \mu, \tau}
   P_{\beta \alpha}
   \Phi_{\nu_\beta, {\rm S}}(E_\nu) \;
  \end{equation}
  and similarly for the flux $\Phi_{\bar{\nu}_\alpha}$ of $\bar{\nu}_\alpha$, but changing $\Phi_{\nu_\beta, {\rm S}} \to \Phi_{\bar{\nu}_\beta, {\rm S}}$.  Because in all of our benchmark flux models neutrinos are produced by pion, kaon, and neutron decays, only $\nu_e$, $\bar{\nu}_e$, $\nu_\mu$, and $\bar{\nu}_\mu$ exist pre-oscillation; however, after oscillations, all six species are populated in the flux at Earth.
 \item
  {\it Flux models for which we only have available the sum of the oscillated fluxes of all neutrino species at Earth (models 1, 8--11 below).}   In this case, we consider the flavor composition to be energy-independent and split the flux of each flavor evenly between $\nu_\alpha$ to $\bar{\nu}_\alpha$ at all energies.  The latter assumption holds approximately, but can have large deviations at high energy, depending on model-dependent details of the neutrino production; see, \eg, \Refe~\cite{Hummer:2010vx} and \figu{benchmark_spectra_per_species} below.  To estimate the flavor composition, we assume that all neutrinos are made in pion decays, \ie, $\pi^+ \to \mu^+ + \nu_\mu$, followed by $\mu^+ \to e^+ + \nu_e + \bar{\nu}_\mu$, and their charge-conjugated processes.  Hence, pre-oscillation, the flavor composition is $\left(f_{e, {\rm S}}^\pi, f_{\mu, {\rm S}}^\pi, f_{\tau, {\rm S}}^\pi \right) \equiv \left(\frac{1}{3}, \frac{2}{3}, 0 \right)$, where $f_{\beta, {\rm S}}^\pi$ is the ratio of $\nu_\beta + \bar{\nu}_\beta$ to the total.  After oscillations, at Earth, the flavor ratios become
  \begin{equation}
   \label{equ:flavor_ratios_earth}
   f_{\alpha, \oplus}^\pi
   =
   \sum_{\beta=e,\mu,\tau} P_{\beta\alpha} f_{\beta, {\rm S}}^\pi \;.
  \end{equation}
  Thus, starting, from the all-species oscillated flux at Earth that we have available, $\Phi_{6\nu}$, we estimate the oscillated fluxes of $\nu_\alpha$ and $\bar{\nu}_\alpha$ at Earth as
  \begin{equation}
   \label{equ:flux_earth_cat_b}
   \Phi_{\nu_\alpha}(E_\nu)
   =
   \Phi_{\bar{\nu}_\alpha}(E_\nu)
   =
   \frac{1}{2}
   f_{\alpha, \oplus}^\pi
   \Phi_{6\nu}(E_\nu) \;,
  \end{equation}
  where the factor of $1/2$ splits the flux of $\nu_\alpha + \bar{\nu}_\alpha$ evenly between them.
 \item
  {\it Flux models for which we only have available the $\nu_\mu + \bar{\nu}_\mu$ oscillated flux at Earth (model 2 below).}  Like with fluxes in category (b), we consider the flavor composition to be energy-independent and split the flux of each flavor evenly between $\nu_\alpha$ to $\bar{\nu}_\alpha$ at all energies.  Starting from the flux of $\nu_\mu + \bar{\nu_\mu}$ at Earth that we have available, we estimate the oscillated fluxes of $\nu_\alpha$ and $\bar{\nu}_\alpha$ at Earth as
  \begin{equation}
   \label{equ:flux_earth_cat_c}
   \Phi_{\nu_\alpha}(E_\nu)
   =
   \Phi_{\bar{\nu}_\alpha}(E_\nu)
   =
   \frac{1}{2}
   \frac{f_{\alpha, \oplus}^\pi}{f_{\mu, \oplus}^\pi}
   \Phi_{\nu_\mu+\bar{\nu}_\mu}(E_\nu) \;,
  \end{equation}
  where the factor of $1/2$ splits the flux of $\nu_\alpha + \bar{\nu}_\alpha$ evenly between them.
\end{enumerate}
For benchmark flux model 12, the flux of each neutrino species at Earth is directly available~\cite{Muzio:PrivateComm}; we adopt them without modification.

In our analysis, we forecast measurements in IceCube-Gen2 in the 2030s.  By then, the values of the mixing parameters are expected to be known more precisely than today~\cite{Esteban:2020cvm, NuFit_5.0}, thanks to the upcoming oscillation experiments DUNE~\cite{DUNE:2020lwj}, Hyper-Kamiokande~\cite{Hyper-Kamiokande:2018ofw}, and JUNO~\cite{JUNO:2015zny}.  Assuming that the true values of the mixing parameters are equal to their present-day best-fit values from the {\sc NuFit~5.0} global fit to oscillation data~\cite{Esteban:2020cvm, NuFit_5.0}, and that the neutrino mass ordering is normal, by 2030 we expect that~\cite{Song:2020nfh} $\sin^2 \theta_{12} = 0.304 \pm 0.00164$, $\sin^2 \theta_{23} = 0.573_{-0.00659}^{+0.006288}$, $\sin^2 \theta_{13} = 0.02219_{-0.00063}^{+0.00062}$, and $\delta_{\rm CP} = (197_{-10.10}^{+11.22})^\circ$.  Thus, for neutrinos produced in pion decays, as in categories (b) and (c) above, the flavor ratios at Earth, computed with \equ{flavor_ratios_earth}, are close to equipartition, \ie,
\begin{eqnarray}
 f_{e,\oplus}^\pi &=& 0.298_{-0.003}^{+0.001} \;, \\
 f_{\mu,\oplus}^\pi &=&  0.359_{-0.006}^{+0.004} \;, \\
 f_{\tau,\oplus}^\pi &=& 0.342_{-0.005}^{+0.009} \;,
\end{eqnarray}
ignoring correlations between the mixing parameters.  The uncertainties in $f_{\beta,\oplus}^\pi$ are tiny; accounting for correlations, they would be even smaller.  Thus, we can safely neglect the uncertainty in the future values of $f_{\beta,\oplus}^\pi$, and just use their best-fit above when computing Eqs.~(\ref{equ:flux_earth_cat_b}) and (\ref{equ:flux_earth_cat_c}) henceforth.  (If the mass ordering is inverted, the best-fit values of $f_{\beta,\oplus}^{\pi}$ change only slightly~\cite{Song:2020nfh}, so we do not explore that case separately.)


\subsection{Benchmark flux models}
\label{section:uhe_neutrinos_benchmarks}

Figure~\ref{fig:fluxes} shows the twelve UHE neutrino diffuse flux models~\cite{Fang:2013vla, Padovani:2015mba, Fang:2017zjf, Heinze:2019jou, Muzio:2019leu, Rodrigues:2020pli, Anker:2020lre, IceCube:2020wum, Muzio:2021zud, IceCube:2021uhz} that we use to benchmark the sensitivity of IceCube-Gen2.  They include extrapolations of the flux of TeV--PeV neutrinos discovered by IceCube to ultra-high energies ({\Large $\bullet$}, models 1 and 2), cosmogenic neutrinos ($\blacksquare$, models 3--5, 7), source neutrinos (\rotatebox[origin=c]{45}{$\blacksquare$}, models 6, 9, 10), and joint predictions of cosmogenic plus source neutrinos ($\rotatebox[origin=c]{90}{\HexaSteel}$, models 8, 11, 12). 

Our selection of benchmark flux models is representative of the breadth of theoretical predictions available in the literature at the time of writing.
The highest of our benchmark fluxes---models 4, 6, and 12---saturate the present upper limits on the UHE neutrino flux.  The lowest---models 1, 3, and 5---lie below the 10-year differential sensitivity of IceCube-Gen2.  The remaining flux models lie in-between these two extremes.  Later, in Section~\ref{section:results}, we find that measuring the UHE $\nu N$ cross section should be possible for all but the lowest flux models.

\begin{figure*}[t]
 \centering
 \includegraphics[width=\textwidth]{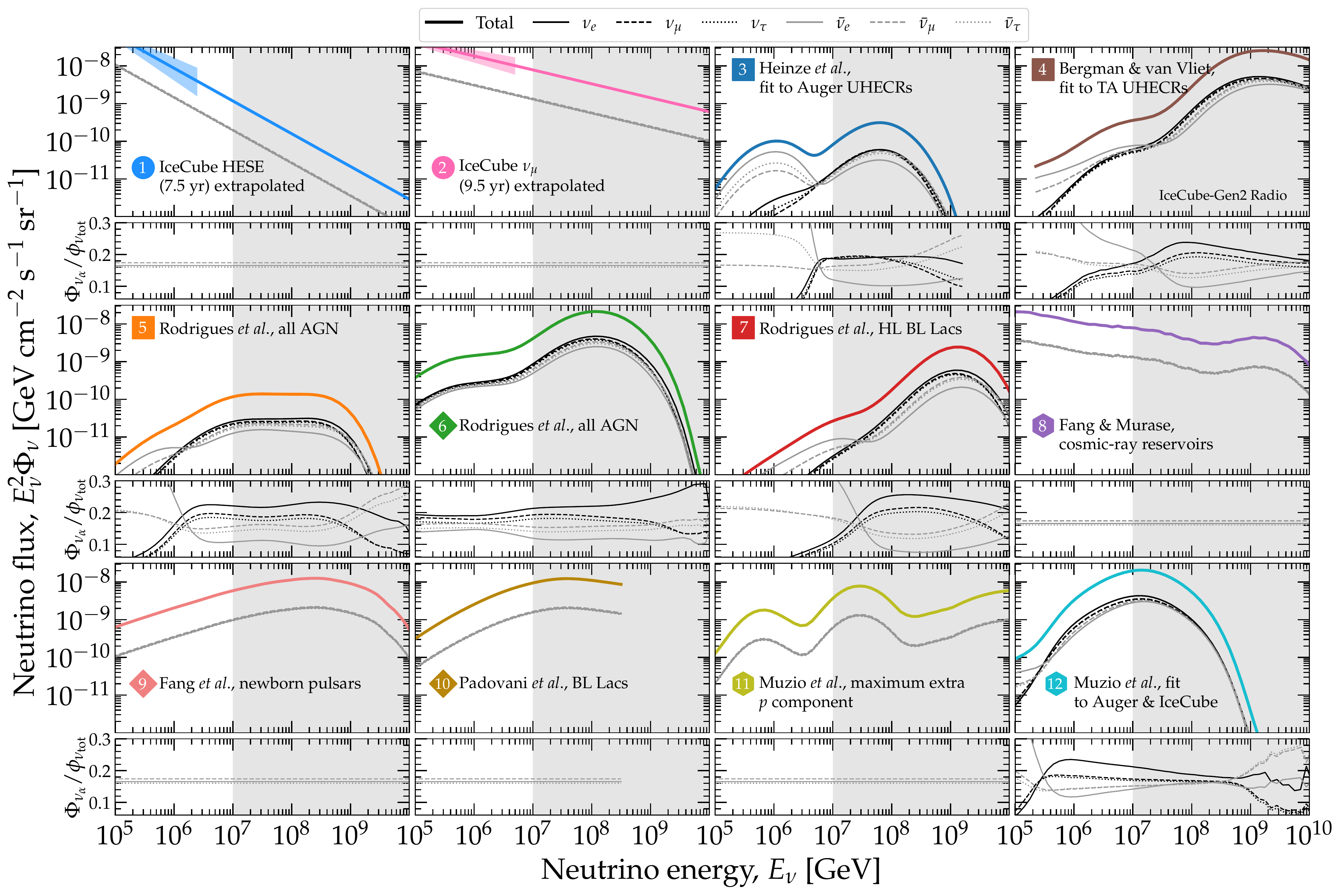}
 \caption{\label{fig:benchmark_spectra_per_species}Flux of each neutrino species for each of the benchmark ultra-high-energy neutrino fluxes that we consider (see \figu{fluxes}).  Panels under the fluxes show the ratio $\Phi_{\nu_\alpha}/\Phi_{\nu_{\rm tot}}$ of the flux of $\nu_\alpha$ to the total flux, \ie, the sum of the fluxes of all species.  For flux models 1, 2, and 7, the ratios are constant in neutrino energy and approximately equal to 1/6; the exact values are in Section~\ref{section:uhe_neutrinos_flavor_composition}.  For all the other flux models, the ratios change with neutrino energy.  See the main text for details.  The shaded region indicates the approximate neutrino energy range to which the radio array of IceCube-Gen2 will be sensitive.}
\end{figure*}

Below we present an overview of the benchmark flux models.  We defer to their original publications for details (see also \Refe~\cite{BustamanteUHENuFlux}).
\begin{enumerate}
 \item
  {\it IceCube HESE (7.5~yr) extrapolated~\cite{IceCube:2020wum}:}  Using 102 High Energy Starting Events (HESE) collected over 7.5~yr, IceCube fit the all-species astrophysical neutrino flux using a power law, $\Phi_{6 \nu}(E_\nu) = \Phi_{6 \nu, 0} (E_\nu/100~{\rm TeV})^{-\gamma_{6 \nu}}$, with $\Phi_{6 \nu, 0} = 6.37_{-1.62}^{+1.47} \times 10^{-18}$~GeV$^{-1}$~cm$^{-2}$~s$^{-1}$~sr$^{-1}$ and $\gamma_{6 \nu} = 2.87_{-0.19}^{+0.20}$, valid from 60~TeV to 10~PeV.  To build our benchmark flux model 2, we extend this flux to $10^{10}$~GeV, unbroken, using the best-fit values of $\Phi_{6 \nu, 0}$ and $\gamma$.  We assume that the flux of each neutrino species shares this common value of $\gamma_{6 \nu}$ and estimate it using \equ{flux_earth_cat_b}.
 \item
  {\it IceCube $\nu_\mu$ (9.5~yr) extrapolated~\cite{IceCube:2021uhz}:}  Using $6.5 \times 10^5$ through-going muon tracks collected over 9.5~yr, IceCube fit the $\nu_\mu+\bar{\nu}_\mu$ astrophysical neutrino flux using a power law, $\Phi_{\nu_\mu+\bar{\nu}_\mu}(E_\nu) = \Phi_{\nu_\mu+\bar{\nu}_\mu, 0} (E_\nu/100~{\rm TeV})^{-\gamma_{\nu_\mu+\bar{\nu}_\mu}}$, with $\Phi_{\nu_\mu+\bar{\nu}_\mu, 0} = 1.44_{-0.26}^{+0.25} \times 10^{-18}$~GeV$^{-1}$~cm$^{-2}$~s$^{-1}$~sr$^{-1}$ and $\gamma_{\nu_\mu+\bar{\nu}_\mu} = 2.37 \pm 0.19$, valid from 15~TeV to 5~PeV.  To build our benchmark flux model 3, we extend this flux to $10^{10}$~GeV, unbroken, using the best-fit values of $\Phi_{\nu_\mu+\bar{\nu}_\mu, 0}$ and $\gamma_{\nu_\mu+\bar{\nu}_\mu}$.  We assume that the flux of each neutrino species shares this common value of $\gamma_{\nu_\mu+\bar{\nu}_\mu}$, and estimate it using \equ{flux_earth_cat_c}.
 \item
  {\it Heinze {\it et al.}, fit to Auger UHECRs (cosmogenic)~\cite{Heinze:2019jou}:}  Cosmogenic neutrinos are generated by UHECRs emitted by a population of nondescript sources, distributed in redshift, and their flux is normalized by fitting the predicted UHECR energy spectrum and mass composition at Earth to recent data from the Pierre Auger Observatory~\cite{Fenu:2017hlc, Bellido:2017cgf}.  References~\cite{Romero-Wolf:2017xqe, AlvesBatista:2018zui} predict similar fluxes, also from fits to Auger data. UHECR interactions on the CMB and EBL, including photodisintegration and photohadronic processes, are computed using {\sc PriNCe}~\cite{PriNCe}.  Because the best-fit value of the maximum UHECR rigidity is low, $2.5 \times 10^9$~GV, there are relatively few UHECRs at the highest energies, and so the neutrino yield is low.  To build our benchmark model 3, we compute the pre-oscillation fluxes of $\nu_e$, $\bar{\nu}_e$, $\nu_\mu$, and $\bar{\nu}_\mu$ as functions of energy using {\sc PriNCe}, and use \equ{flux_earth_cat_a} to transform them into the oscillated fluxes of all species at Earth.  
 \item
  {\it Bergman \& van Vliet, fit to TA UHECRs (cosmogenic)~\cite{Anker:2020lre}:}  Cosmogenic neutrinos are generated in the same way as for the benchmark flux model 3, but instead fitting the UHECR energy spectrum and mass composition at Earth to recent data from TA~\cite{Tsunesada:2017aaq, Bergman:2017ikd}.  Reference~\cite{Bergman:2021djm} predicts a similar flux.  Because the TA data is compatible with a lighter UHECR mass composition and a higher maximum rigidity, the neutrino flux inferred using TA data is larger than with Auger data (model 3).  To build our benchmark model 4, the pre-oscillation fluxes of $\nu_e$, $\bar{\nu}_e$, $\nu_\mu$, and $\bar{\nu}_\mu$ as functions of energy are computed using {\sc CRPropa3}~\cite{AlvesBatista:2016vpy, vanVliet:PrivateComm}, and use \equ{flux_earth_cat_a} to transform them into the oscillated fluxes of all species at Earth.  
 \item
  {\it Rodrigues {\it et al.}, all AGN (cosmogenic)~\cite{Rodrigues:2020pli}:}  Neutrinos are produced by the entire population of active galactic nuclei (AGN), which serve as UHECR accelerators.  AGN are divided into three sub-populations: low-luminosity BL Lacs, high-luminosity BL Lacs, and flat-spectrum radio quasars (FSRQs).  The number density of each sub-population evolves differently with redshift and luminosity~\cite{Ajello:2011zi, Ajello:2013lka}.  Before escaping,  UHECRs interact in the AGN jets, via photodisintegration and photohadronic processes~\cite{Boncioli:2016lkt, Rodrigues:2017fmu}.  Cosmogenic neutrinos are produced by the UHECRs that escape, and their flux is computed using {\sc PriNCe}~\cite{PriNCe}.  Source neutrinos are produced inside the jets, and their flux is computed using {\sc NeuCosmA}~\cite{Hummer:2010vx, Baerwald:2010fk, Morejon:2019pfu}.  The predicted UHECR energy spectrum and mass composition at Earth agree with Auger data~\cite{Fenu:2017hlc}, while the UHE neutrino flux satisfies the IceCube upper limit~\cite{IceCube:2018fhm}.  Low-luminosity BL Lacs explain the UHECR flux, while FSRQs dominate neutrino production.  To build our benchmark model 5, we adopt the maximum allowed cosmogenic neutrino flux from the entire population of AGN (Fig.~2 in \Refe~\cite{Rodrigues:2020pli}).   We take the pre-oscillation fluxes of $\nu_e$, $\bar{\nu}_e$, $\nu_\mu$, and $\bar{\nu}_\mu$ as functions of energy~\cite{Rodrigues:PrivateComm}, and use \equ{flux_earth_cat_a} to transform them into the oscillated fluxes of all species at Earth.  
 \item
  {\it Rodrigues {\it et al.}, all AGN (source)~\cite{Rodrigues:2020pli}:}  We consider the flux of source neutrinos that is the counterpart to the cosmogenic flux of model 5.  To build our benchmark model 6, we adopt the maximum allowed source neutrino flux from the entire population of AGN (Fig.~2 in \Refe~\cite{Rodrigues:2020pli}).  We take the pre-oscillation fluxes of $\nu_e$, $\bar{\nu}_e$, $\nu_\mu$, and $\bar{\nu}_\mu$ as functions of energy~\cite{Rodrigues:PrivateComm}, and use \equ{flux_earth_cat_a} to transform them into the oscillated fluxes of all species at Earth. 
 \item
  {\it Rodrigues {\it et al.}, HL BL Lacs (cosmogenic)~\cite{Rodrigues:2020pli}:}  UHECRs and neutrinos are produced only by high-luminosity (HL) BL Lacs.  The predicted UHECRs agree with the Auger energy spectrum above the ankle, but are lighter than the Auger mass composition above a few EeV.  We adopt the cosmogenic neutrino spectrum from HL BL Lacs (Fig.~5 in \Refe~\cite{Rodrigues:2020pli}) as benchmark because it peaks at energies higher than the benchmark models 5 and 6, and has a normalization in-between theirs.   We take the pre-oscillation fluxes of $\nu_e$, $\bar{\nu}_e$, $\nu_\mu$, and $\bar{\nu}_\mu$ as functions of energy~\cite{Rodrigues:PrivateComm}, and use \equ{flux_earth_cat_a} to transform them into oscillated fluxes of all species at Earth. 
 \item
  {\it Fang \& Murase, cosmic-ray reservoirs (cosmogenic + source)~\cite{Fang:2017zjf}:}  Neutrinos are produced in a grand-unified multi-messenger model of high-energy emission where UHECRs are accelerated in the jets of supermassive black holes of radio-loud AGN embedded in galaxy clusters that act as cosmic-ray reservoirs.  There, UHECRs remain confined for 1--10~Gyr and produce UHE neutrinos via $p\gamma$ and $pp$ interactions.  From 100~TeV to 100~PeV, neutrinos are primarily made inside the clusters, in UHECR interactions on the intra-cluster medium; above $10^9$~GeV, neutrinos are primarily cosmogenic.  The neutrino flux normalization results from fitting the predicted UHECR energy spectrum and mass composition to Auger data~\cite{PierreAuger:2015fol}, and the predicted TeV--PeV neutrino flux to IceCube data~\cite{IceCube:2016umi, IceCube:2017zho}.   Reference~\cite{Fang:2017zjf} provided the all-species neutrino flux, $\Phi_{6 \nu}$.  To build our benchmark flux model 8, we use it to estimate the flux of each neutrino species using \equ{flux_earth_cat_b}.
 \item
  {\it Fang {\it et al.}, newborn pulsars (source)~\cite{Fang:2013vla}:}  Fast-spinning newborn pulsars that harbor intense surface magnetic fields, of up to $10^{13}$~G, may efficiently accelerate charged particles in the pulsar wind during pulsar spin-down.  Accelerated particles propagate through the expanding supernova ejecta that surrounds the pulsar; as they do, $pp$ interactions on the ejecta produce neutrinos.  Reference~\cite{Fang:2013vla} computed the diffuse flux of neutrinos produced by the cosmological population of newborn pulsars, integrated over their neutrino-producing lifetimes, with a spread in magnetic field intensity and spin period, and distributed in redshift following the star formation rate (SFR).  (We consider only the neutrino contribution from the sources, not the contribution of cosmogenic neutrinos produced by UHECRs emitted by the pulsars, which is of the same order~\cite{Fang:2013vla}.)  Reference~\cite{Fang:2013vla} provided the all-species neutrino flux, $\Phi_{6 \nu}$.  To build our benchmark flux model 9, we use it to estimate the flux of each neutrino species using \equ{flux_earth_cat_b}.
 \item
  {\it Padovani {\it et al.}, BL Lacs (source)~\cite{Padovani:2015mba}:}  The neutrino flux is obtained within the framework of the simplified view of blazars~\cite{Giommi:2011sn}.  Neutrinos are produced in photohadronic interactions inside the jets of BL Lacs, whose population is simulated using a spread of redshifts and source features like synchrotron peak energies and X-ray flux.  The flux prediction was originally constructed to explain the TeV--PeV neutrino range; we adopt it because it spills into the UHE regime.  A key parameter of the model is $Y_{\nu\gamma}$, the ratio of the neutrino intensity to the gamma-ray intensity.  Following \Refe~\cite{IceCube:2016uab}, to satisfy the IceCube upper limit on the UHE neutrino flux, we set $Y_{\nu\gamma} = 0.13$.  Reference~\cite{Padovani:2015mba} provided the all-species neutrino flux, $\Phi_{6 \nu}$.  To build our benchmark flux model 10, we use it to estimate the flux of each neutrino species using \equ{flux_earth_cat_b}.
 \item
  {\it Muzio {\it et al.}, maximum extra $p$ component (cosmogenic + source)~\cite{Muzio:2019leu}:}  Cosmogenic and source neutrinos are produced via photohadronic interactions within the UFA15 multi-messenger framework~\cite{Unger:2015laa}, where sources emit UHECRs whose energy spectrum and mass composition at Earth are fit to Auger data.   Reference~\cite{Muzio:2019leu} added a sub-dominant UHECR pure-proton component that escapes the sources with energies above $10^9$~GeV, motivated in part by the observation, in Auger, of a slowdown in the increase of average nuclear mass with energy~\cite{PierreAuger:2017tlx}, and that enhances the neutrino flux.  We adopt the maximum allowed neutrino flux that results from the joint single-mass UFA15 plus pure-proton components, using {\sc Sybill2.3c}~\cite{Fedynitch:2018cbl} for the hadronic interaction of UHECRs in the atmosphere (Fig.~9 in \Refe~\cite{Muzio:2019leu}).  Reference~\cite{Muzio:2019leu} provided the all-species neutrino flux, $\Phi_{6 \nu}$.  To build our benchmark flux model 11, we use it to estimate the flux of each neutrino species using \equ{flux_earth_cat_b}.
 \item
  {\it Muzio {\it et al.},  fit to Auger $\&$ IceCube (cosmogenic + source)~\cite{Muzio:2021zud}:}  Cosmogenic and source neutrinos are produced via photohadronic interactions within the UFA15 multi-messenger framework (see above); in addition, source neutrinos are produced via $pp$ interactions of UHECRs in the source environment.  Neutrinos from hadronic interactions dominate at low energies, below the $\Delta$ resonance energy.  UHECR predictions are fit to Auger data.  The total neutrino flux from \Refe~\cite{Muzio:2021zud} includes contributions from UHECR sources and non-UHECR sources; the total flux is fit to the IceCube TeV--PeV neutrino flux measurement~\cite{IceCube:2020acn, IceCube:2021rpz}.  We adopt the best-fit total neutrino flux (``UHECR $\nu$" plus ``Non-UHECR $\nu$ from Fig.~1 in \Refe~\cite{Muzio:2021zud}).  To build our benchmark model 12, we use the oscillated fluxes of each neutrino species at Earth as a function of energy, which are available directly from the calculation~\cite{Muzio:PrivateComm}.
\end{enumerate}

Figure~\ref{fig:benchmark_spectra_per_species} shows the breakdown into the flux of each neutrino species for the benchmark models; see Section~\ref{section:uhe_neutrinos_flavor_composition}.  For models 1, 2, 8--11, the ratio of each species to the total flux is constant in energy.  For models 3--7 and 12, for which non-trivial energy evolution of the flux of each species separately is available,  common trends are evident.  At low energies, typically below the energy range of the IceCube-Gen2 radio component, $\bar{\nu}_\alpha$ dominate due to the presence and oscillation of $\bar{\nu}_e$ produced in the beta-decay of neutrons and neutron-rich isotopes created in UHECR interactions, mainly during their extragalactic propagation.  At higher energies, neutrinos are produced by pion decay; throughout the IceCube-Gen2 energy range, flavor equipartition holds approximately.  Roughly within the range $10^7$--$10^9$~GeV, there is a slight excess of neutrinos over anti-neutrinos, because more $\pi^+$ than $\pi^-$ are produced.  At the highest energies, multi-pion production dominates, and the excess flips.  Later, in \figu{flavor_ratios}, we show how the flavor composition is affected by neutrino propagation through the Earth.


\section{Neutrino propagation inside Earth}
\label{section:neutrino_propagation}

\begin{figure*}[t!]
 \centering
 \includegraphics[trim=0 1.5cm 1.5cm 4.0cm, clip=true, width=0.49\textwidth]{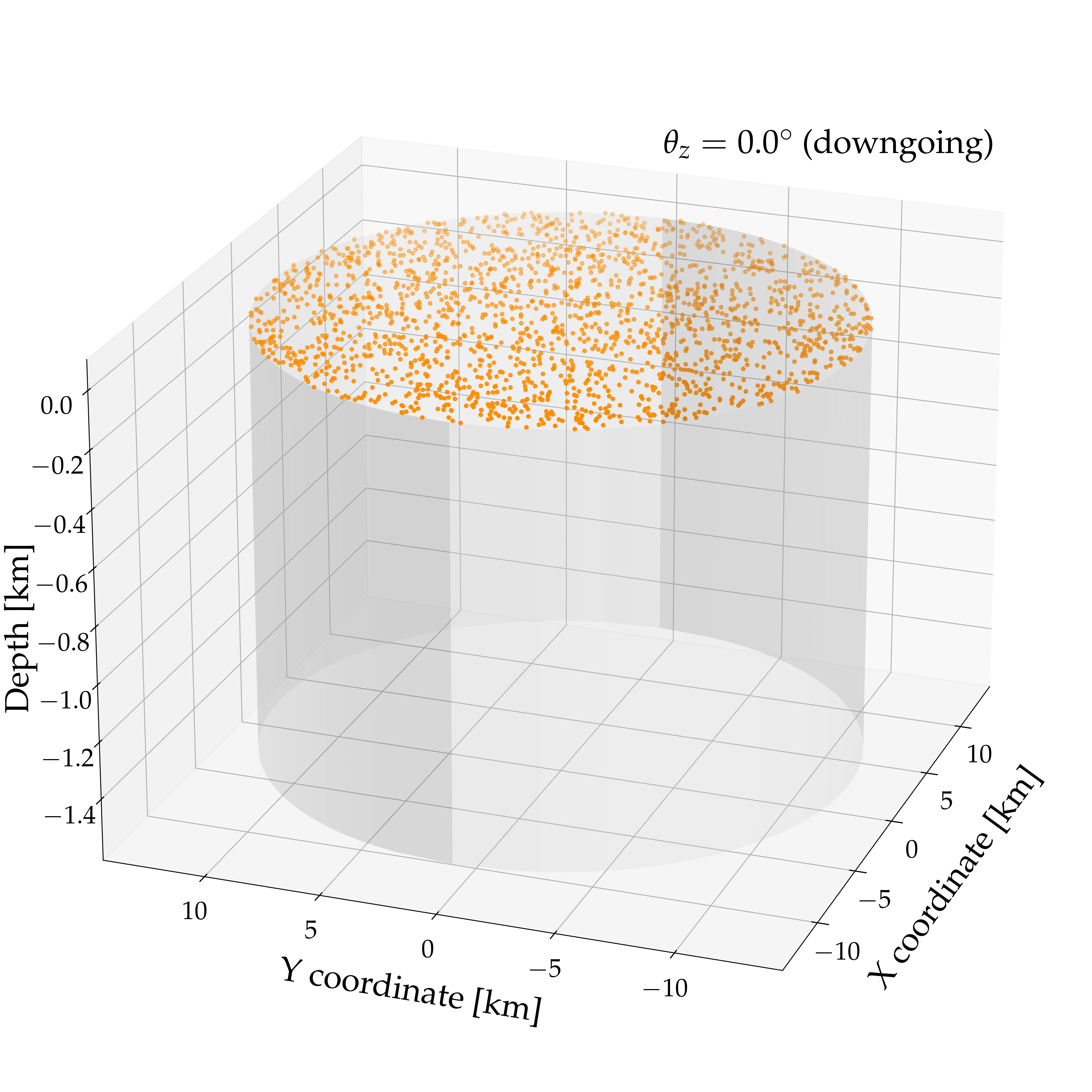}
 \includegraphics[trim=0 1.5cm 1.5cm 4.0cm, clip=true, width=0.49\textwidth]{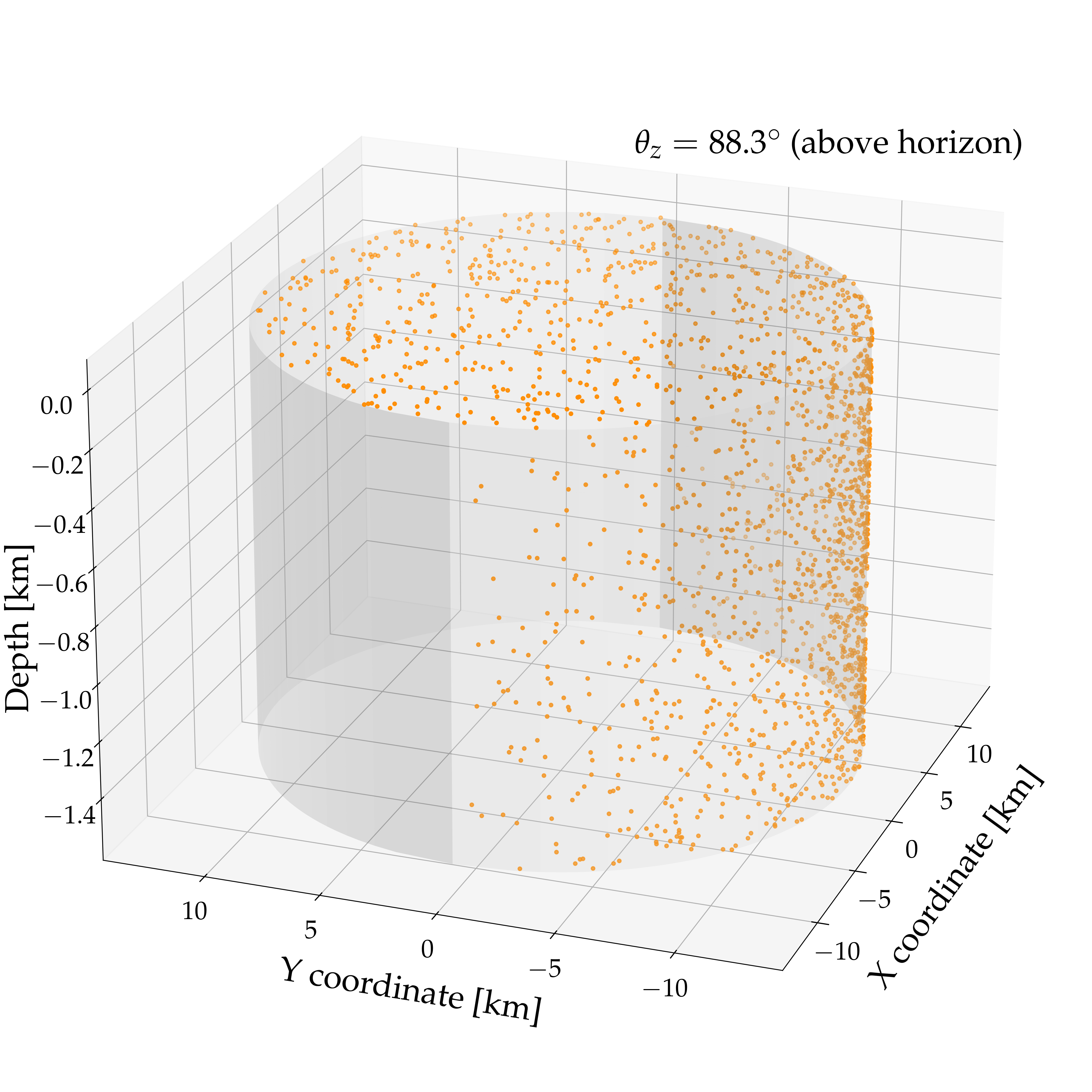}
 \includegraphics[trim=0 1.5cm 1.5cm 4.0cm, clip=true, width=0.49\textwidth]{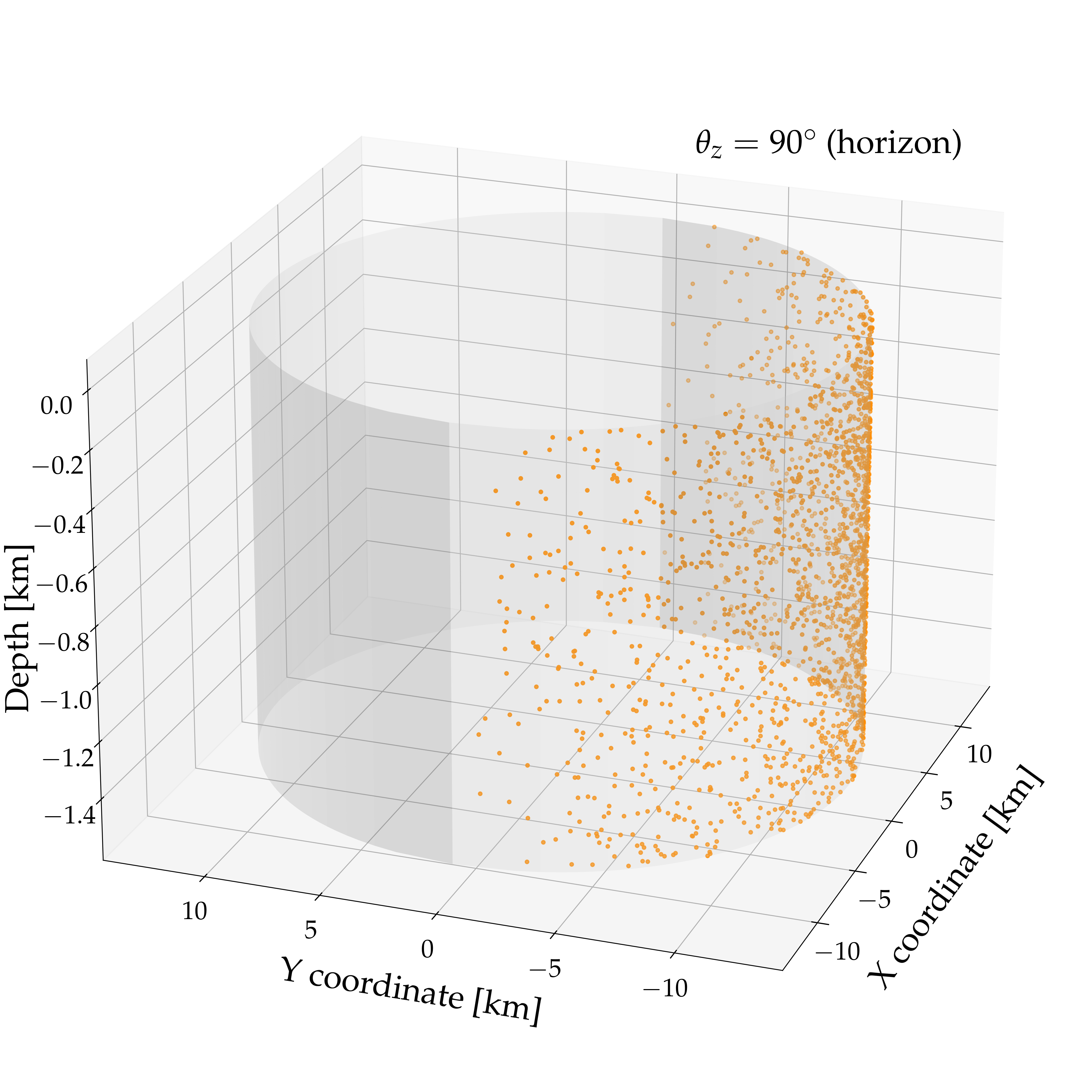}
 \includegraphics[trim=0 1.5cm 1.5cm 4.0cm, clip=true, width=0.49\textwidth]{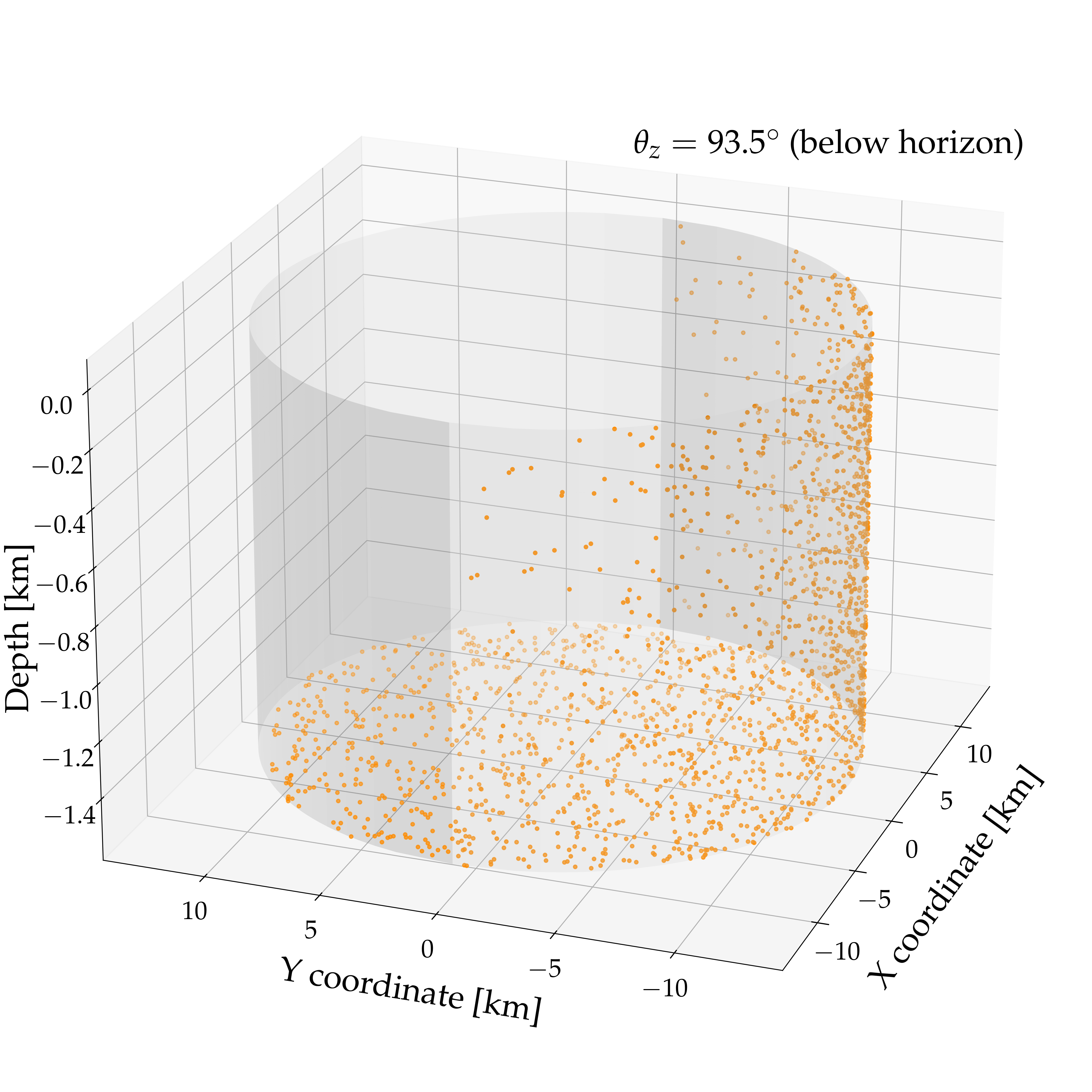}
 \caption{\label{fig:detector_hits}Distribution of neutrinos hitting or simulated IceCube-Gen2 detector volume, from four illustrative directions.  See the main text for details.  For this figure only, neutrinos are not attenuated while propagating inside the Earth.}
\end{figure*}

To compute the expected rate of neutrino-initiated showers at a neutrino telescope, first we compute the flux of neutrinos that reaches it, after propagating through the Earth across different directions.


\subsection{Computing neutrino propagation}

Above TeV energies, the dominant interaction that neutrinos undergo while propagating inside the Earth is $\nu$N DIS, NC and CC; see Section~\ref{section:cross_section}.  NC scatterings pile up originally high-energy neutrinos at low energies, while CC scatterings remove neutrinos from the flux altogether.  The exception is the CC scattering of $\nu_\tau$, where the final-state tauon may propagate some distance before decaying into a $\nu_\tau$, albeit with an energy lower than that of the original $\nu_\tau$; this is known as ``$\nu_\tau$ regeneration".  Because of it, the flux of  $\nu_\tau$ is less attenuated than that of other flavors.  This becomes especially important above 10~PeV, where the final-state $\nu_\tau$ are still high-energy.  

We propagate UHE neutrinos towards the detector along different directions, parametrized as $\cos \theta_z$, where $\theta_z$ is the zenith angle measured from the location of the detector.  For us, this is the South Pole ($\cos\theta_z = 1$), where IceCube-Gen2 will be located; see \figu{cartoon}.  We compute the energy spectra of each neutrino species that reach the detector from $-1 \leq \cos\theta_z \leq +1$.

Figure~\ref{fig:detector_hits} shows examples of neutrinos hitting the surface of the simulated IceCube-Gen2 detector volume from different directions.  We model the detector volume as a cylinder of radius $12.6$~km and height $1.50$~km, buried underground at a distance of $1.51$~km from the surface of the Earth to the bottom of the cylinder.  The top surface area of the cylinder is 500~km$^2$~\cite{IceCube-Gen2:2020qha}. Once a neutrino reaches the surface of the cylinder, we stop its propagation.  Inside the cylinder, the propagation of the neutrino is computed separately, in the detection step of our calculation, where any further interaction that occurs within the detector volume initiates a particle shower that emits a radio signal that might be detected by underground antennas; see Section~\ref{section:radio_detection}.  Modeling the detector volume as a cylinder {\it vs.}~as a point impacts by up to 10\% the attenuation of neutrinos that reach it from directions around the horizon, \ie, Earth-skimming neutrinos, for which the detector is of comparable size to the distance traveled inside the Earth.  This is especially relevant because these neutrinos offer the greatest sensitivity to the $\nu N$ cross section; see Section~\ref{section:cs_measurement}.

Depending on the direction of the neutrino, it will encounter a different matter column density.  To account for this, for the internal matter density of Earth, we adopt the Preliminary Reference Earth Model~\cite{Dziewonski:1981xy}, built from seismographic data, which models the density radially out from the center of a spherical Earth, as a series of concentric layers of increasing density towards the center.  For our calculations, since IceCube-Gen2 will be embedded in the Antarctic ice, we add a layer of ice of thickness 3~km at the surface of the Earth.  In addition, the composition of matter inside the Earth changes with radial distance: deeper layers contain heavier elements---iron, nickel---than shallower layers.  Further, matter is, in general, not isoscalar, though this affects mainly neutrino energies below 1~TeV~\cite{Garcia:2020jwr}.  When propagating neutrinos inside the Earth, we account for the changes in density and composition as a function of position inside Earth.

As an illustration only, and not accounting for $\nu_\tau$ regeneration, the exponential dampening in \equ{event_rate_simple} describes the attenuation of the neutrino flux inside Earth.  (We {\it do not} use those simplified expressions to produce our results, but more sophisticated methods; see Section~\ref{section:radio_detection_rates}.)  The attenuation due to CC interactions is stronger the higher the neutrino energy and the longer the length of the path traveled by the neutrinos inside the Earth.  The low-energy pile-up due to NC interactions has a similar dependence on energy and direction

As a result of the interactions inside the Earth, while the neutrino flux is isotropic at the surface of the Earth, it has become anisotropic by the time it reaches the detector.  At EeV energies, no detectable flux reaches the detector from below; instead, the flux comes from above, where it is only lightly attenuated by the detector overburden, and from around the horizon, where the attenuation is significant to modify the shape of the spectrum, but not enough to eliminate it.

We use the state-of-the-art neutrino propagation code {\sc NuPropEarth}~\cite{Garcia:2020jwr, NuPropEarth} to compute the fluxes of $\nu_\alpha$ and $\bar{\nu}_\alpha$, $\Phi_{\nu_\alpha}^{\rm det}(\cos\theta_z, E_\nu)$ and $\Phi_{\bar{\nu}_\alpha}^{\rm det}(\cos\theta_z, E_\nu)$.  We propagate $\nu_e$, $\bar{\nu}_e$, $\nu_\mu$, $\bar{\nu}_\mu$, $\nu_\tau$, and $\bar{\nu}_\tau$ separately, along different directions.  In $\nu_\tau$ regeneration, {\sc NuPropEarth} accounts for the energy losses due to electromagnetic interactions during the propagation of intermediate tauons, via {\sc TAUSIC}~\cite{Kudryavtsev:2008qh}, and computes the distribution of decay products in tauon decays, via {\sc TAUOLA}~\cite{Davidson:2010rw}.  

In {\sc NuPropEarth}, for the $\nu N$ DIS cross section, we use the central value of the BGR18 calculation~\cite{Bertone:2018dse} (see Section~\ref{section:cs_basics}), as implemented in the {\sc HEDIS}~\cite{Garcia:2020jwr} module of the {\sc GENIE}~\cite{Andreopoulos:2009rq} neutrino event generator.  {\sc HEDIS} uses {\sc PYTHIA6}~\cite{Sjostrand:2006za} to compute the hadronization of final-state particles~\cite{Garcia:2019hze}.  We have modified {\sc NuPropEarth} and {\sc HEDIS} to be able to use versions of the BGR18 cross section that are scaled up and down by a constant factor, as part of our method of measuring the cross section; see Section~\ref{section:cs_extracting}.  For the cross sections of the sub-dominant neutrino interactions (see Section~\ref{section:nu_other_int}), we use their implementations in {\sc HEDIS}, unmodified.

\begin{figure}[t!]
 \centering
 \includegraphics[width=\columnwidth]{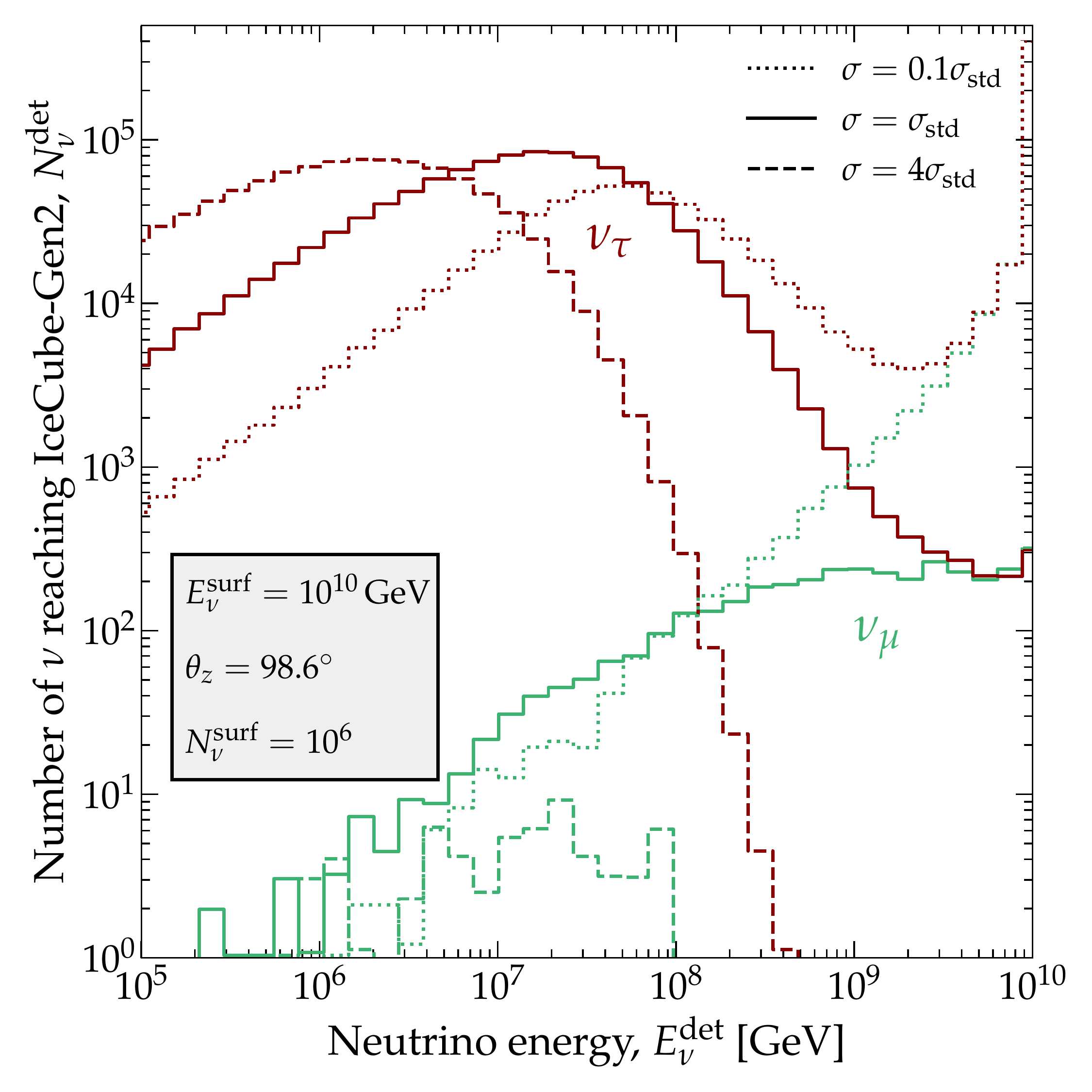}
 \caption{\label{fig:attenuation}Sample histograms of the number of UHE $\nu_\mu$ and $\nu_\tau$ that reach the detector after propagating through the Earth.  At the surface of the Earth, opposite to the detector, we inject a mono-energetic beam of $N_{\nu}^{\rm surf} = 10^6$ neutrinos, each with energy $E_\nu^{\rm surf} = 10^{10}$~GeV, pointed toward the detector along a zenith angle $\theta_z = 98.6^\circ$.  We show cases of neutrinos propagated using the central value of the BGR18 $\nu N$ DIS cross section ($\sigma_{\rm std}$)~\cite{Bertone:2018dse}, and sample scaled-up and scaled-down versions of it.  For our analysis, we pre-compute histograms for many more values of $E_\nu^{\rm surf}$, $\theta_z$, and the cross section.  Results for $\nu_e$, not shown, are similar to $\nu_\mu$.  See Section~\ref{section:neutrino_propagation} for details.}
\end{figure}

Figure~\ref{fig:attenuation} shows energy histograms at the detector resulting from propagating a mono-energetic Earth-skimming neutrino beam inside the Earth.  For $\nu_\mu$, the cascading down to lower energies due to multiple interactions inside the Earth is evident.  For $\nu_\tau$, the effect of regeneration is evident as a pile-up at lower energies.  Figure~\ref{fig:attenuation} shows that changes in the $\nu N$ DIS cross section affect the propagation inside Earth significantly.  For $\nu_\mu$, and also for $\nu_e$, not shown, the main effect of a higher cross section is to attenuate the flux further.  For $\nu_\tau$, a higher cross section shifts the peak of the pile-up to lower energies, due to a larger number of neutrino interactions.


\subsection{Computational speed-ups}

As part of our statistical analysis below (see Section~\ref{section:cs_extracting_stat_proc}), we need to propagate many different UHE spectra of $\nu_e$, $\bar{\nu}_e$, $\nu_\mu$, $\bar{\nu}_\mu$, $\nu_\tau$, and $\bar{\nu}_\tau$ separately through the Earth, for different values of the $\nu N$ DIS cross section, along different directions, and with high accuracy.  This is a computationally taxing task.  

We circumvent this limitation as follows.  First, we inject a large number $N_{\nu_\alpha}^{{\rm surf}} = 10^6$ of $\nu_\alpha$ at the surface of the Earth, each with initial energy $E_\nu^{\rm surf}$, and propagate it along different directions towards the detector, using {\sc NuPropEarth}.  Upon arriving at the detector, the neutrinos are no longer mono-energetic, but their final energies, $E_\nu^{\rm det}$, are spread out as a result of interactions inside Earth, \ie, $N_{\nu_\alpha}^{\rm det} \equiv N_{\nu_\alpha}^{\rm det}(\cos\theta_z, E_\nu^{\rm surf}, E_\nu^{\rm det})$.  Second, we compute the transmission coefficients as
\begin{equation}
 T_{\nu_\alpha}(\cos\theta_z, E_\nu^{\rm surf}, E_\nu^{\rm det})
 \equiv
 \frac{N_{\nu_\alpha}^{\rm det}(\cos\theta_z, E_\nu^{\rm surf}, E_\nu^{\rm det})}
 {N_{\nu_\alpha}^{\rm surf}} \;,
\end{equation}
bin them in bins of $E_\nu^{\rm det}$ of width $\Delta E_\nu^{\rm det}$, \ie,
\begin{eqnarray}
 && 
 \mathcal{T}_{\nu_\alpha}(\cos\theta_z, E_\nu^{\rm surf}, E_\nu^{\rm det}, \Delta E_\nu^{\rm det}) 
 \nonumber \\
 &&~\qquad
 \equiv 
 \sum_{x = E_\nu^{\rm det}}^{E_\nu^{\rm det}+\Delta E_\nu^{\rm det}}
 \frac{T_{\nu_\alpha}(\cos\theta_z, E_\nu^{\rm surf}, x)}{\Delta E_\nu^{\rm det}} \;,
\end{eqnarray}
and save $\mathcal{T}_{\nu_\alpha}$ as look-up tables.  We do this separately for $\nu_\alpha$ and $\bar{\nu}_\alpha$ of all flavors.  Third, given any UHE neutrino spectrum at the surface of the Earth, $\Phi_{\nu_\alpha}$, we use the pre-computed $\mathcal{T}_{\nu_\alpha}$ to estimate the average flux at the detector within an interval of final energy $[E_\nu, E_\nu+\Delta E_\nu]$ as
\begin{eqnarray}
 &&
 \Phi_{\nu_\alpha}
 ^{\rm det}(\cos\theta_z, E_\nu, \Delta E_\nu)
 \simeq
 \Delta E_\nu
 \nonumber \\
 \label{equ:spectrum_det_approx}
 && \times
 \sum_{E_\nu^{\rm surf} > E_\nu}
 \Phi_{\nu_\alpha}(E_\nu^{\rm surf})
 \mathcal{T}_{\nu_\alpha}(\cos\theta_z, E_\nu^{\rm surf}, E_\nu, \Delta E_\nu) \;,
\end{eqnarray}
and similarly for $\bar{\nu}_\alpha$.  Finally, we approximate the true spectrum by its binned average, \ie, $\Phi_{\nu_\alpha}^{\rm det}(\cos\theta_z, E_\nu) \approx \Phi_{\nu_\alpha}^{\rm det}(\cos\theta_z, E_\nu, \Delta E_\nu)$.  We pre-compute the look-up coefficients $\mathcal{T}_{\nu_\alpha}$ in fine grids of $E_\nu^{\rm surf}$ and $E_\nu^{\rm det}$.  We repeat the above procedure to generate look-up tables for different values of the $\nu N$ cross section, since we need them for our statistical procedure; see Section~\ref{section:cs_extracting_stat_proc}.

\begin{figure}[t]
 \centering
 \includegraphics[width=\columnwidth]{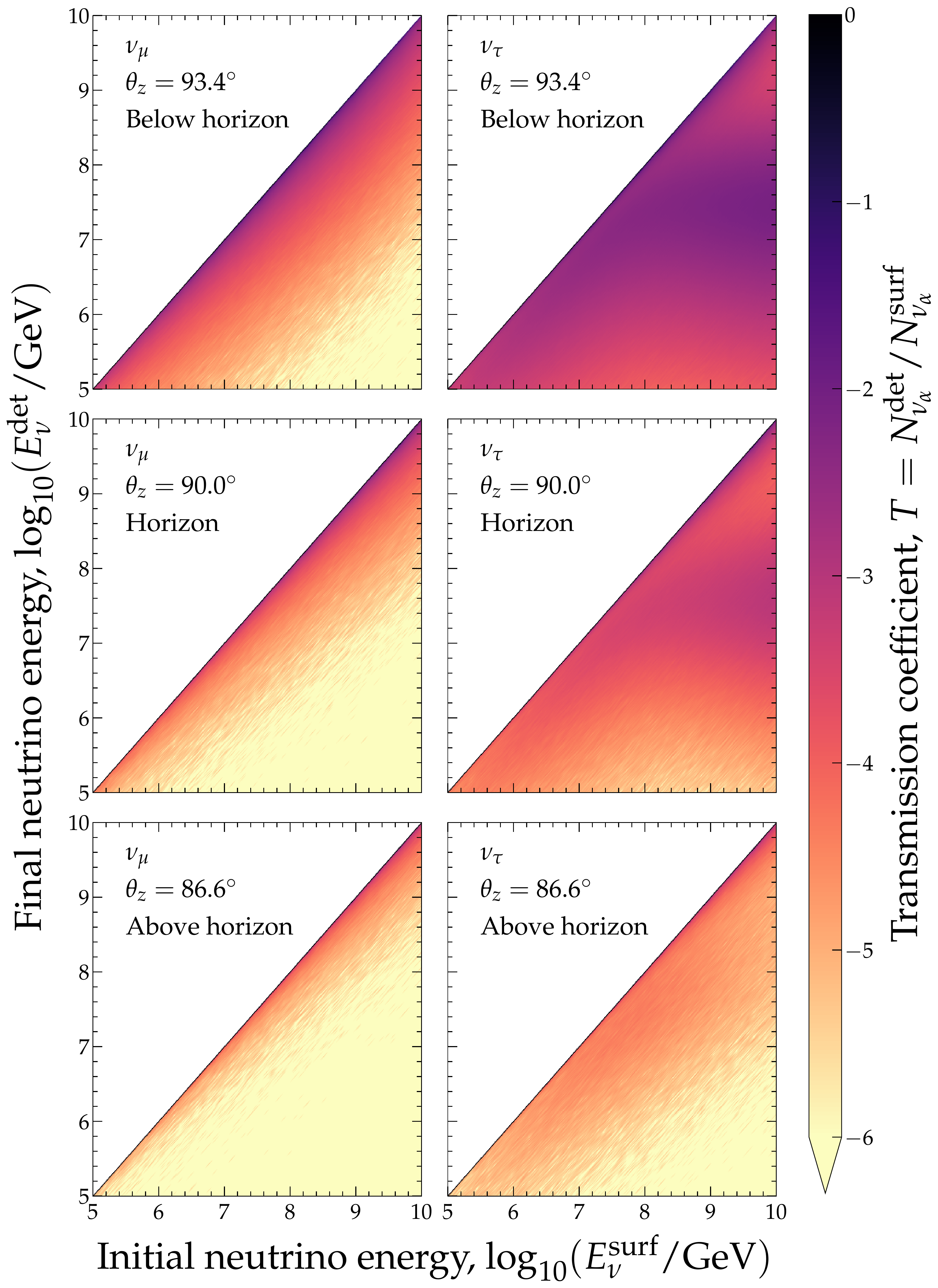}
 \caption{\label{fig:transmission}Examples of the transmission coefficient $T \equiv N_{\nu_\alpha}^{\rm det}(\cos\theta_z, E_\nu^{\rm surf}, E_\nu^{\rm det}) / N_{\nu_\alpha}^{\rm surf}$ of ultra-high-energy $\nu_\mu$ and $\nu_\tau$, from the surface of the Earth, across it, and up to the location of IceCube-Gen2, for illustrative directions $\cos\theta_z$.  At every value of the initial neutrino energy $E_\nu^{\rm surf}$, a number $N_{\nu_\alpha}^{\rm surf} = 10^6$ of mono-energetic neutrinos, each with energy $E_\nu^{\rm surf}$, is injected at the surface of the Earth and, after propagating inside it, arrive at the detector with a spread of final energies $E_\nu^{\rm det}$.  For this plot, we use the central value of the BGR18 $\nu N$ cross section ($\sigma_{\rm std}$, in the notation of Section~\ref{section:cs_extracting_overview}).  See the main text for details.}
\end{figure}

Figure \ref{fig:transmission} shows sample transmission coefficients $T_{\nu_\mu}$ and $T_{\nu_\tau}$, for directions at and around the horizon.  For $\nu_\mu$, and also for $\nu_e$, not shown, because the $\nu N$ cross section grows with energy, the cascading down to lower energies becomes more important the higher the injected energy $E_\nu^{\rm surf}$.  It is most significant when neutrinos arrive from below the horizon, due to the larger matter column density that they traverse.  For $\nu_\tau$, in addition, the presence of regeneration is evident for $E_\nu^{\rm surf} \gtrsim 10$~PeV, for neutrinos coming from the horizon and below it.  At the detector, regenerated $\nu_\tau$ are concentrated in a band centered around $E_\nu^{\rm det} \approx 10$~PeV.

When producing our results, we use \equ{spectrum_det_approx} to compute neutrino spectra.  By doing this, we circumvent the computationally intensive need to propagate every time each benchmark neutrino flux from scratch, for each value of the $\nu N$ cross section.


\subsection{UHE neutrino flux at the detector}

\begin{figure}[t!]
 \centering
 \includegraphics[width=\columnwidth]{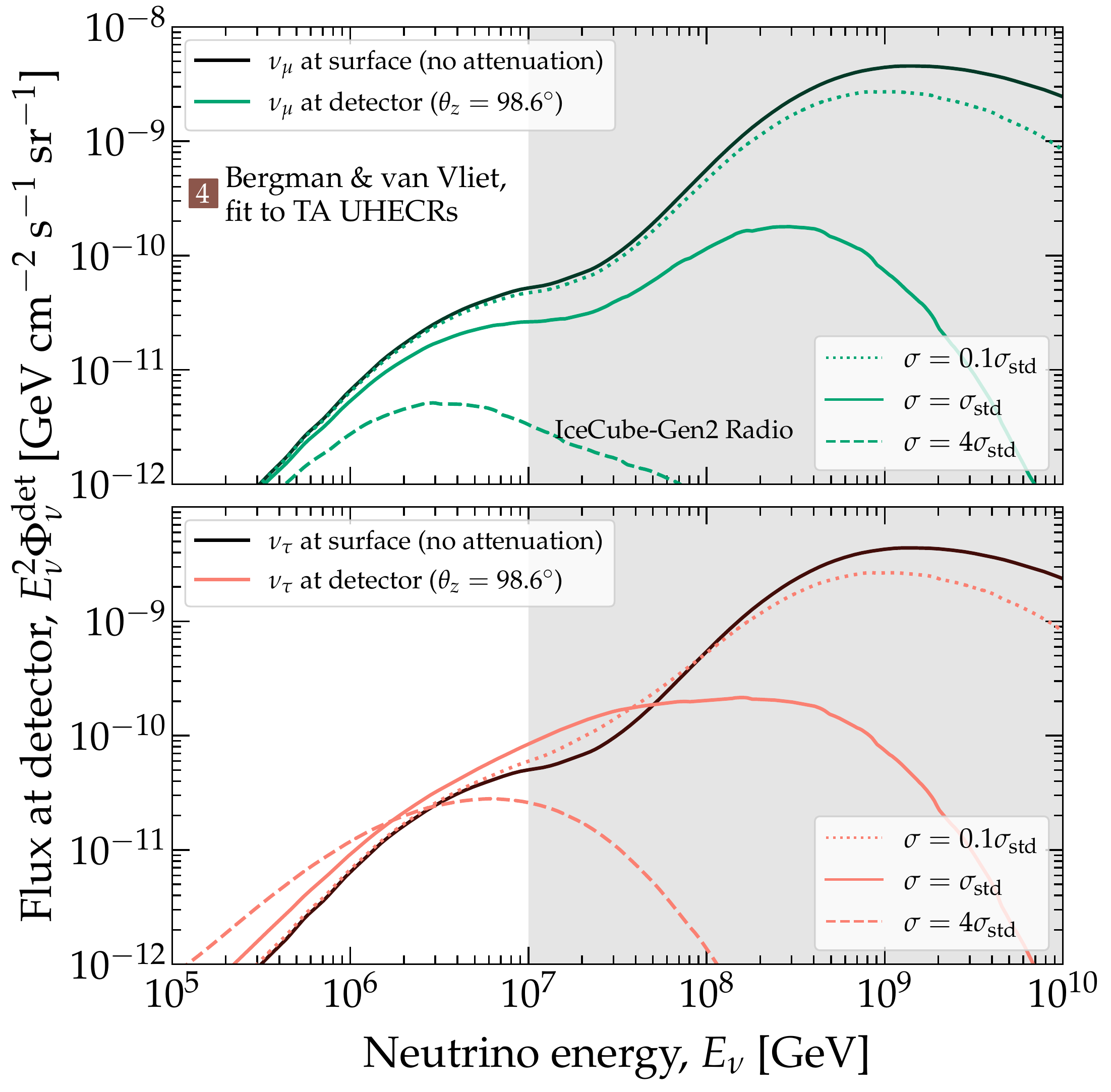}
 \caption{\label{fig:flux_at_det}Benchmark UHE $\nu_\mu$ and $\nu_\tau$ flux model 4~\cite{Anker:2020lre} (see Section~\ref{section:uhe_neutrinos}) at the surface of the Earth and at IceCube-Gen2, after propagating through the Earth along an illustrative direction, $\theta_z = 98.6^\circ$.  The flux of $\nu_e$, not shown, is similar to that of $\nu_\mu$.  The effect of in-Earth propagation on the fluxes of $\bar{\nu}_\alpha$ ($\alpha = e, \mu, \tau$), not shown, is similar to the effect $\nu_\alpha$.  Fluxes at the detector are computed using {\sc NuPropEarth}~\cite{Garcia:2020jwr, NuPropEarth}.  We show fluxes propagated using the central value of the BGR18 $\nu N$ DIS cross section ($\sigma_{\rm std}$)~\cite{Bertone:2018dse}, and scaled-up and scaled-down versions of it.  See Section~\ref{section:neutrino_propagation} for details.}
\end{figure}

Figure~\ref{fig:flux_at_det} illustrates the effect of the propagation through Earth on the benchmark flux model 4, for an example arrival direction.  We choose a direction from below the horizon because the column density traversed along it is large enough that changes in the $\nu N$ cross section imprint sizeable changes in the flux that reaches the detector.  For $\nu_\mu$, and also for $\nu_e$, not shown, even the central value of the BGR18 $\nu N$ cross section ($\sigma_{\rm std}$) is large enough to suppress the flux at the detector by more that one order of magnitude relative to the flux at the surface of the Earth, at the highest energies.  Larger cross sections vanish the flux altogether.  For $\nu_\tau$, the suppression is mitigated, though not counterbalanced, by the pile-up of low-energy, regenerated $\nu_\tau$.  For downgoing neutrinos ($\theta_z \lesssim 80^\circ$), not shown, the fluxes are unaffected even by large cross sections, due to the small column densities.  Conversely, for upgoing neutrinos ($\theta_z \gtrsim 100^\circ$), not shown, the fluxes vanish even if the cross section is small, due to the large column densities.

\begin{figure}[t!]
 \centering
 \includegraphics[width=\columnwidth]{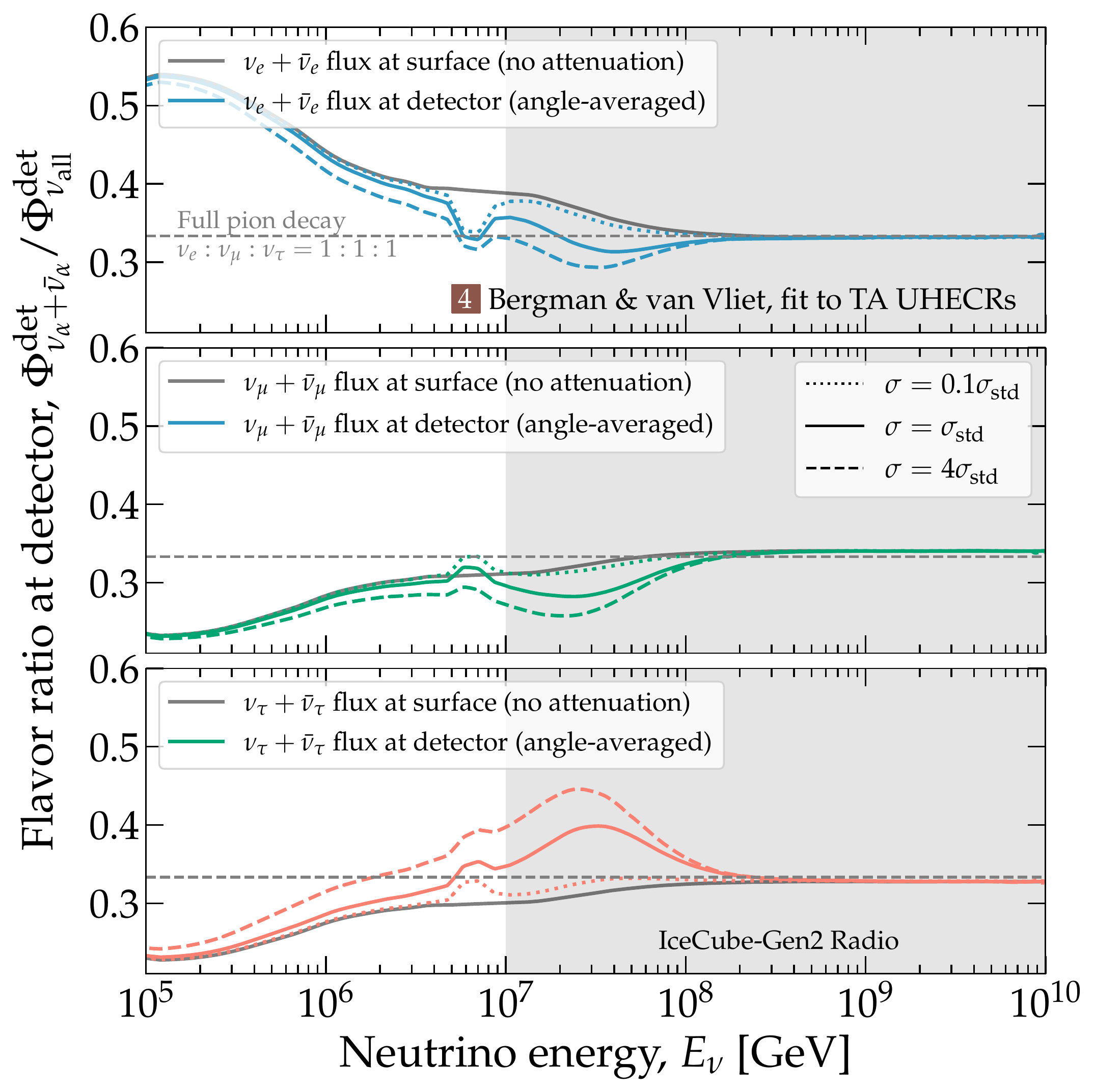}
 \caption{\label{fig:flavor_ratios}Flavor ratios of neutrinos, \ie, ratios of the $\nu_\alpha + \bar{\nu}_\alpha$ flux compared to the all-flavor flux, at the surface of the Earth and at IceCube-Gen2, for the benchmark UHE flux model 4~\cite{Anker:2020lre} (see Section~\ref{section:uhe_neutrinos}).  In this plot, the flavor ratios at IceCube-Gen2 are the average for Earth-skimming neutrinos with $\theta_z \in [85^\circ, 95^\circ]$.  Like in \figu{flux_at_det}, we show flavor ratios of fluxes propagated using the central value of the BGR18 $\nu N$ DIS cross section ($\sigma_{\rm std}$)~\cite{Bertone:2018dse}, and scaled-up and scaled-down versions of it.  For comparison, we show the nominal expectation of flavor equipartition at the surface of the Earth from neutrino production via the full pion decay chain.}
\end{figure}

Figure~\ref{fig:flavor_ratios} shows how the flavor ratios, \ie, the proportion of $\nu_\alpha + \bar{\nu}_\alpha$ to the all-flavor flux, are affected by the propagation through Earth, also for benchmark flux model 4.  Above $10^8$~GeV, neutrinos of all flavors are attenuated equally, and so their flavor ratios at the detector are equal; at these energies, they also match the flavor ratios at the surface of the Earth; see \figu{benchmark_spectra_per_species}.  The effects of the propagation become apparent at lower energies.  There, the neutrinos of all flavors regenerated from NC interactions plus the $\nu_\tau$ regenerated from CC interactions pile up.  As a result of the latter, the tau-flavor ratio dominates over the electron- and muon-flavor ratios.  The dip in the electron-flavor ratio around 6.3~PeV is due to the Glashow resonance experienced only by $\bar{\nu}_e$.  There are corresponding bumps at the same energy in the muon- and tau-flavor ratios.  The above features become more prominent the higher the $\nu N$ DIS cross section.  While these features are an implicit part of our analysis, we do not make use of flavor identification, since the capabilities of IceCube-Gen2 in this direction are still under study~\cite{Garcia-Fernandez:2020dhb, Glaser:2021hfi, Stjarnholm:2021xpj}. Section~\ref{section:limitations} comments on this possibility for future work.


\section{Radio-detecting UHE neutrinos}
\label{section:radio_detection}

We gear our forecasts to the case of UHE neutrino radio-detection in the planned radio array of IceCube-Gen2 \cite{IceCube-Gen2:2020qha}.  Once neutrinos have propagated through the Earth and reached the detector, they might interact inside the detector volume via NC and CC $\nu N$ DIS.  See  Section~\ref{section:cross_section} for details on the $\nu N$ DIS process.  In most of these interactions, a substantial fraction of the neutrino energy is transferred into a high-energy particle shower.  As the shower develops, it produces coherent radio emission that might be detected by antennas of the array.


\subsection{Neutrino-induced radio emission in ice}
\label{section:radio_detection_basics}

In the NC or CC DIS interaction of a neutrino or anti-neutrino of energy $E_\nu$ of any flavor, the final-state hadrons initiate a hadronic shower, rich in pions and muons~\cite{Li:2016kra}.  The energy of the hadronic shower is $E_{\rm sh} = (1-y) E_\nu$, where $y$ is the inelasticity; see Section~\ref{section:cross_section}.  In the CC DIS interaction of a $\nu_e$ or $\bar{\nu}_e$, the final-state electron or positron initiates an additional electromagnetic shower, rich in photons, electrons, and positrons, co-located with the hadronic shower. The energy of the electromagnetic shower is $E_{\rm sh} = y E_\nu$.  

At neutrino energies below roughly $10^9$~GeV, the hadronic and electromagnetic showers develop in phase and appear as a single shower with energy $E_\mathrm{sh} = E_\nu$; see, \eg, \Refe~\cite{Glaser:2019cws,  Stjarnholm:2021xpj}).  At higher energies, the electromagnetic shower is subject to the Landau–Pomeranchuk–Migdal (LPM) effect~\cite{Landau:1953um, Migdal:1956tc}, which reduces the cross section of the high-energy electrons and positrons.  The precise role of the LPM effect in the radio-detection of neutrinos is under study~\cite{Stjarnholm:2021xpj}, but seems to be significant: if present, it delays the first interactions of electrons and positrons in the shower or leads to multiple spatially displaced sub-showers.  As a result, the hadronic and electromagnetic showers may develop differently, and the shower energy might not match the neutrino energy anymore.  The simulations of $\nu_e$-induced CC showers that we perform to describe the detector response, described in Section~\ref{section:radio_detection_gen2}, include the LPM effect.  Nevertheless, when computing $\nu_e$-induced CC event rates, as described in Section~\ref{section:radio_detection_rates}, we maintain the relation $E_\mathrm{sh} = E_\nu$ across all energies, since further work is needed to find an equivalent form of it that accounts for the changing dominance of the LPM effect with energy.

Separately, in the CC DIS interactions of $\nu_\mu$ and $\nu_\tau$, we also ignore the contribution of secondary interactions of final-state muons and tauons to the event rates, because they are challenging to simulate.  Since their inclusion would increase the event rates by up to 25\%~\cite{Garcia-Fernandez:2020dhb, Glaser:2021hfi}, depending on energy and spectral shape, ignoring them makes our forecasts conservative.  Future revised estimates might isolate the contribution of the LPM effect and include secondary leptons, via changes to the relation between shower and neutrino energies, \equ{energy_nu}, and to the simulated detector effective volume; see below.

IceCube-Gen2 will be built in the Antarctic ice; see below for details.  Because ice is dielectric, as a neutrino-induced shower develops inside it, it builds up a time-varying negative charge excess in the shower front which produces coherent radio emission.  This {\it Askaryan radiation}~\cite{Askaryan:1961pfb} is strongest when the shower is observed along a cone with half-angle of $\arccos(1/n) \approx 56^\circ$, centered on the shower axis, where $n = 1.78$ is the index of refraction of deep ice.  Along this ``Cherenkov angle", the radiation emitted by the shower interferes constructively.  The radio signal is a broadband, bipolar pulse a few nanoseconds long, predominantly in the frequency range 100~MHz--1~GHz.   While the shower track itself is only a few tens of meters long, the radio emission can propagate over kilometers.  However, the emission strength decreases quickly if the shower is observed from angles smaller or larger than the Cherenkov angle, even from only a couple of degrees away from it.  Reference~\cite{Zas:1991jv} first pointed out that Askaryan radiation can be used to detect neutrinos. Reference~\cite{Schroder:2016hrv} contains an in-depth description of radio emission from high-energy particles.

Accelerator measurements have demonstrated the existence of in-ice Askaryan emission, and found agreement with theoretical predictions~\cite{Saltzberg:2000bk, Miocinovic:2006it, ANITA:2006nif}.  Additional evidence comes from the observation of radio emission from extensive air showers---\ie, particle showers in the atmosphere---to which Askaryan radiation contributes in a sub-dominant capacity~\cite{PierreAuger:2014ldh, Schellart:2014oaa, Schroder:2016hrv, Huege:2016veh}. 


\subsection{The radio component of IceCube-Gen2}
\label{section:radio_detection_gen2}

\begin{figure}[t]
 \centering
 \includegraphics[width=\columnwidth]{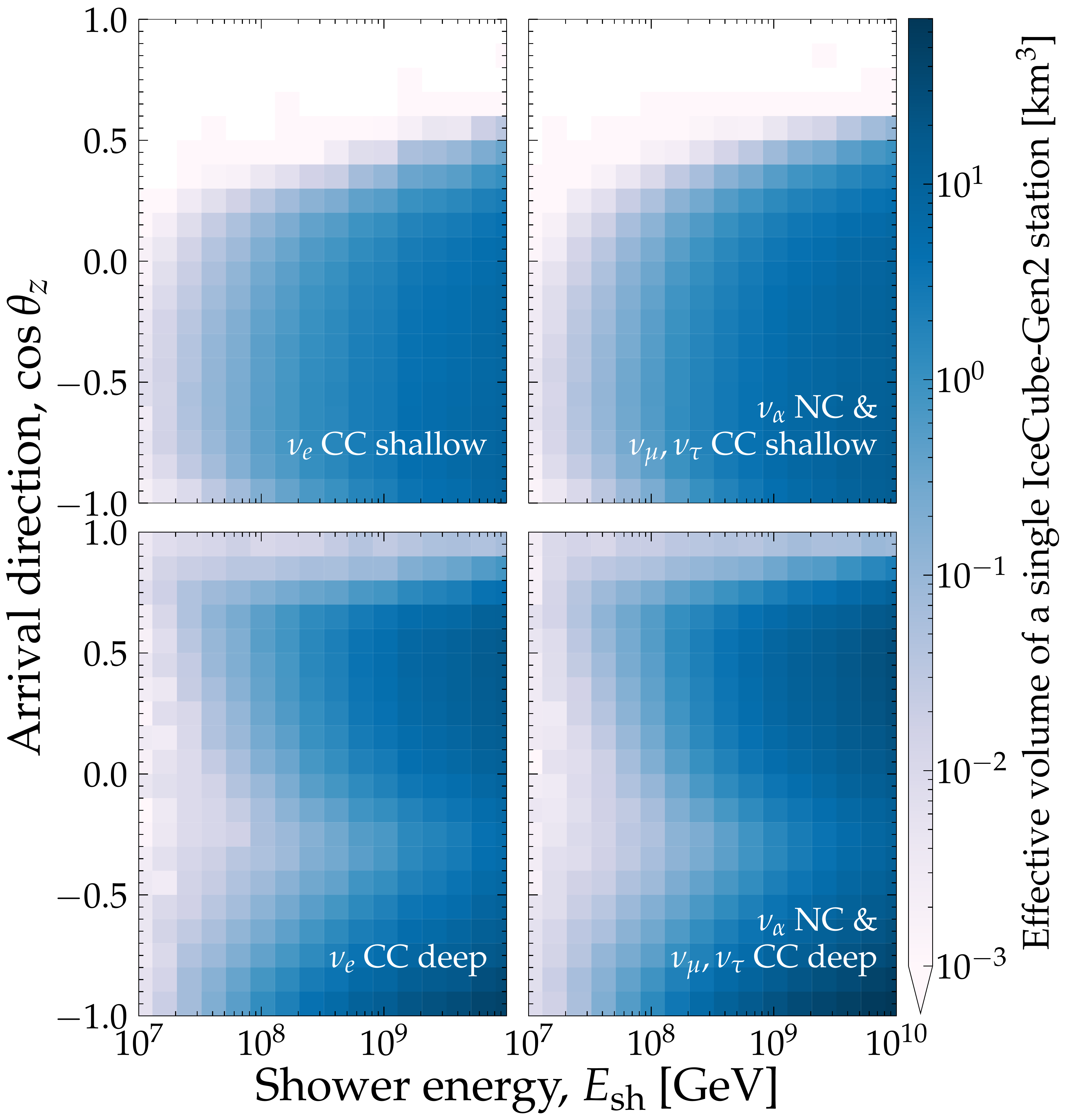}
 \caption{\label{fig:veff_single}Effective volume of a single detector component of the IceCube-Gen2 radio array, simulated using {\sc NuRadioMC}~\cite{Glaser:2019cws}.  The effective volume does not include the effects of neutrino propagation inside the Earth; its directional dependence is due only to the station geometry and to the radio propagation in the Antarctic ice. The directional dependence resulting from neutrino propagation through the Earth is computed separately on the neutrino flux; see Section~\ref{section:neutrino_propagation}. {\it Top row:} Volume for a single shallow detector component.  {\it Bottom row:} Volume for a single deep detector component.  {\it Left column:} Volume for $\nu_e$ CC interactions.  {\it Right column:} Volume for NC interactions of all flavors and CC interactions of $\nu_\mu$ and $\nu_\tau$.  See \figu{veff_total} for the effective volume of the full array and the main text for details.}
\end{figure}

\begin{figure}[t]
 \centering
 \includegraphics[width=\columnwidth]{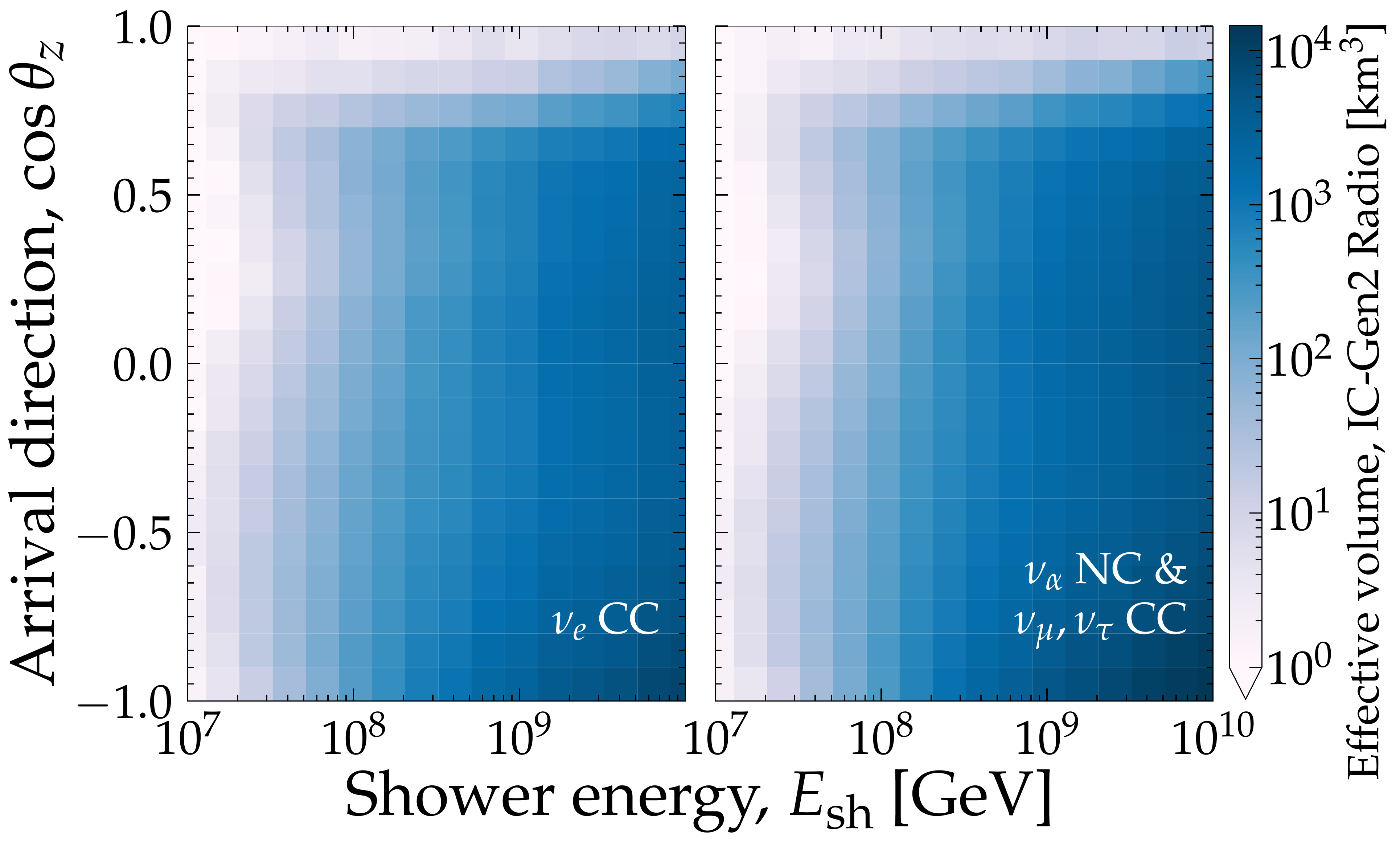}
 \caption{\label{fig:veff_total}Effective volume of the baseline design of the full IceCube-Gen2 radio array.  This is the sum of the effective volumes of 144 hybrid (shallow + deep) stations plus 169 shallow-only stations.  {\it Left:} Volume for $\nu_e$ CC interactions, $V_{{\rm eff}, {\nu_e}}^{\rm CC}$.  {\it Right:} Volume for NC interactions of all flavors and CC interactions of $\nu_\mu$ and $\nu_\tau$, $V_{{\rm eff}, {\nu_\alpha}}^{\rm NC}$.  See \figu{veff_single} for the effective volumes of single stations and the main text for details.}
\end{figure}

The main advantage of radio-detection of UHE neutrinos is the large attenuation length of radio signals in polar ice, of roughly 1~km~\cite{Barwick:2005zz}, versus the attenuation and scattering lengths of optical Cherenkov light, of roughly 100~m.  This allows for a cost-efficient instrumentation of huge detector volumes with a sparse array of compact radio-detection stations.  Each station contains several antennas buried in the ice at depths of up to 200~m and acts as an autonomous unit.  To maximize the overall sensitivity of the array, the detector stations are separated by more than 1~km, so that coincidences between stations are rare. Each detector stations measures enough information to determine the shower energy and neutrino arrival direction; see, \eg, \Refs~\cite{Glaser:2019kjh, RNO-G:2021zfm, Glaser:2021jss, ARIANNA:2021pzm, Stjarnholm:2021xpj, Aguilar:2021uzt}.

The envisioned design of IceCube-Gen2~\cite{IceCube-Gen2:2020qha} capitalizes on this opportunity by including an array of radio antennas covering a total surface area of 500~km$^2$~\cite{Hallmann:2021kqk}.  The radio array will be of critical importance in the quest for the discovery of UHE neutrinos.  Its design combines the advantages found in the pathfinder experiments ARA~\cite{ARA:2015wxq}, ARIANNA~\cite{ARIANNA:2019scz}, and RNO-G~\cite{Aguilar:2019jay} into a hybrid array.
The IceCube-Gen2 array consists of two types of radio detector stations that measure and reconstruct neutrino properties with complementary accuracy, intended to maximize the discovery potential by mitigating risks and adding multiple handles for the rejection of rare backgrounds~\cite{Hallmann:2021kqk}.  The design blends the hybrid stations explored by RNO-G \cite{Aguilar:2019jay}---narrow bicone and quad-slot antennas on three strings up to a depth of 200~m plus high-gain log-periodic-dipole-array (LPDA) antennas close to the surface---with shallow-only stations explored by ARIANNA---LPDA antennas close to the surface with one additional dipole antenna at a depth of 15~m to aid event reconstruction~\cite{Anker:2019zcx}. 

A key ingredient in our estimates of the neutrino-induced event rates below is the detector response of the IceCube-Gen2 radio component.  We compute it via numerical simulations using the same open-source tools that are used by the IceCube-Gen2 Collaboration, {\sc NuRadioMC}~\cite{Glaser:2019cws} and {\sc NuRadioReco}~\cite{Glaser:2019rxw}.  Below we describe how we simulate the response of one radio station.  After, we scale up its response, expressed in terms of the effective volume, to the size of the full radio array.  

{\sc NuRadioMC} simulates the neutrino interaction in the ice, the generation of the radio emission and its propagation to the antennas, and performs a full detector and trigger simulation.  We simulate showers of varying energy, $E_{\rm sh}$, that enter the detector from different directions, $\cos \theta_z$.  To compute the Askaryan emission, we adopt the ARZ prescription~\cite{Alvarez-Muniz:2020ary}, combined with a representative shower library of charge-excess profiles~\cite{Glaser:2019cws}, which provides realistic modelling of the LPM effect for $\nu_e$ CC interactions.  The deep antennas of the station are triggered by a four-channel phased array~\cite{Allison:2018ynt, Aguilar:2019jay} and the shallow antennas by a high/low-threshold trigger with a time-coincidence trigger requiring coincident detection of two out of the four LPDA antennas, using an optimized trigger bandwidth~\cite{Glaser:2020pot}. We use the exact same simulation settings as in \Refe~\cite{Hallmann:2021kqk}.  

We report the detector response via its effective volume, \ie, the simulation volume multiplied by the ratio of the number of detected showers to the number of simulated showers; see, \eg, \Refe~\cite{vanSanten:2022wss}.  We do this separately for the shallow and deep detector components of the simulated station, and separately for the NC interactions of all neutrino flavors and CC interactions of $\nu_\mu$ and $\nu_\tau$, and for the CC interactions of $\nu_e$.  Results for $\nu_\alpha$ and $\bar{\nu}_\alpha$ are the same, since their UHE $\nu N$ DIS cross sections and inelasticity distributions are nearly equal; see Section~\ref{section:cross_section} and \Refs~\cite{Connolly:2011vc, Cooper-Sarkar:2011jtt, Bertone:2018dse}.  The resulting effective volumes are functions of $E_{\rm sh}$ and $\cos \theta_z$.

There are two notable differences in the effective volumes that we have generated compared to previously reported results.  First, we do not fold in the effects of in-Earth propagation into the effective volumes.  Those effects are instead imprinted on the neutrino flux that arrives at the detector; see Section~\ref{section:neutrino_propagation}.  As a result, the directional dependence of the effective volumes is due solely to the geometry of the detector components and the propagation of radio signals in the Antarctic ice.  This facilitates exploring the effect of using different values of the cross section during neutrino in-Earth propagation and detection.  Second, the effective volumes that we use are functions of shower energy, not of neutrino energy.  In our calculation of shower rates below, this choice allows us to account for all possible combinations of $E_\nu$ and $y$ that produce a shower of a given energy $E_{\rm sh}$ (see Section~\ref{section:cs_basics}), and to assess the detectability of that shower.

\begin{figure}[t]
 \centering
 \includegraphics[width=\columnwidth]{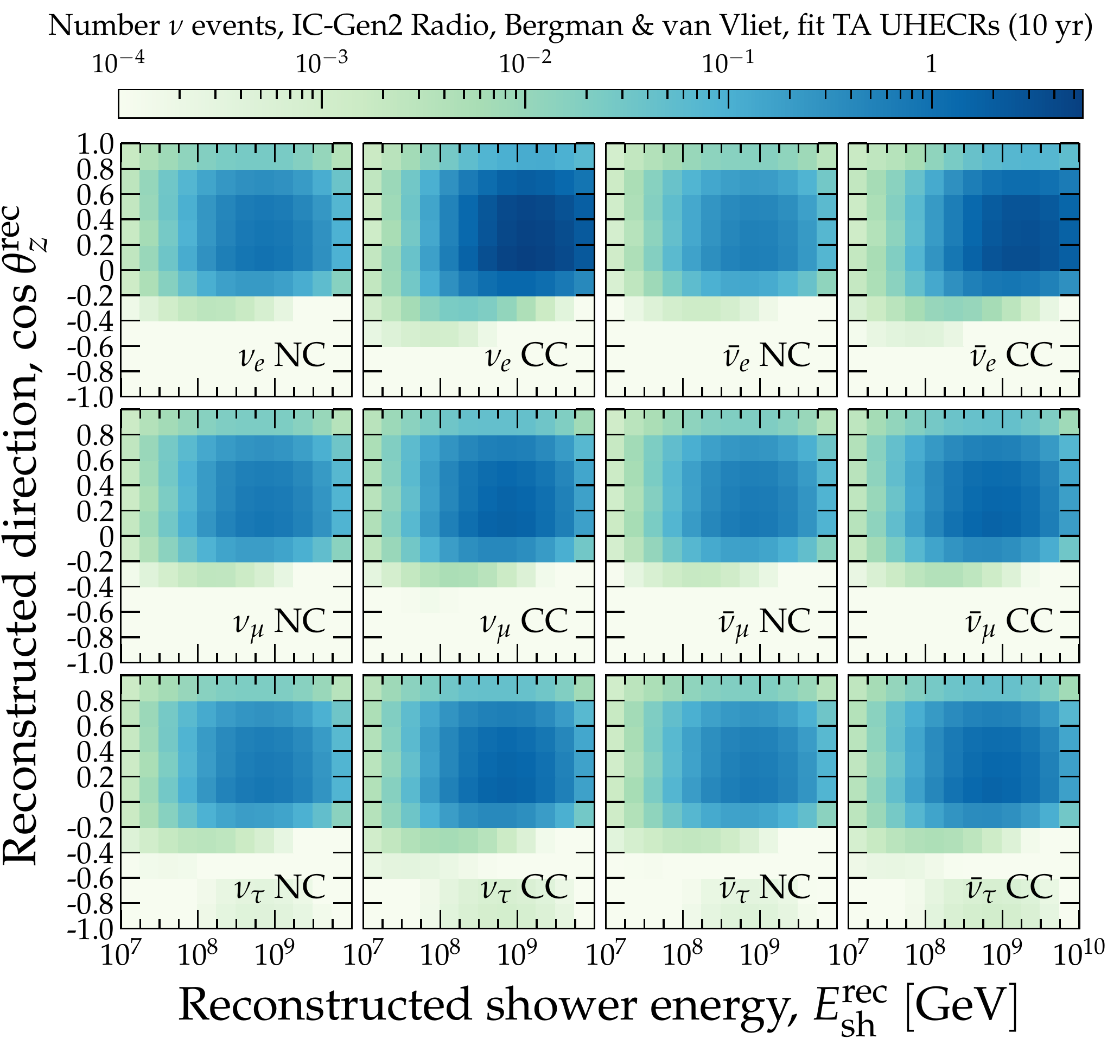}
 \caption{\label{fig:binned_events_per_channel}Mean expected number of neutrino-induced events detected in the IceCube-Gen2 radio array, for the benchmark UHE neutrino flux model 4~\cite{Anker:2020lre} (see Section~\ref{section:uhe_neutrinos}), separately for NC and CC $\nu N$ DIS, and for each neutrino species.  For this plot, for Table~\ref{tab:event_rates_total} and \figu{event_rate}, and in our main results, we use the baseline choices for the energy and angular resolution, $\sigma_\epsilon = 0.1$ and $\sigma_{\theta_z} = 2^\circ$, respectively.  See Section~\ref{section:radio_detection} for details.}
\end{figure}

Figure~\ref{fig:veff_single} shows the resulting effective volumes for single stations and their dependence on shower energy and direction.  The effective volumes are higher for higher shower energy, since the radio signal intensifies with energy.  The directional dependence of the effective volumes is more nuanced.  Broadly stated, shallow antennas have a higher effective volume around the horizon ($\cos \theta_z \approx 0$) and deep antennas have a higher effective volume for downgoing directions ($\cos \theta_z \gtrsim 0$).  This is because deeper antennas are less affected by shadowing due to the changing index of refraction in the 200~m of overhead ice that leads to a downward bending of signal trajectories; see, \eg, \Refe~\cite{Glaser:2019cws}.  Since the capability to measure the UHE $\nu N$ cross section is contingent on the observation of near-horizontal events (see Section~\ref{section:cs_measurement}), shallow antennas are key.  Later, in \figu{deep_vs_shallow}, we quantify their importance.  (In \figu{veff_single}, the effective volume for the deep detector component dips at $\cos\theta_z \approx -0.3$ because the beams of the phased-array trigger system have not been optimized for upgoing neutrinos, as they will anyway be attenuated by propagating through the Earth; see Section~\ref{section:neutrino_propagation} and \figu{binned_events_per_channel}.)

Figure~\ref{fig:veff_total} shows the effective volume of the full IceCube-Gen2 radio array.  We adopt the baseline array design from \Refe~\cite{Hallmann:2021kqk}: 144 hybrid stations, containing shallow and deep detector components, plus 169 shallow-only stations.  Because few showers are expected to be detected simultaneously by multiple radio stations~\cite{Hallmann:2021kqk}, it is straightforward to scale the single-station effective volumes up to the size of the full array: it is equal to the single-station volume of shallow antennas times the number of stations that contain shallow antennas (313) plus the single-station volume of deep antennas times the number of stations that contain deep antennas (144).  Below, when forecasting event rates, we use the full-array effective volumes from \figu{veff_total}, computed separately for the NC interaction of all neutrino flavors or the CC interaction of $\nu_\mu$ and $\nu_\tau$, $V_{{\rm eff}, {\nu_\alpha}}^{\rm NC}$, and for the CC interaction of $\nu_e$, $V_{{\rm eff}, {\nu_e}}^{\rm CC}$.  At the time of writing, the number of shallow and deep stations in the IceCube-Gen2 array is under consideration; our results might provide guidance to optimize it for cross-section measurements.

\subsection{Reconstruction of energy and direction}
\label{section:radio_reconstruction}

The reconstruction of the shower direction and energy requires the measurement of the signal arrival direction,  polarization, viewing angle, \ie, the angle under which the shower is observed, and the distance to the neutrino interaction vertex, as well as accurate modelling of the ice properties to correct for the bending of the signal trajectories due to the changing index of refraction~\cite{Glaser:2019kjh}.  

The signal arrival direction can be determined to sub-degree precision via the signal arrival times to the different antennas of a detector station~\cite{ARIANNA:2020zrg, Kim:2020qyd, ARIANNA:2021pzm}.  The index-of-refraction profile is known well enough to achieve sub-degree precision on the signal direction, as tested in measurements at the South Pole~\cite{ARIANNA:2020zrg}.  The viewing angle can typically be reconstructed to within $1^\circ$~\cite{Gaswint:2021smu, Aguilar:2021uzt, RNO-G:2021zfm}.  The dominant uncertainty on the direction comes from the measurement of the signal polarization~\cite{ARIANNA:2020zrg, RNO-G:2021zfm, Gaswint:2021smu}.  Polarization is generally measured better in shallow detector components because their orthogonal LPDA pairs provide equal sensitivity to both polarization states.  For a shallow station, the capability to measure the polarization has been tested in-situ~\cite{ARIANNA:2020zrg} and via measurements of cosmic rays~\cite{Arianna:2021lnr}. 

The ability to measure the viewing angle and to combine all individual measurements to estimate the arrival direction was quantified in simulation studies using the forward folding technique~\cite{Glaser:2019rxw, Gaswint:2021smu, ARIANNA:2021pzm, RNO-G:2021zfm} and deep neural networks~\cite{Stjarnholm:2021xpj, ARA:2021bss}.  The best resolution of the direction was obtained with a shallow detector station with a 68\% quantile of $3^\circ$, which translates into an uncertainty of the zenith angle of $\sigma_{\theta_z} = 2^\circ$.  We adopt this angular resolution as the baseline in our forecasts below.  Later, in \figu{angular_resolution}, we explore how our results depend on the angular resolution.  

Similar studies have estimated the uncertainty of the reconstructed shower energy~\cite{Anker:2019zcx, Aguilar:2021uzt, Gaswint:2021smu}.  They yield a resolution of approximately 0.1 in $\log_{10}(E_{\rm sh}^{\rm rec} / E_{\rm sh})$, where $E_{\rm sh}^{\rm rec}$ is the reconstructed shower energy and $E_{\rm sh}$ is the real shower energy, or roughly 30\% on a linear scale in the energy range of interest.  We adopt this energy resolution as the baseline in our forecasts below.
\medskip


\subsection{Neutrino-induced event rates in IceCube-Gen2}
\label{section:radio_detection_rates}

In a realistic experimental setting, only the shower energy is measured, not the energy of the neutrino that initiated it.  Thus, in predicting the detected shower rate, we must account for the fact that a shower measured with a certain energy $E_{\rm sh}$ could have been initiated by any combination of neutrino energy, $E_\nu$, and inelasticity, $y$, that satisfies $E_{\rm sh} = E_\nu y$ for NC interactions of all neutrino flavors and CC interactions of $\nu_\mu$ and $\nu_\tau$, or $E_{\rm sh} = E_\nu$ for CC interactions of $\nu_e$.  See Section~\ref{section:radio_detection_basics} for details.  The relative contribution of each possible combination is weighed by the neutrino flux---which determines how many neutrinos of energy $E_\nu$ reach the detector---and by the differential cross section---which determines the chances that the interaction of a neutrino of energy $E_\nu$ has inelasticity $y$.  (While the neutrino energy may be reconstructed from the measured shower energy, doing so requires making additional assumptions~\cite{Glaser:2019kjh, Aguilar:2021uzt}, and is unnecessary for our goals, so we do not attempt that.)

Thus, the differential rate of showers induced in the radio array of IceCube-Gen2 by $\nu_\alpha$ arriving at the detector with flux $\Phi_{\nu_\alpha}^{\rm det}$ after propagating inside Earth is
\begin{widetext}
 \begin{equation}
  \label{equ:spectrum_true}
  \frac{d^2N_{\nu_\alpha}}{dE_{\rm sh} d\cos\theta_z}
  =
  2 \pi T n_t 
  \int_0^1 dy
  \left(
  \left.
  \frac{E_{\nu_\alpha}^{\rm NC}(E_{\rm sh}, y)}{E_{\rm sh}}
  V_{{\rm eff}, \nu_\alpha}^{\rm NC}(E_{\rm sh}, \cos\theta_z)
  \frac{d\sigma_{\nu_\alpha{\rm w}}^{\rm NC}(E_\nu, y)}{dy}
  \Phi^{\rm det}_{\nu_\alpha}(E_\nu,\cos\theta_z)
  \right\vert_{E_\nu = E_{\nu_\alpha}^{\rm NC}(E_{\rm sh}, y)}
  +
  {\rm NC} \to {\rm CC}
  \right)
   \;,
 \end{equation}
\end{widetext}
where $T$ is the exposure time of the detector, $n_t \equiv N_{\rm Av}\rho_{\rm ice} / M_{\rm ice}$ is the number density of water molecules in ice, $N_{\rm Av}$ is Avogadro's number, $\rho_{\rm ice} = 0.9168$~g~cm$^{-3}$ is the density of ice, and $M_{\rm ice} = 18.01528$~g~mol$^{-1}$ is the molar mass of water.  Equation~(\ref{equ:spectrum_true}) accounts for the contributions of NC and CC interactions.  On the right-hand side, the term $E_{\nu_\alpha}^{\rm NC}(E_{\rm sh}, y) / E_{\rm sh}$ re-scales the units of the flux from neutrino energy to shower energy.  The effective volume, $V_{{\rm eff}, \nu_\alpha}^{\rm NC}$, is the full-array volume described in Section~\ref{section:radio_detection_gen2}. The cross section, $\sigma_{\nu_\alpha {\rm w}}^{\rm NC}$, is for neutrino DIS interaction on a water molecule (H$_{2}$O), \ie, $\sigma_{\nu_\alpha {\rm w}}^{\rm NC} = 10 \sigma_{\nu_\alpha p}^{\rm NC} + 8 \sigma_{\nu_\alpha n}^{\rm NC}$, where $ \sigma_{\nu_\alpha p}^{\rm NC}$ and $\sigma_{\nu_\alpha n}^{\rm NC}$ are the BGR18 $\nu_\alpha p$ and $\nu_\alpha n$ cross sections, respectively, and similarly for the CC cross sections~\cite{Bertone:2018dse}.  In \equ{spectrum_true}, the differential shower rate on the left-hand side is evaluated at a shower energy $E_{\rm sh}$.  This determines, on the right-hand side, the neutrino energy at which the integrand is evaluated, \ie,
\begin{equation}
 \label{equ:energy_nu}
 E_{\nu_\alpha}^{i}(E_{\rm sh}, y)
 =
 \left\{
  \begin{array}{cl}
   E_{\rm sh}/y,                    &  {\rm for}~\nu_\alpha~{\rm NC} \\
   E_{\rm sh},                      &  {\rm for}~\nu_e~{\rm CC}      \\
   E_{\rm sh}/y,                     &  {\rm for}~\nu_\mu~{\rm and}~\nu_\tau~{\rm CC}      \\
  \end{array}
 \right. \;,
\end{equation}
for $i = {\rm NC}, {\rm CC}$.  Equation~(\ref{equ:energy_nu}) applies also to anti-neutrinos.  (In our numerical solution of \equ{spectrum_true}, we set the lower limit of the $y$ integral to $10^{-8}$ instead of 0 to prevent the integral from diverging.)

The differential shower rate in \equ{spectrum_true} is the ``true" shower rate.  Because the detector measures energy and direction imperfectly, the true rate is unobservable.  Instead, to produce our results, we use exclusively the ``reconstructed" shower rate, \ie, the rate expressed in terms of the reconstructed shower energy, $E_{\rm sh}^{\rm rec}$, and the reconstructed direction, $\theta_z^{\rm rec}$.  These are the shower quantities that would be observed in a realistic experiment, after accounting for the measurement uncertainties.  To do this, we fold the true shower rate with resolution functions in shower energy and direction, $R_{E_{\rm sh}}$ and $R_{\theta_{z}}$, respectively, that capture the measurement precision of the detector.

\begingroup
\begin{table*}[t!]
 \begin{ruledtabular}
  \caption{\label{tab:event_rates_total}Expected rates of neutrino-induced showers in the radio component of IceCube-Gen2, after $T = 10$~years of exposure time, for the benchmark UHE diffuse neutrino fluxes used in this analysis (see Section~\ref{section:uhe_neutrinos}), and varying the $\nu N$ DIS cross section $\sigma \equiv \sigma_{\nu_\alpha N}$, where $\alpha = e, \mu, \tau$.  The cross section $\sigma_{\rm std}$ is the central BGR18 prediction~\cite{Bertone:2018dse} (see Section~\ref{section:cross_section}), against which we measure variations.  The variation is the same for all flavors of $\nu_\alpha$ and $\bar{\nu}_\alpha$.  Possible flux types are: extrapolation to ultra-high energies ({\Large $\bullet$}), cosmogenic ($\blacksquare$), source (\rotatebox[origin=c]{45}{$\blacksquare$}), and cosmogenic + source ($\rotatebox[origin=c]{90}{\HexaSteel}$).  Results are obtained using the baseline choices for the detector resolution: shower energy resolution of $\sigma_\epsilon = 0.1$, with $\epsilon \equiv \log_{10}(E_{\rm sh}^{\rm rec}/E_{\rm sh})$, and angular resolution of $\sigma_{\theta_{z}} = 2^\circ$; see Section~\ref{section:radio_detection_rates} for details.  In this table, the shower rates shown are binned in a single bin of reconstructed shower energy, $10^7 \leq E_{\rm sh}^{\rm rec}/{\rm GeV} \leq 10^{10}$.  Rates are all-sky, \ie, summed over all values of reconstructed direction, $-1 \leq \cos \theta_z^{\rm rec} \leq 1$; we show separately the fraction of showers that are upgoing or near-horizontal, \ie, with $\theta_z^{\rm rec} > 80^\circ$.  (To produce our main results, instead we bin finely in reconstructed energy and direction; see Section~\ref{section:cs_extracting_binning}.)}
  \centering
  \renewcommand{\arraystretch}{1.4}
  \setlength{\tabcolsep}{1pt}
  \begin{tabular}{ccccccccc}
   \multirow{3}{*}{\#} &
   \multirow{3}{*}{Type} &
   \multirow{3}{*}{UHE $\nu$ flux model} & 
   \multicolumn{6}{c}{Shower rate in IceCube-Gen2 radio array (10~yr)}  \\
   &
   &
   &
   \multicolumn{2}{c}{$\sigma = 0.1\sigma_{\rm std}$} &
   \multicolumn{2}{c}{$\sigma = \sigma_{\rm std}$} &
   \multicolumn{2}{c}{$\sigma = 4\sigma_{\rm std}$} \\
   &
   &
   &
   All-sky &
   $\theta_z^{\rm rec} > 80^\circ$ &
   All-sky &
   $\theta_z^{\rm rec} > 80^\circ$ &
   All-sky &
   $\theta_z^{\rm rec} > 80^\circ$ \\
   \hline
   1 &
   {\Large $\bullet$} &
   IceCube HESE (7.5~yr) extrapolated~\cite{IceCube:2020wum} &
   0.12 &
   62.58 $\%$ &
   0.73 &
   36.86 $\%$ &
   2.57 &
   29.23 $\%$ \\
   2 &
   {\Large $\bullet$} &
   IceCube $\nu_\mu$ (9.5~yr) extrapolated~\cite{IceCube:2021uhz} &
   4.09 &
   57.42 $\%$ &
   26.90 &
   35.34 $\%$ &
   98.13 &
   29.41 $\%$ \\
   3 &
   $\blacksquare$ &
   Heinze {\it et al.}, fit to Auger UHECRs~\cite{Heinze:2019jou} &
   0.12 &
   64.19 $\%$ &
   0.71 &
   37.33 $\%$ &
   2.52 &
   29.65 $\%$ \\
   4 &
   $\blacksquare$ &
   Bergman \& van Vliet, fit to TA UHECRs~\cite{Anker:2020lre} &
   45.29 &
   51.00 $\%$ &
   332.34 &
   33.74 $\%$ &
   1243.57 &
   29.13 $\%$ \\
   5 &
   $\blacksquare$ &
   Rodrigues {\it et al.}, all AGN~\cite{Rodrigues:2020pli} &
   0.14 &
   60.30 $\%$ &
   0.89 &
   36.43 $\%$ &
   3.20 &
   30.04 $\%$ \\
   6 &
   \rotatebox[origin=c]{45}{$\blacksquare$} &
   Rodrigues {\it et al.}, all AGN~\cite{Rodrigues:2020pli} &
   17.33 &
   60.94 $\%$ &
   107.16 &
   36.62 $\%$ &
   385.97 &
   30.11 $\%$ \\
   7 &
   $\blacksquare$ &
   Rodrigues {\it et al.}, HL BL Lacs~\cite{Rodrigues:2020pli} &
   3.32 &
   51.90 $\%$ &
   24.24 &
   34.08 $\%$ &
   89.89 &
   29.40 $\%$ \\
   8 &
   \rotatebox[origin=c]{90}{\HexaSteel} &
   Fang \& Murase, cosmic-ray reservoirs~\cite{Fang:2017zjf} &
   8.25 &
   54.48 $\%$ &
   57.41 &
   34.59 $\%$ &
   211.27 &
   29.33 $\%$ \\
   9 &
   \rotatebox[origin=c]{45}{$\blacksquare$} &
   Fang {\it et al.}, newborn pulsars~\cite{Fang:2013vla} &
   18.66 &
   56.60 $\%$ &
   125.38 &
   35.28 $\%$ &
   485.65 &
   29.73 $\%$ \\
   10 &
   \rotatebox[origin=c]{45}{$\blacksquare$} &
   Padovani {\it et al.}, BL Lacs~\cite{Padovani:2015mba} &
   9.41 &
   61.28 $\%$ &
   57.85 &
   36.67 $\%$ &
   207.58 &
   30.00 $\%$ \\
   11 &
   \rotatebox[origin=c]{90}{\HexaSteel} &
   Muzio {\it et al.}, max. extra $p$ component~\cite{Muzio:2019leu} &
   7.93 &
   51.62 $\%$ &
   56.55 &
   33.57 $\%$ &
   213.28 &
   28.60 $\%$ \\
   12 &
   \rotatebox[origin=c]{90}{\HexaSteel} &
   Muzio {\it et al.}, fit to Auger $\&$ IceCube~\cite{Muzio:2021zud} &
   3.13 &
   66.46 $\%$ &
   17.12 &
   38.06 $\%$ &
   58.93 &
   28.78 $\%$ \\
  \end{tabular}
 \end{ruledtabular}
\end{table*}
\endgroup

\begin{figure*}[t!]
 \centering
 \includegraphics[width=\textwidth]{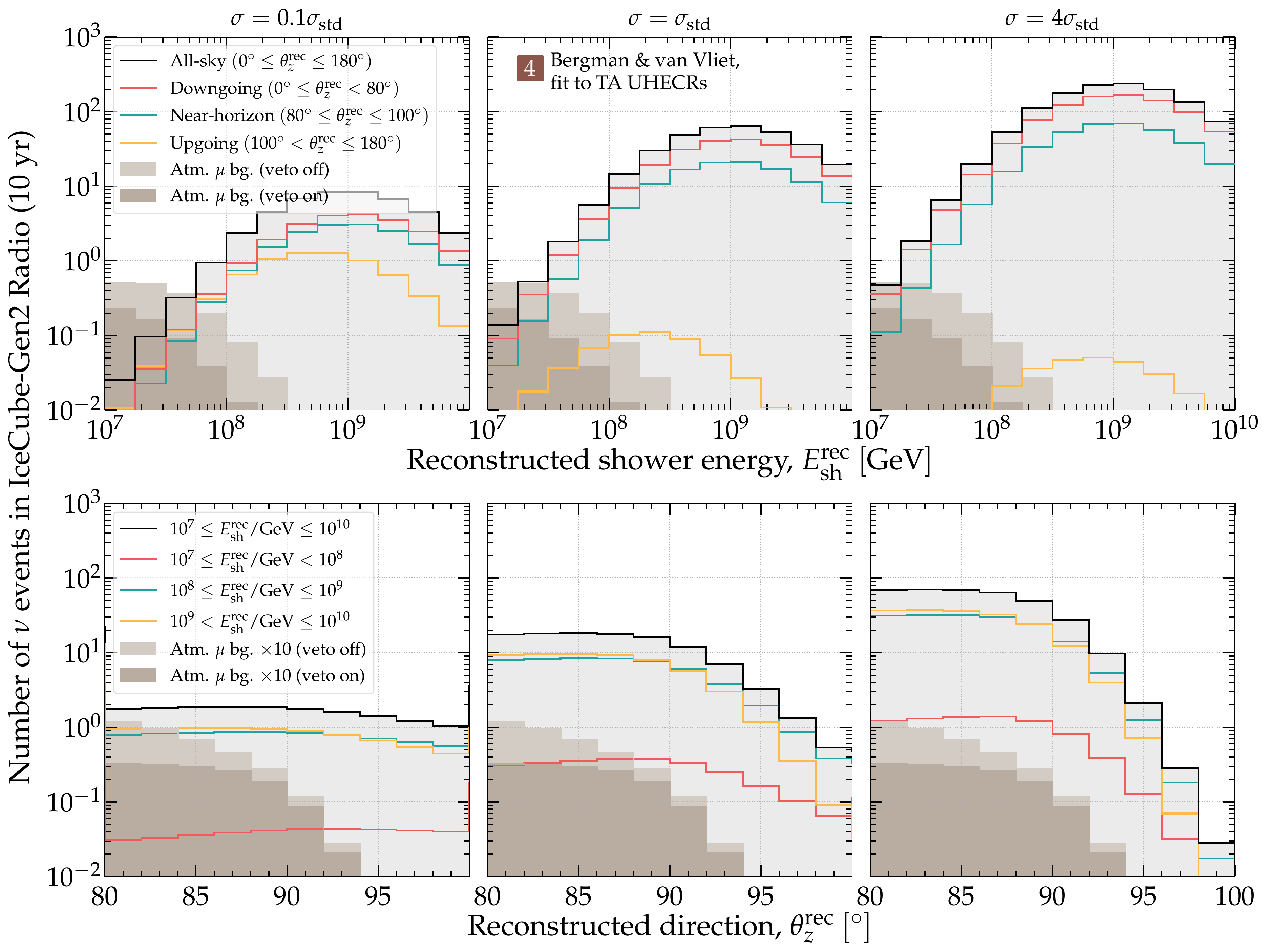}
 \caption{\label{fig:event_rate}Mean expected distribution of neutrino-initiated showers in the radio component of IceCube-Gen2 after 10 years of exposure.  For this plot, we assume the benchmark model 4 for the UHE diffuse neutrino flux~\cite{Anker:2020lre} (see \figu{fluxes}); to produce our results in the main text, we explore all flux models, 1--12 (see Section~\ref{section:uhe_neutrinos}, Tables~\ref{tab:event_rates_total} and \ref{tab:sensitivities}).  Each column shows a different choice of the $\nu N$ DIS cross section that affects the propagation of neutrinos through the Earth and their detection.  The cross section $\sigma_{\rm std}$ is the central value of the BGR18 calculation~\cite{Bertone:2018dse}, which we take as our baseline.  See Section~\ref{section:radio_detection_rates} for details about the calculation of shower rates.  We include the background of showers due to atmospheric muons, which factors into our statistical analysis, computed using the hadronic interaction model {\sc Sybill 2.3c}~\cite{Fedynitch:2018cbl}.  (In the lower row of panels, the background is multiplied by a factor of 10 to make it visible.)  In our analysis, we use exclusively the background after the application of the surface veto (``veto on"); in this plot, we show the case without the veto (``veto off") only for comparison.}
\end{figure*}

For the resolution function in energy, we adopt a Gaussian function centered at the real shower energy, $E_{\rm sh}$, \ie,
\begin{equation}
 \label{equ:energy_res_function}
 R_{E_{\rm sh}}(E_{\rm sh}^{\rm rec}, E_{\rm sh})
 =
 \frac{\mathcal{N}_{E_{\rm sh}}}{\sqrt{2\pi}\sigma_{E_{\rm sh}}}
 e^{-\frac{(E_{\rm sh}^{\rm rec}-E_{\rm sh})^2}{2\sigma_{E_{\rm sh}}^2}} \;,
\end{equation}
where the normalization constant,
\begin{equation}
 \label{equ:energy_res_function_norm}
 \mathcal{N}_{E_{\rm sh}}(E_{\rm sh}^{\rm rec}) 
 \equiv 
 \frac{2} 
 {
 1
 +
 {\rm erf}
 \left(
 \frac{E_{\rm sh}^{\rm rec}}{\sqrt{2}\sigma_{E_{\rm sh}}}
 \right)
 } \;,
\end{equation}
ensures that the integral of $R_{E_{\rm sh}}$ from $E_{\rm sh} = 0$ to $\infty$ equals 1.  The width of the resolution function is $\sigma_{E_{\rm sh}} = 10^{\sigma_\epsilon} E_{\rm sh}$, where $\sigma_\epsilon$ is the spread in the ratio $\epsilon \equiv \log_{10}(E_{\rm sh}^{\rm rec}/E_{\rm sh})$.  Reference~\cite{Aguilar:2021uzt} reported approximately $\sigma_\epsilon = 0.1$, based on simulations performed for RNO-G; we take this value to be representative also of IceCube-Gen2, and use it to produce our results; see Section~\ref{section:radio_reconstruction}.  References~\cite{Anker:2019zcx, Gaswint:2021smu} reported similar results for a shallow detector component in ARIANNA~\cite{Anker:2019zcx, Gaswint:2021smu}.  The choice of $\sigma_\epsilon$ has little effect on our results, as long as it is under control, since the cross section is extracted mainly from the angular distribution of the showers, rather than from their energy distribution; see Section~\ref{section:cs_measurement}.

For the resolution function in direction, we adopt a Gaussian function centered at the real direction, $\theta_z$, \ie,
\begin{equation}
 \label{equ:angle_res_function}
 R_{\theta_{z}}(\theta_{z}^{\rm rec}, \theta_z)
 =
 \frac{ \mathcal{N}_{\theta_z}}{\sqrt{2\pi}\sigma_{\theta_z}}
 e^{-\frac{(\theta_{z}^{\rm rec}-\theta_z)^2}{2\sigma_{\theta_z}^2}} \;,
\end{equation}
where the normalization constant, 
\begin{equation}
 \label{equ:angle_res_function_norm}
 \mathcal{N}_{\theta_z}(\theta_z^{\rm rec}) 
 \equiv 
 \frac{2}
 {
 {\rm erf}
 \left(
 \frac{\pi-\theta_z^{\rm rec}}{\sqrt{2}\sigma_{\theta_z}}
 \right)
 +
 {\rm erf}
 \left(
 \frac{\theta_z^{\rm rec}}{\sqrt{2}\sigma_{\theta_z}}
 \right)
 } \;,
\end{equation}
ensures that the integral of $R_{\theta_{z}}$ from $\theta_z = 0$ to $\pi$ equals 1.  We use $\sigma_{\theta_z} = 2^\circ$ to produce our main results; see Section~\ref{section:cs_measurement}.  In Section~\ref{section:results}, we study the effect of varying $\sigma_{\theta_z}$ on measuring the $\nu N$ cross section.

Folding the resolution functions, Eqs.~(\ref{equ:energy_res_function})--(\ref{equ:angle_res_function_norm}), with the true shower rate, \equ{spectrum_true}, yields the reconstructed shower rate, \ie,
\begin{eqnarray}
 \label{equ:spectrum_rec}
 \frac{d^2N_{\nu_\alpha}}
 {dE_{\rm sh}^{\rm rec} d\theta_{z}^{\rm rec}}
 &=&
 \int_{-1}^{+1} d\cos\theta_z
 \int_{0}^{\infty} dE_{\rm sh}
 \frac{d^2N_{\nu_\alpha}}{dE_{\rm sh} d\cos\theta_z} 
 \nonumber \\
 &&
 \times~
 \mathcal{R}_{E_{\rm sh}}(E_{\rm sh}^{\rm rec}, E_{{\rm sh}})~
 \mathcal{R}_{\theta_{z}}(\theta_{z}^{\rm rec}, \theta_z) \;.
\end{eqnarray}
Finally, the total shower rate is the contribution from $\nu_\alpha$ and $\bar{\nu}_\alpha$ of all flavors, \ie, 
\begin{equation}
 \label{equ:shower_rate_rec_total}
 \frac{d^2N_\nu}
 {dE_{\rm sh}^{\rm rec} d\theta_{z}^{\rm rec}}
 =
 \sum_{\alpha = e, \mu, \tau}
 \frac{d^2N_{\nu_\alpha}}
 {dE_{\rm sh}^{\rm rec} d\theta_{z}^{\rm rec}}
 +
 \nu_\alpha \to \bar{\nu}_\alpha \;.
\end{equation}
We use \equ{shower_rate_rec_total} to forecast shower rates throughout our analysis, integrated in bins of $E_{\rm sh}^{\rm rec}$ and $\cos \theta_z^{\rm rec}$.  In Section~\ref{section:cs_extracting_binning}, we comment on the effect of the choice of binning on the measurement of the cross section.

Figure~\ref{fig:binned_events_per_channel} shows the mean expected number of neutrino-induced showers for benchmark UHE neutrino flux model 4~\cite{Anker:2020lre} as illustration, broken down into the contribution of each interaction type and neutrino species.  Because flux model 4 has comparable fluxes of each neutrino species at the surface of the Earth (see \figu{benchmark_spectra_per_species}), the differences between the rates of different channels in \figu{binned_events_per_channel} are due mainly to differences in the in-Earth propagation of the different species, the connection between neutrino and shower energy, \equ{energy_nu}, and the effective volumes for $\nu_e$ CC interactions and for all other interaction channels (see \figu{veff_total}).

Figure~\ref{fig:binned_events_per_channel} displays features that are common to all interaction channels and all benchmark flux models.  Barring differences in the flux of the different species that reach the detector, the following observations hold generally.  First, the dominant contribution comes from the CC interactions of $\nu_e$ and $\bar{\nu}_e$, the only two cases for which $E_{\rm sh} = E_\nu$; see \equ{energy_nu}.  (In \figu{binned_events_per_channel}, there are more showers in the range $10^9$--$10^{10}$~GeV because those showers receive the full neutrino energy, and the flux model 4 peaks within that range; see \figu{fluxes}.)  Second, the NC interactions of all species, and the CC interactions of $\nu_\mu$, $\bar{\nu}_\mu$, $\nu_\tau$, and $\bar{\nu}_\tau$, contribute at about the same level, since $E_{\rm sh} = y E_\nu$ for all of them; see \equ{energy_nu}.  Third, in all channels, because of in-Earth attenuation (see Section~\ref{section:neutrino_propagation}), events are mainly downgoing, \ie, $0.2 \leq \cos \theta_z^{\rm rec} \lesssim 1$, and, to a lesser extent, near-horizontal, \ie, $-0.2 \leq \cos \theta_z^{\rm rec} \leq 0.2$.  Fourth, all channels have a deficit around $\cos \theta_z^{\rm rec} = 1$, because the neutrino interaction takes place almost always below the deepest antenna of the station, and all radio signals are emitted downward for vertical neutrino directions; see \figu{veff_total}.  Fifth, and finally, all channels have a deficit around $E_{\rm sh}^{\rm rec} = 10^7$~GeV, because the detector loses sensitivity quickly at low shower energies, where radio emission is weak; see \figu{veff_total}.

Table~\ref{tab:event_rates_total} shows that the IceCube-Gen2 all-sky shower rate grows with the $\nu N$ cross section, regardless of the choice of benchmark flux model.  On the one hand, for all values of the cross section, the shower rate is dominated by downgoing showers, with $\theta_z^{\rm rec} \leq 80^\circ$, as expected from Section~\ref{section:cs_measurement}. On the other hand, the fraction of near-horizontal and upgoing showers ($\theta_z^{\rm rec} > 80^\circ$) decreases with the cross section.  These two observations reveal that, in the all-sky shower rate, the main dependence on the cross section comes from its role in the detector, rather than in the in-Earth attenuation of the neutrino flux; see \equ{event_rate_simple} for a simplified picture.  Table~\ref{tab:event_rates_total} confirms what \equ{event_rate_simple} professed: as the attenuation becomes more important, with rising cross section, the rate of near-horizontal and upgoing showers decreases.  This is where the sensitivity to the cross section comes from.  Table~\ref{tab:event_rates_total} shows that the changes in the shower rate with the cross section are roughly comparable for all benchmark flux models; differences between them are due to their different energy spectra. 

Figure~\ref{fig:event_rate} illustrates the behavior seen in Table~\ref{tab:event_rates_total} for one  benchmark flux model; other models behave similarly.  The shape of the shower distribution in $E_{\rm sh}^{\rm rec}$ is largely insensitive to the value of the cross section: changes in the cross section merely scale the distribution down---if the cross section is smaller---or up---if the cross section is larger.  This is because the all-sky shower rate is dominated by downgoing showers, initiated by neutrinos whose flux is unattenuated due to traveling a short distance inside Earth.  Equation~(\ref{equ:event_rate_simple}) shows that for downgoing showers, because there is little attenuation, the rate scales roughly linearly with the cross section.  In contrast, the shape of the shower distribution in $\cos \theta_z^{\rm rec}$ is strongly affected by the value of the cross section.  Equation~(\ref{equ:event_rate_simple}) shows that, due to the exponential attenuation, a smaller cross section mitigates in-Earth attenuation and flattens the distribution in $\cos \theta_z^{\rm rec}$, while a larger cross section enhances the attenuation and tilts the distribution in $\cos \theta_z^{\rm rec}$.  This exponential angular dependence of the shower rate on the cross section---rather than the linear dependence of the all-sky shower rate---is where most of the sensitivity to the cross section comes from.  The tilting of the angular distribution becomes stronger the larger the value of $\theta_z^{\rm rec}$, \ie, the longer the distance neutrinos travel inside the Earth.  However, for upgoing neutrinos, the flux is nearly fully attenuated, even for the most optimistic diffuse flux models.  Therefore, in practice, the sensitivity to the cross section comes from the angular distribution of showers coming from around the horizon.  We pay special attention to them later, in Section~\ref{section:cs_extracting_binning}, by binning them in fine angular bins.


\section{Measuring the cross section}
\label{section:cs_extracting}

We forecast the capability of IceCube-Gen2 to measure the UHE $\nu N$ DIS cross section by means of its effect on the predicted shower rates.  Because of the degeneracy that exists between the cross section and the neutrino flux normalization when computing shower rates (see Section~\ref{section:cs_measurement_overview}), our forecasts are always for their simultaneous measurement.
We use the methods introduced in Section~\ref{section:radio_detection_rates} to generate samples of showers for different choices of their values.  We perform the analysis below separately for each of the UHE neutrino flux models 1--12 introduced in Section~\ref{section:uhe_neutrinos}.


\subsection{Overview}
\label{section:cs_extracting_overview}

For a choice of flux model, we start by generating the ``real" shower sample, \ie, a sample of showers distributed and binned in $E_{\rm sh}^{\rm rec}$ and $\cos \theta_z^{\rm rec}$, computed using the ``real" values of the cross section and flux normalization.  For the cross section, the real value is given by the central value of the BGR18 DIS calculation, $\sigma_{\rm std}$.  For the flux normalization, the real value, $\Phi_{0, {\rm std}}$, is given by the model flux evaluated at a reference energy of $E_{\nu,0} = 10^8$~GeV.  The resulting real shower sample is the mean expected sample predicted for IceCube-Gen2.  Later, in Section~\ref{section:cs_extracting_stat_proc}, we use a statistical procedure to assess how well we can recover the real values of the cross section and flux normalization in a realistic experimental setting.  As part of that procedure, we generate test shower samples using values of the cross section, $\sigma_{\nu_\alpha N}$, and flux normalization, $\Phi_0$, different from the real values, and compare those shower samples to the real one, including the presence of a non-neutrino background.

To generate test shower samples, we parametrize deviations from the real value of the $\nu N$ DIS cross section, within the range $E_\nu = 10^7$--$10^{10}$~GeV, via
\begin{equation}
 \label{equ:cs_norm}
 f_\sigma
 \equiv
 \sigma / \sigma_{\rm std} \;.
\end{equation}
We consider only energy-independent modifications of the cross section, as in \Refe~\cite{IceCube:2017roe}, so $f_\sigma$ is constant in energy.  Changing $f_\sigma$ shifts the value of the cross section up or down from the central BGR18 value.  The same value of $f_\sigma$ applies to all flavors of $\nu_\alpha$ and $\bar{\nu}_\alpha$, and to interactions on protons ($N = p$) and neutrons ($N = n$).  The real shower sample is computed using $f_\sigma = 1$. Later, in Section~\ref{section:cs_extracting_stat_proc},we allow its value to float generously, from 0.01 to 100, and generate a large number of test shower distributions for its different values.  

Changing $f_\sigma$ affects the neutrino flux that reaches the detector---by changing the propagation through the Earth (see Section~\ref{section:neutrino_propagation})---and the cross section at the detection (see Section~\ref{section:radio_detection_rates}).  For each choice of $f_\sigma$, we compute both effects.  In our analysis, changing $f_\sigma$ affects only the $\nu N$ DIS cross section off the nucleon partons; the remaining, sub-leading neutrino interaction channels that act during neutrino propagation through the Earth (see Section~\ref{section:neutrino_propagation}) remain unchanged.  Section~\ref{section:limitations} comments on possible generalizations. 

Similarly, for each choice of flux model out of models 1--12, we parametrize deviations from the real value of the flux normalization, via
\begin{equation}
 \label{equ:def_f_Phi}
 f_\Phi
 \equiv
 \Phi_0 / \Phi_{0, {\rm std}} \;,
\end{equation}
where the value of $\Phi_{0, {\rm std}}$ is unique to each flux model.
Changing $f_\Phi$ shifts the flux up or down from its central value, but the shape of the neutrino spectrum, flavor composition, and $\nu$ {\it vs.}~$\bar{\nu}$ content are unchanged.  The real shower sample is computed using $f_\Phi = 1$.  The same value of $f_\Phi$ applies to all flavors of $\nu_\alpha$ and $\bar{\nu}_\alpha$.  Later, in Section~\ref{section:cs_extracting_stat_proc}, we allow its value to float generously, and generate a large number of test shower distributions for its different values.  We vary $f_\Phi$ indirectly: for a given choice of flux model out of 1--12, we allow the value of $E_{\nu,0}^2\Phi_0$ to float from $10^{-12}$ to $10^{-7}$~GeV~cm$^{-2}$~s$^{-1}$~sr$^{-1}$, and compute $f_\Phi$ for each value of $E_{\nu,0}^2\Phi_0$.  Section~\ref{section:limitations} comments on possible generalizations of this procedure.


\subsection{Non-neutrino backgrounds}
\label{section:background}

In the radio component of IceCube-Gen2, in addition to neutrino-initiated events, we expect a background of irreducible non-neutrino-induced events that could mimic a neutrino signal.  There are two main such backgrounds: in-ice showers induced by high-energy atmospheric muons created by cosmic-ray interactions in the atmosphere~\cite{Garcia-Fernandez:2020dhb}, and air-shower cores, \ie, cores of showers that start in the atmosphere, induced mainly by cosmic rays, and that continue developing downwards in the ice~\cite{DeKockere:2022bto}.  The latter lead to in-ice Askaryan radiation that can be reflected back upwards by reflection layers that are known to be present in deeper ice.  In our forecasts, we account only for the background of showers due to atmospheric muons, by generating their expected distribution in reconstructed energy and direction, as described below.  We do not account for the background of cosmic-ray air-shower cores, since their estimates are presently uncertain, but comment on them in Section~\ref{section:limitations}.

\begin{figure}[t]
 \centering
 \includegraphics[width=\columnwidth]{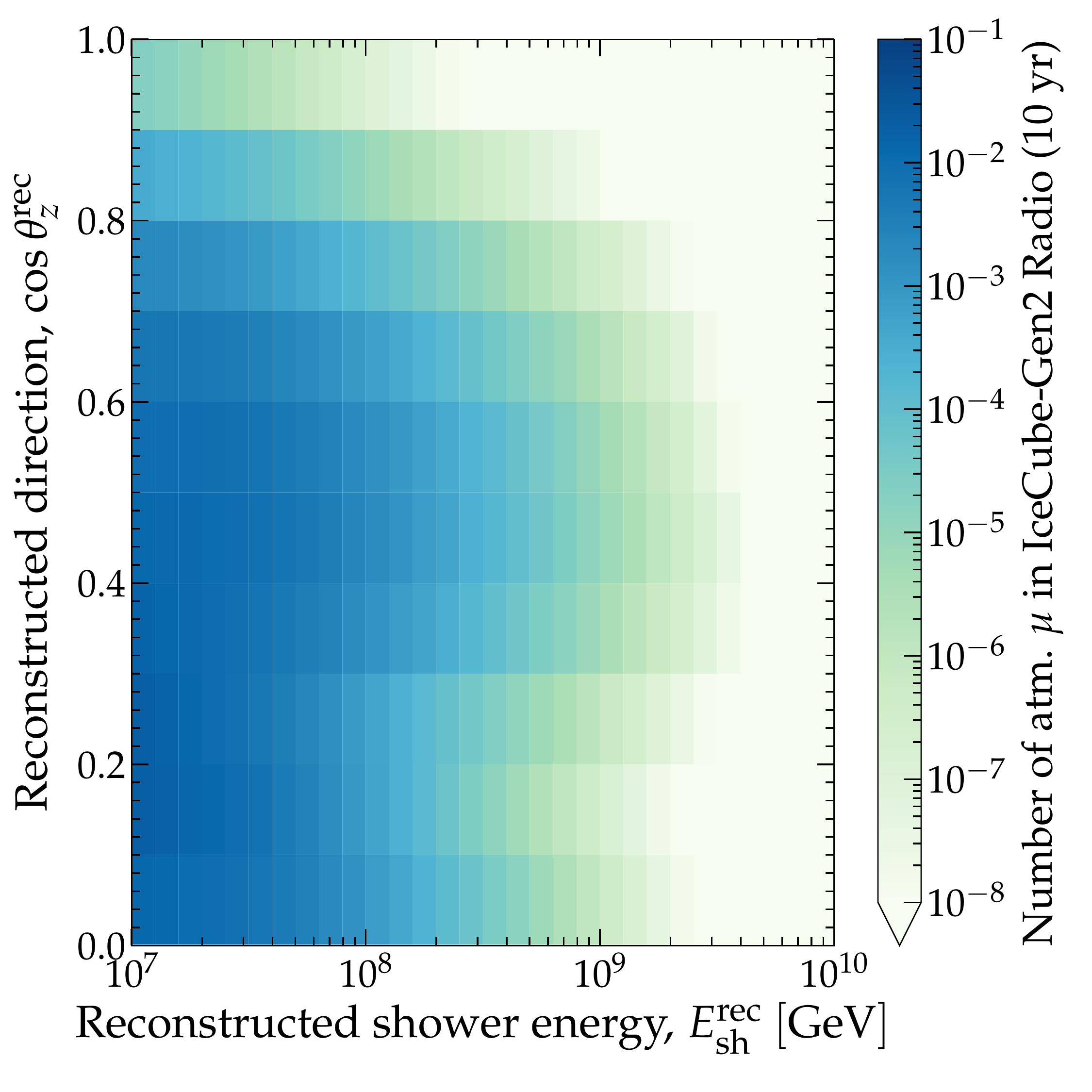}
 \caption{\label{fig:muon_background}Mean expected number of background showers initiated by high-energy atmospheric muons in the baseline design of the IceCube-Gen2 radio array in 10~years of exposure time.  The energy and angular resolution are the same as the baseline choices for neutrino-initiated events, \ie, $\sigma_\epsilon = 0.1$ and $\sigma_{\theta_z} = 2^\circ$, respectively.  The hadronic interaction model used is {\sc Sybill 2.3c}~\cite{Fedynitch:2018cbl}.  The background rates in this plot and in our analysis are after the application of a cosmic-ray veto used to suppress the background.  See the main text for details.}
\end{figure}

Figure~\ref{fig:muon_background} shows the mean distribution of muon-induced shower events in the radio component of IceCube-Gen2 \cite{Garcia-Fernandez:2020dhb, Glaser:2021hfi, Hallmann:2021kqk}; see also \figu{event_rate}.  We estimate it using the hadronic interaction model {\sc Sybill 2.3c}~\cite{Fedynitch:2018cbl} and apply a veto coming from the detection of atmospheric muons by a surface array of cosmic-ray detector tanks, that helps to mitigates the rate.  We expect roughly 0.2 showers per year in the range $10^4$--$10^{11}$~GeV, and $< 0.1$ showers per year in the range $10^7$--$10^{10}$~GeV relevant for our forecasts.  These rates are subject to significant uncertainties, due mainly to the high-energy muon flux, but the ability to measure the cross section in IceCube-Gen2 will not be affected by the uncertainties; we comment on this below.  The background of showers from atmospheric muons is direction- and energy-dependent.  Figure~\ref{fig:event_rate} shows that most of the showers are in the range $10^7$--$10^9$~GeV; thus, the highest-energy end of the shower spectrum is nearly free from this background.  Since it is unlikely that a muon produced in the atmosphere reaches the detector from below the horizon, the background for $\theta_z > 90^\circ$ is zero (this changes slightly after applying angular smearing below).  

Similarly to how we use the reconstructed neutrino-induced shower rate (see Section~\ref{section:radio_detection_rates}) in our analysis, we also use the reconstructed rate of muon-induced showers; this is what \figu{muon_background} shows.  We compute it in the same way: by convolving the true rate of muon-induced showers with the resolution functions in Eqs.~(\ref{equ:energy_res_function})--(\ref{equ:angle_res_function_norm}), and expressing the result in terms of $E_{\rm sh}^{\rm rec}$ and $\cos \theta_z^{\rm rec}$.  At all times, we use the same energy and angular resolution for muon-induced showers as for neutrino-induced showers.  As a result, a tiny fraction of the muon-induced showers leak into the highest energies and into directions just below the horizon, \ie, for $\theta_z^{\rm rec} \gtrsim 90^\circ$; see \figu{event_rate}.


\subsection{Choice of shower binning}
\label{section:cs_extracting_binning}

To produce our forecasts, we bin the real and test shower samples (see above) in fine bins of $E_{\rm sh}^{\rm rec}$ and $\cos \theta_z^{\rm rec}$.  Accordingly, in our statistical analysis we use a binned likelihood function; see Section~\ref{section:cs_extracting_stat_proc}.  The bins of reconstructed shower energy and direction that we use are equal in size to or larger than the width of the detector energy and angular resolution, \ie, $\sigma_{\theta_z}$ and $\sigma_{E_{\rm sh}}$, respectively, from Eqs.~(\ref{equ:energy_res_function}) and (\ref{equ:angle_res_function}).  Section~\ref{section:limitations} comments on the possible future use of an unbinned likelihood. 

When binning in $\cos \theta_z^{\rm rec}$, we pay special attention to Earth-skimming neutrinos, which have the strongest sensitivity to the cross section; see Section~\ref{section:cs_measurement}.  For downgoing showers, we expect a large rate; but, since they are largely unattenuated by their passage through Earth, they do not carry a significant angular dependence induced by the cross section. Thus, for them we use a single bin covering $\theta_z^{\rm rec} \in [0^\circ, 80^\circ]$.  For horizontal and near-horizontal showers, our region of interest, we use ten bins, equally spaced between $\theta_z^{\rm rec} = 80^\circ$ and $100^\circ$, each with a bin size equal to the angular resolution of the detector, $\sigma_{\theta_z}$.  To produce our main results, we use the baseline angular resolution of $\sigma_{\theta_z} = 2^\circ$; later, we explore alternative choices (see \figu{angular_resolution}). For upgoing showers, of which there are few due to the attenuation of the flux inside Earth, we use two large angular bins: one for $\theta_z^{\rm rec} \in [100^\circ, 110^\circ]$, which is populated only if the flux is high and the cross section is low (see \figu{event_rate}), and one for $\theta_z^{\rm rec} \in [110^\circ, 180^\circ]$, which is typically nearly empty.  For upgoing showers, using finer binning would lead to a large number of empty, uninformative bins. 

When binning in $E_{\rm sh}^{\rm rec}$, we use twelve bins equally spaced in logarithmic scale from $\log_{10}(E_{\rm sh}^{\rm rec}/{\rm GeV}) = 7$ to 10.  Even though in our analysis we allow only for energy-independent re-scaling of the cross section from its predicted central value (see Section~\ref{section:cs_extracting_overview}), the distribution of showers in energy still matters when extracting the cross section, for three reasons.  First, during neutrino propagation inside the Earth an energy-independent re-scaling of the cross section affects the flux attenuation differently at different energies, due to the non-linear dependence on energy of the regeneration of neutrinos in NC interactions and $\nu_\tau$ regeneration; see \figu{attenuation}.  Second, the distribution of showers in energy reflects the distribution of neutrino energies in the diffuse flux.  Third, showers from the background of atmospheric muons are concentrated at low energies: above a few times $10^7$~GeV, there are nearly none, so binning in energy enhances the statistical power of the neutrino-induced showers in that range.  Naturally, the choice of energy binning is relevant only when the predicted number of showers is relatively large.  If shower rate is low, like for benchmark flux models 1, 3, and 5, we have checked that the choice of $E_{\rm sh}^{\rm rec}$ binning does not affect cross-section measurements.


\subsection{Statistical analysis}
\label{section:cs_extracting_stat_proc}

To produce our forecasts, we use a Bayesian statistical analysis.  Section~\ref{section:cs_extracting_overview} outlined the procedure; below, we present it in detail.  We carry the procedure separately for each UHE neutrino flux model, 1--12 (see Section~\ref{section:uhe_neutrinos}). 

First, for a given choice of flux model we generate the real shower sample ($r$), \ie, the mean expected number of showers assuming the real values $f_\sigma = 1$ and $f_\Phi = 1$ (see Section~\ref{section:cs_extracting_overview}), binned in $E_{\rm sh}^{\rm rec}$ and $\cos \theta_z^{\rm rec}$ as described in Section~\ref{section:cs_extracting_binning}.  In the $i$-th bin in $E_{\rm sh}^{\rm rec}$ and the $j$-th bin in $\cos \theta_z^{\rm rec}$ of the real shower sample, the mean number of neutrino-induced showers is $\bar{N}_{\nu, ij}^{(r)}$.  To this bin, we associate a Poisson probability distribution, with mean $\bar{N}_{\nu, ij}^{(r)}$, of observing $N_{\nu, ij}^{(r)}$ showers in a particular realization of the shower distribution, \ie,
\begin{equation}
 \label{equ:poisson_nu_real}
 \mathcal{P}_{\nu, ij}\left(N_{\nu, ij}^{(r)}\right)
 =
 \frac{
 \left(\bar{N}_{\nu, ij}^{(r)}\right)^{N_{\nu, ij}^{(r)}} e^{-\bar{N}_{\nu, ij}^{(r)}} 
 }
 {N_{\nu, ij}^{(r)}!} \;.
\end{equation}
Similarly, in each bin, the mean number of background muon-induced showers is $\bar{N}_{\mu, ij}^{(r)}$, and we associate to it a Poisson probability distribution, given by \equ{poisson_nu_real} but with $\bar{N}_{\nu, ij}^{(r)} \to \bar{N}_{\mu, ij}^{(r)}$ and $N_{\nu, ij}^{(r)} \to N_{\mu, ij}^{(r)}$, of observing $N_{\mu, ij}^{(r)}$ muon-induced showers in a particular realization of the shower distribution.  Thus, in a particular {\it observed} realization of the real shower distribution, the total number of showers in the bin is $N_{{\rm obs}, ij} = N_{\nu, ij}^{(r)} + N_{\mu, ij}^{(r)}$, where the neutrino and muon contributions are sampled from their respective probability distributions.

Next, for the same choice of flux model, we generate test shower samples ($t$) for values of $f_\sigma$ and $f_\Phi$ that are in general different from their real values.  In each bin, the mean expected numbers of neutrino-induced and muon-induced showers are, respectively, $\bar{N}_{\nu, ij}^{(t)}(\boldsymbol\theta)$ and $\bar{N}_{\mu, ij}^{(t)}(\boldsymbol\theta)$, where we use $\boldsymbol\theta \equiv (f_\sigma, f_\Phi)$ as shorthand.  The total mean expected number of showers in this bin is $\bar{N}_{{\rm test}, ij}(\boldsymbol\theta) = \bar{N}_{\nu, ij}^{(t)}(\boldsymbol\theta) + \bar{N}_{\mu, ij}^{(t)}(\boldsymbol\theta)$.

Given a specific random realization of the real shower distribution, generated following the above procedure, we quantify its compatibility with the mean test shower distribution expected for a particular choice of $\boldsymbol\theta$, in each bin, via the likelihood function
\begin{equation}
 \label{equ:likelihood_partial}
 \mathcal{L}_{ij}(\boldsymbol\theta)
 =
 \frac{
 \bar{N}_{{\rm test},ij}(\boldsymbol\theta)
 ^{N_{{\rm obs},ij}}
 e^{-\bar{N}_{{\rm test},ij}(\boldsymbol\theta)}
 }
 {
 N_{{\rm obs},ij}!
 } \;.
\end{equation}
The full likelihood function is the product of all the partial likelihood functions, \ie,
\begin{equation}
 \label{equ:likelihood}
 \mathcal{L}(\boldsymbol\theta)
 =
 \prod_{i=1}^{N_{E_{\rm sh}^{\rm rec}}} \prod_{j=1}^{N_{\theta_z^{\rm rec}}} \mathcal{L}_{ij}(\boldsymbol\theta) \;,
\end{equation}
where $N_{E_{\rm sh}^{\rm rec}}$ and $N_{\theta_z^{\rm rec}}$ are, respectively, the number of bins in $E_{\rm sh}^{\rm rec}$ and $\theta_z^{\rm rec}$; see Section~\ref{section:cs_extracting_binning} for details.  The posterior probability distribution is
\begin{equation}
 \label{equ:posterior}
 \mathcal{P}(\boldsymbol\theta)
 =
 \mathcal{L}(\boldsymbol\theta)
 \pi(\boldsymbol\theta) \;,
\end{equation}
where $\pi(\boldsymbol\theta) \equiv \pi(f_\sigma) \pi(f_\Phi)$, $\pi(f_\sigma)$ is the prior on $f_\sigma$, and $\pi(f_\Phi)$ is the prior on $f_\Phi$.  

We use wide priors in $\log_{10} f_\sigma$ and $\log_{10} f_\Phi$ to facilitate exploring large possible deviations from the real values of $f_\sigma = 1$ and $f_\Phi = 1$.  The priors are flat to avoid introducing unnecessary bias.  For $\log_{10} f_\sigma$, the prior spans the interval from -2 and 2; this range covers all reasonable modifications of the cross section.  For $\log_{10} f_\Phi$, the prior spans a different interval for each flux model.  For a given flux model, it is defined by varying the flux at $E_{\nu,0} = 10^8$~GeV from $E_{\nu,0}^2 \Phi(E_{\nu,0}) = 10^{-12}$ to $10^{-7}$~GeV~cm$^{-2}$~s$^{-1}$~sr$^{-1}$ and using \equ{def_f_Phi} to compute $f_\Phi$.  The low end of this interval corresponds, in practice, to an unobservable flux.  The high end exceeds the upper limits on the flux from IceCube and Auger (see \figu{fluxes}).  However, this does not put our analysis in tension: the IceCube and Auger limits were obtained under the assumption that the $\nu N$ cross section has its nominal value ($f_\sigma = 1$) while, in our analysis, when we explore large reductions of the cross section ($f_\sigma \ll 1$), they can be traded off for large enhancements of the flux while keeping the number of showers unchanged; see Section~\ref{section:cs_measurement}.

For a particular random realization of the real shower distribution, we compute the posterior, \equ{posterior}, as a function of $f_\sigma$ and $f_\Phi$.  We generate a large number of different random realizations of the real shower distribution using the above prescription; for each, we compute the associated posterior.  Since our goal is to produce forecasts for the average sensitivity of IceCube-Gen2, we take the arithmetic average, $\langle\mathcal{P}\rangle$, of all the posteriors thus generated.  This is equivalent to marginalizing over the Poisson distribution in each bin of the real shower distribution.  In this way, our forecasts reflect the average sensitivity of the detector, taking into account statistical fluctuations.  In an actual detection in IceCube-Gen2, and assuming that $f_\sigma = 1$ and $f_\Phi = 1$ are indeed the real values, the posterior computed with real data would be one of the random realizations that we averaged over.

Below, to produce our forecasts, we maximize $\langle \mathcal{P} \rangle$ with respect to $f_\sigma$ and $f_\Phi$, and find their best-fit values and credible intervals.  We report results either as two-dimensional posteriors (see Figs.~\ref{fig:sensitivity} and \ref{fig:all_2dposteriors}), or as one-dimensional posteriors obtained by marginalizing the full posterior over one of the two parameters (see Table~\ref{tab:sensitivities} and \figu{violin}), \ie, $\langle\mathcal{P}(f_\sigma)\rangle_{f_\sigma} = \int df_\Phi \langle\mathcal{P}(f_\sigma, f_\Phi)\rangle$ for $f_\sigma$, and equivalently for $f_\Phi$.


\section{Results}
\label{section:results}

Below, to produce our main results, we use the baseline design of the IceCube-Gen2 radio array, \ie, 144 hybrid (deep + shallow) stations plus 169 shallow-only stations, the baseline detector energy resolution, $\sigma_\epsilon = 0.1$, and angular resolution, $\sigma_{\theta_z} = 2^\circ$, and $T = 10$~years of exposure time.  We state explicitly when we vary these parameters.

\begin{figure}[t]
    \centering
    \includegraphics[width=\columnwidth]{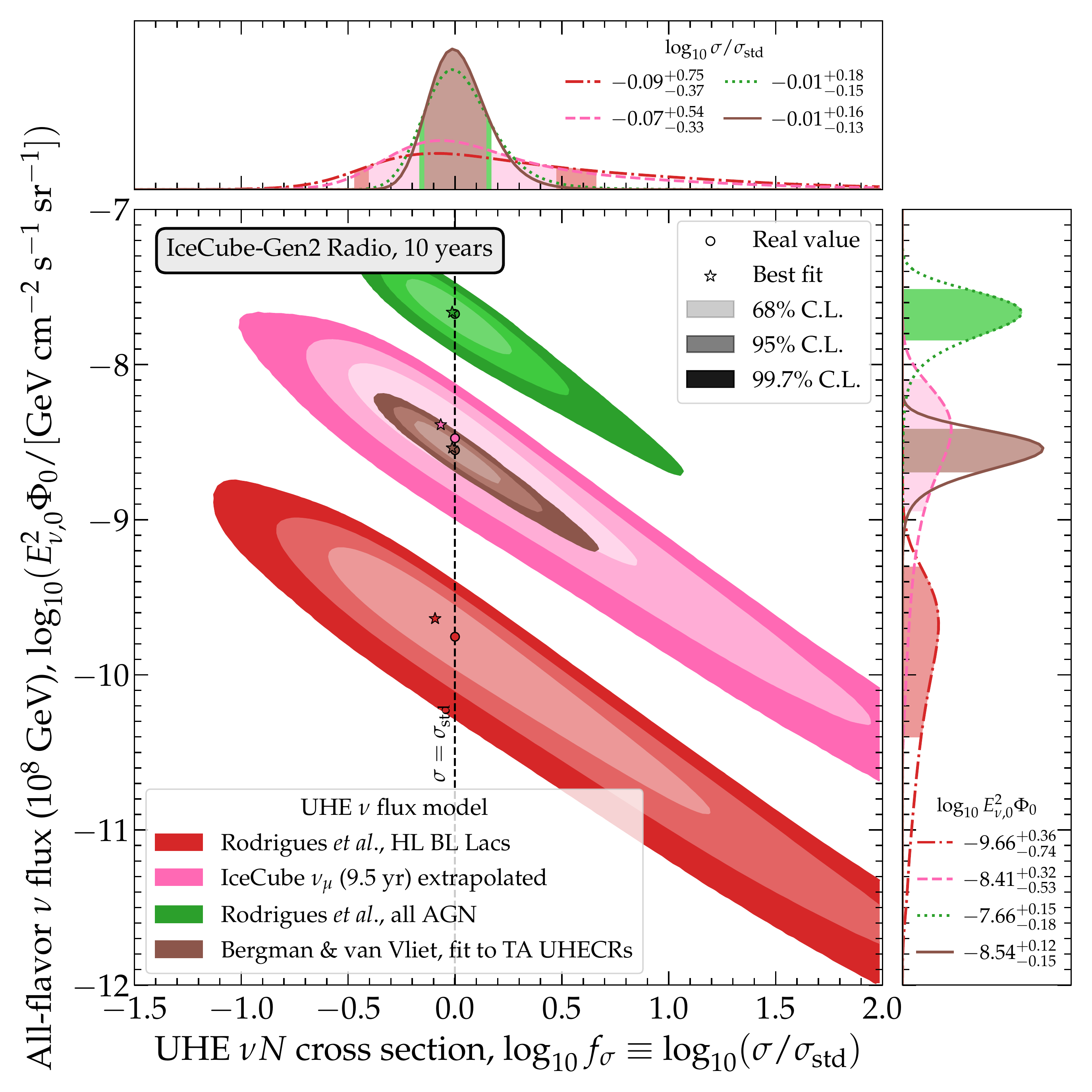}
    \caption{Mean expected capability of the radio component of IceCube-Gen2 to jointly measure the UHE $\nu N$ DIS cross section and the diffuse neutrino flux normalization, for representative benchmark flux models 2, 4, 6, and 7 (see \figu{fluxes}), shown as two-dimensional posteriors.  Results account for the background of showers induced by atmospheric muons; see Section~\ref{section:background}.  Each posterior is the average over $10^6$ random realizations of the shower distribution, binned in reconstructed shower energy and direction; see Section~\ref{section:cs_extracting_stat_proc}.
    Results are for the baseline detector design with 144 hybrid (shallow + deep) stations plus 169 shallow-only stations, angular resolution of $\sigma_{\theta_z} = 2^\circ$, energy resolution of $\sigma_{\epsilon} = 0.1$, and $T = 10$~years of exposure.  See \figu{all_2dposteriors} for similar results for all flux models, Table~\ref{tab:sensitivities} and \figu{violin} for marginalized one-dimensional parameter ranges and posteriors, and the main text for details.}
    \label{fig:sensitivity}
\end{figure}

Figure~\ref{fig:sensitivity} shows our results for the simultaneous measurement of the cross section and flux normalization for a selection of benchmark UHE neutrino flux models that is representative of the full variety explored in our analysis; see \figu{fluxes}.  Appendix~\ref{section:appendix_posteriors} contains similar results for all benchmark flux models.  In \figu{sensitivity}, the elongated shape of the two-dimensional posteriors follows from the partial degeneracy between the cross section and the flux normalization: the number of showers depends on their product; see \equ{event_rate_simple}.  The posteriors are narrower for flux models with a higher associated mean expected shower rate, \ie, in \figu{sensitivity}, for flux models 4 (``Bergman \& van Vliet, fit to TA UHECRs") and 6 (``Rodrigues {\it et al.}, all AGN").  This is because, in those cases, the number of horizontal showers is large enough to partially break the degeneracy between cross section and flux; in \equ{event_rate_simple}, this happens via the exponential attenuation term. 

Figure~\ref{fig:sensitivity} reveals an asymmetry between the region of high flux and small cross section in the upper-left corner---which is disfavored---and the region of low flux and large cross section in the lower-right corner---which is allowed.  The asymmetry becomes more evident for lower-flux models.  This preference for low flux and large cross section stems from the effect of the cross section on the shower distribution, as illustrated in \figu{event_rate} for flux model 4.  A small cross section ($f_\sigma \lesssim 1$) implies that the in-Earth attenuation is small and, therefore, the flux at the detector, and the angular distribution of the associated showers, are closer to isotropic.  In contrast, a high cross section ($f_\sigma \gtrsim 1$) implies a tilted angular distribution, with a sharp drop of the rate around the horizon.
For the real values of the cross section and flux normalization, $f_\sigma = 1$ and $f_\Phi = 1$, the predicted near-horizontal shower rate is low: for most fluxes, it is only about one shower per bin in 10~years (exceptionally, for flux model 4, the highest among the benchmarks, there is a handful of near-horizontal showers).  Therefore, high values of $f_\sigma \gtrsim 1$, which also result in a sharp drop, are favored over $f_\sigma \lesssim 1$; the asymmetry in \figu{sensitivity} reflects this.

\begin{figure}[t]
 \centering
 \includegraphics[width=\columnwidth]{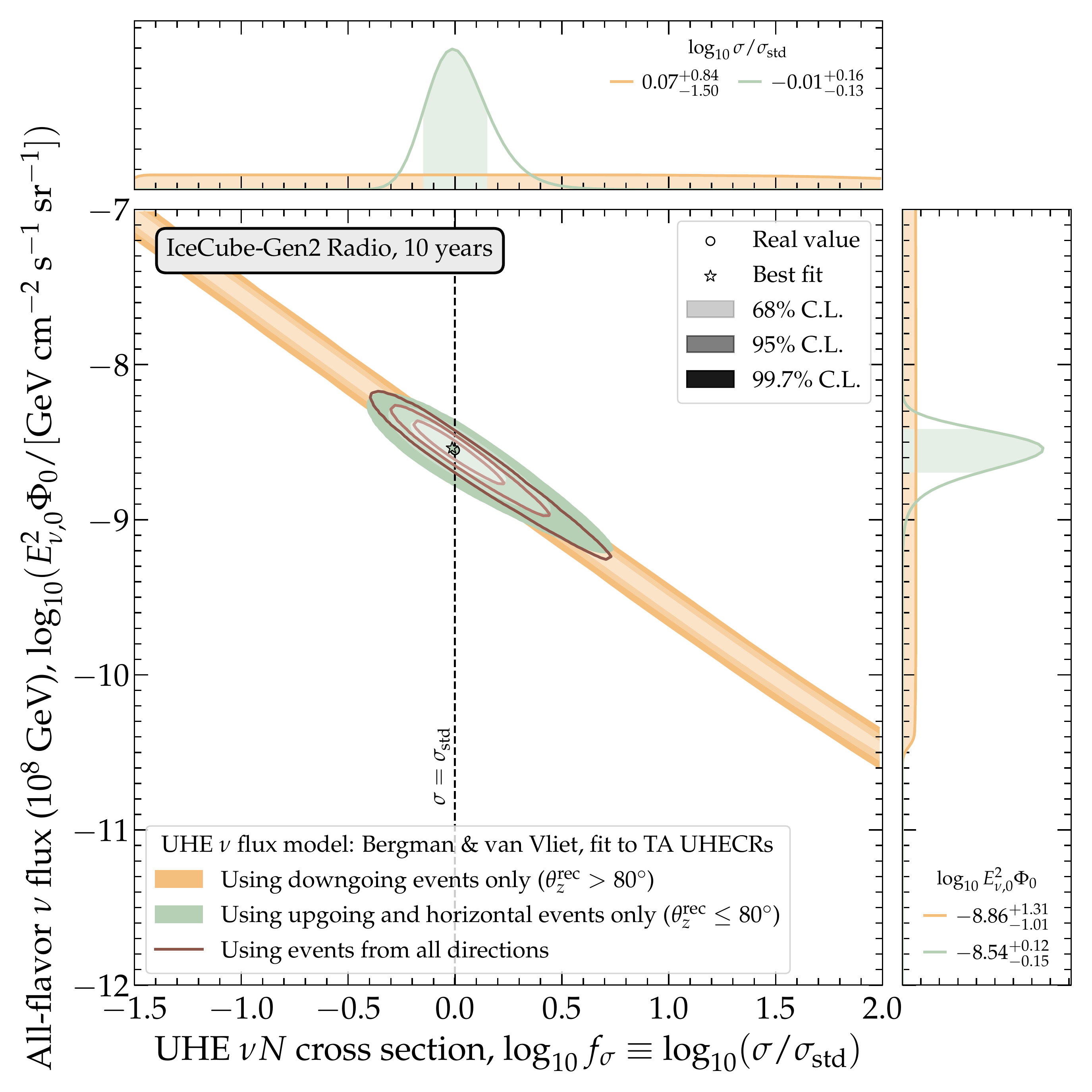}
 \caption{\label{fig:up_vs_down}Same as \Fig~\ref{fig:sensitivity}, but for benchmark flux model 4 alone, and showing results using only downgoing showers and only upgoing plus horizontal showers.}
\end{figure}

Figure~\ref{fig:up_vs_down} shows that the sensitivity to $f_\sigma$ and $f_\Phi$ comes predominantly from near-horizontal and horizontal showers, since they are the ones initiated by neutrinos that undergo significant, but not complete, attenuation inside the Earth.  Using only downgoing showers reveals the underlying degeneracy between cross section and flux in the calculation of shower rates (see Section~\ref{section:cs_measurement}), but is unable to break it.  In short, downgoing showers, more numerous, help in the discovery of the diffuse UHE neutrino flux, while near-horizontal showers, more rare, help to jointly perform studies beyond flux discovery, like cross-section measurements. 

\begin{figure}[t]
 \centering
 \includegraphics[width=\columnwidth]{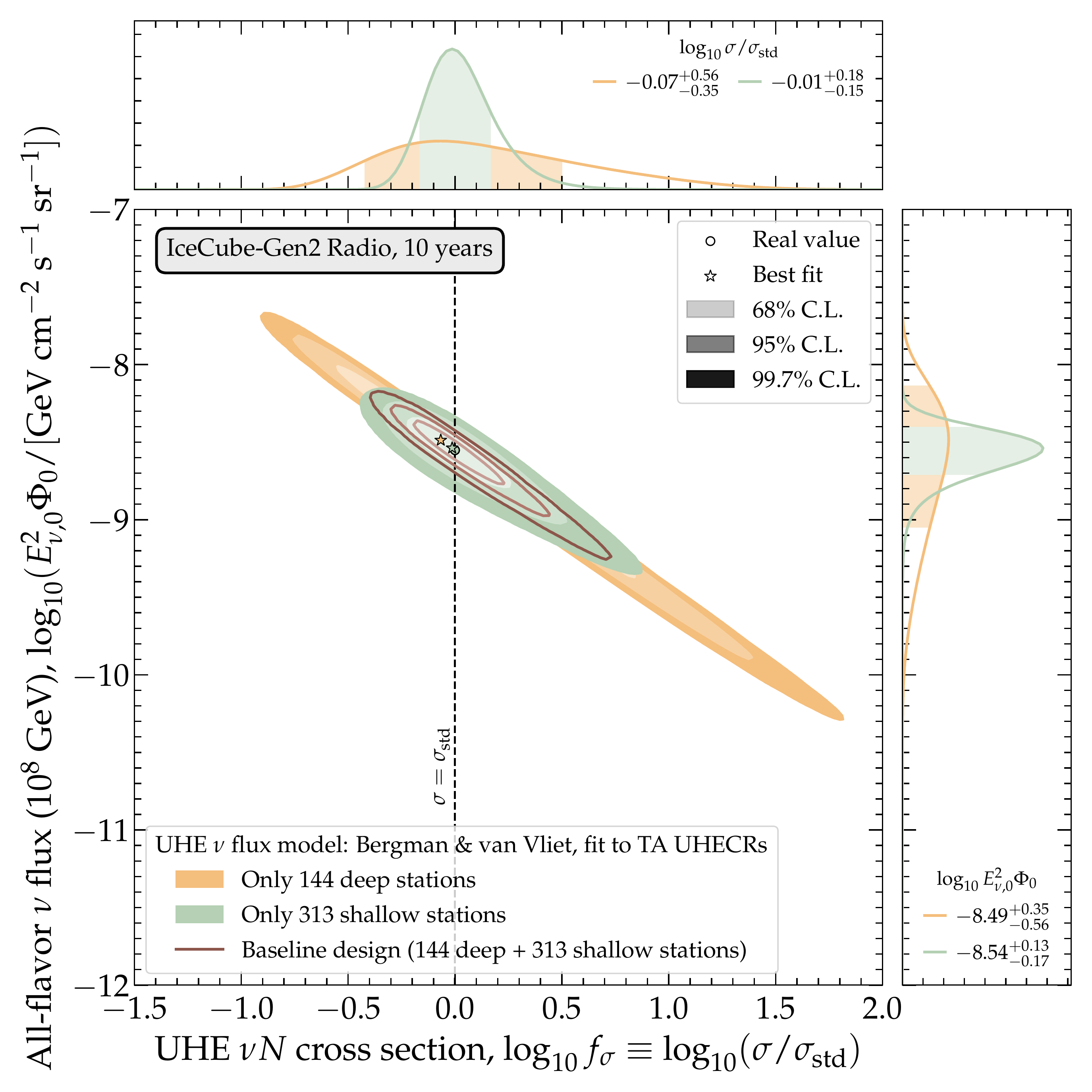}
 \caption{\label{fig:deep_vs_shallow}Same as \Fig~\ref{fig:sensitivity}, but for benchmark flux model 4 alone, and showing results using only the deep and shallow stations of the baseline array design.}
\end{figure}

Figure~\ref{fig:deep_vs_shallow} shows that, using the baseline array design---made up of 144 hybrid (shallow + deep) plus 169 shallow-only stations---the precision in the measured values $f_\sigma$ and $f_\Phi$ is dominated by shallow stations.  The caveat to this is that in our analysis we assume that the angular resolution of all detected showers is the same, whereas in reality there will be differences.  For instance, detection in deep antennas is expected to yield poorer resolution (see Section~\ref{section:radio_reconstruction}), which would further boost the importance of shallow stations.  Section~\ref{section:limitations} comments on future improvements of the analysis in this direction.

Table~\ref{tab:sensitivities} shows the one-dimensional marginalized ranges of $f_\sigma$ and $f_\Phi$ for all the benchmark flux models.  The results show that the radio component of IceCube-Gen2 could indeed jointly measure $f_\sigma$ and $f_\Phi$, to varying precision, as long as at least a few tens of showers are detected in 10~years.  Appendix~\ref{section:appendix_posteriors} shows the corresponding one-dimensional marginalized posteriors.  

We classify the results in Table~\ref{tab:sensitivities} in three groups, according to the size of the flux and its associated shower rate.  First, for low-flux models 1, 3, and 5, we expect fewer than one shower in 10~years and so no measurement is possible.  Second, for medium-flux models 2, 7, 8, 10, 11, and 12, we expect tens of events; with them, we can measure $f_\sigma$ to within 50--180\%  and $f_\Phi$ to within 70--110\%, depending on the flux model.  Third, for high-flux models 4, 6, and 9, we expect 100--300 events;  with them, we can measure $f_\sigma$ and $f_\Phi$ to within $40$\%. Table~\ref{tab:sensitivities} reveals that these groups are not dominated by a single flux type; rather, different flux types---extrapolated, cosmogenic, source, and cosmogenic + source (see Section~\ref{section:uhe_neutrinos})---are represented in all groups.  

Table~\ref{tab:sensitivities} reveals important nuance associated to the shape of the UHE neutrino spectrum.  For instance, even though flux model 12 has a lower shower rate than flux model 7, it leads to more accurate measurements of $f_\sigma$ and $f_\Phi$.  Figure~\ref{fig:fluxes} shows that model 12 peaks at lower energies than model 7.  Since the $\nu N$ cross section grows with energy, this implies that in model 12 neutrinos are less attenuated by their passage through Earth, and so more of them reach the detector from near-horizontal directions. Conversely, in model 7 neutrinos are higher-energy, so they are more strongly attenuated and a relatively larger number of them are downgoing.  This, again, highlights the importance of Earth-skimming showers to perform studies beyond flux discovery.

We have tested that the capability of IceCube-Gen2 to measure the cross section is not limited by the uncertainty in the background shower rate of atmospheric muons.  We did this by repeating our analysis using a rate of muon-induced showers five times higher than the one in \figu{muon_background}.  The results are practically indistinguishable from those in Table~\ref{tab:sensitivities}.  There are three reasons for this.  First, for the benchmark flux models for which the cross section can be measured, the rate of muon-induced showers is tiny compared to the rate of neutrino-induced showers.  Second, muon-induced showers are concentrated at lower energies, leaving the higher energies background-free.  Third, muon-induced showers affect almost exclusively  downgoing bins, whereas the sensitivity to the cross section comes from bins around the horizon.

\begin{figure}[t]
 \centering
 \includegraphics[width=\columnwidth]{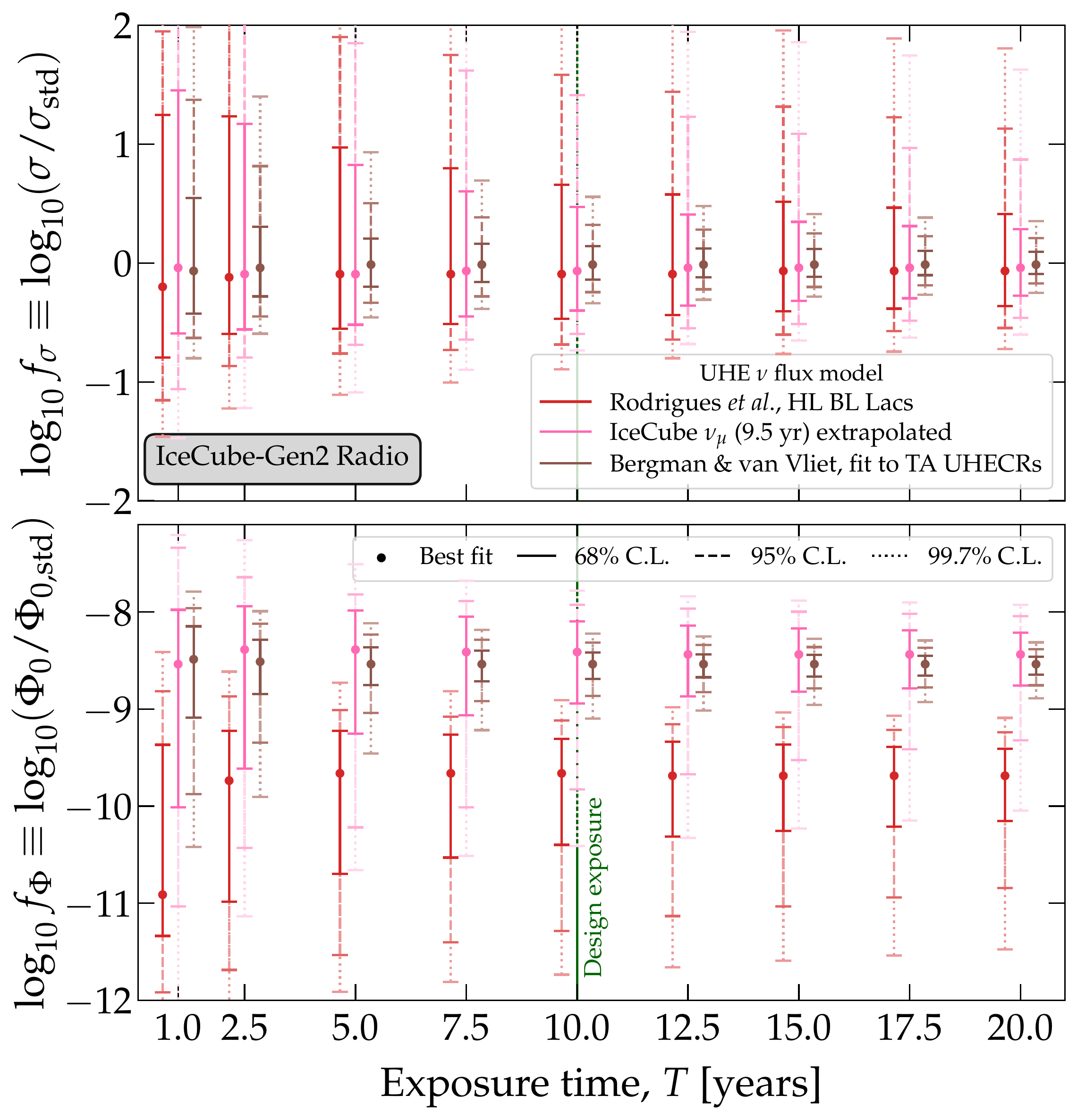}
 \caption{\label{fig:exposure} Evolution with exposure time of the  precision in the forecast measurement of the UHE $\nu N$ cross section ({\it top}) and the UHE neutrino flux normalization ({\it bottom}) in the radio component of IceCube-Gen2, for benchmark flux models 2, 4, and 7; see \figu{fluxes}.  See the main text for details.}
\end{figure}

In our analysis, we set the real values of the cross section and flux normalization to be $f_\sigma = 1$ and $f_\Phi = 1$; see above.  However, Table~\ref{tab:sensitivities} shows that, for all flux models, their average best-fit values have a consistent bias toward $f_\sigma \lesssim 1$ and $f_\Phi \gtrsim 1$ (except for low-flux models 1, 3, and 5, for which $f_\Phi$ is compatible with zero due to the low shower rate).  
The bias toward an average best-fit $f_\sigma \lesssim 1$, though slight, stems from the interplay in shower rates between the single bin that contains downgoing showers---whose mean expected shower rate is relatively high---and the multiple bins that contain horizontal and near-horizontal showers---whose mean expected shower rates are relatively low.  As a result, the Poisson probability distribution of the rate, \equ{poisson_nu_real}, of the single downgoing bin is rather symmetric, while Poisson probability distributions of the horizontal bins are rather asymmetric.  Therefore, in any one of the random realizations of the shower distribution that we generate as part of our statistical analysis (see Section~\ref{section:cs_extracting_stat_proc}), in the downgoing bin under-fluctuations and over-fluctuations relative to the mean shower rate are equally likely, while in the horizontal bins under-fluctuations are more likely than over-fluctuations.  As a result, on average, under-fluctuations in the all-sky shower rate are more likely than over-fluctuations.  Because lower shower rates correspond to smaller values of the cross section, via \equ{event_rate_simple}, the best-fit value of $f_\sigma \lesssim 1$ on average.  Nevertheless, within 68\%~C.L., the inferred values of $f_\sigma$ and $f_\Phi$ are compatible with their real ones (again, except for the low-flux models).  Further, their best-fit values approach their real values closer the higher the mean expected shower rate, and the longer the exposure time; see \figu{exposure}.

Figure~\ref{fig:panorama} places our results in the context of existing cross-section measurements, including in IceCube, by showcasing a selection of three optimistic, representative benchmark flux models, 2, 4, and 5.  In the best-possible scenario, \ie, for flux model 4 (``Bergman \& van Vliet, fit to TA UHECRs"), after 10 years we could be able to measure the cross section to within the theory uncertainty of the BGR18 prediction~\cite{Bertone:2018dse}.  For lower, but still detectable, flux models, \figu{panorama} shows that the measurement worsens, but remains informative.  In particular, for flux model 2 (``IceCube $\nu_\mu$ flux (9.5~yr), extrapolated to UHE")---a conservative benchmark based on the known TeV--PeV diffuse neutrino flux---the cross section could be measured to roughly the same accuracy at ultra-high energies as it is currently measured in the TeV--PeV range.  Overall, medium-flux and high-flux models (see above) offer enough accuracy to test a diversity of proposed deviations of the cross section, due to new physics, \eg, large extra dimensions~\cite{Alvarez-Muniz:2001efi}, electroweak sphalerons~\cite{Ellis:2016dgb}, or to non-perturbative effects in the nucleon structure, \eg, color glass condensate~\cite{Henley:2005ms}; see Fig.~2 in \Refe~\cite{Ackermann:2019cxh} for an example.

\begingroup
\begin{table*}[t]
 \begin{ruledtabular}
  \caption{\label{tab:sensitivities}Expected rates of neutrino-induced showers and measurement capability of the cross section and flux normalization in the radio component of IceCube-Gen2, after $T = 10$~years of exposure time, for the benchmark UHE diffuse neutrino fluxes used in this analysis; see Section~\ref{section:uhe_neutrinos}.  Possible flux types are: extrapolation to ultra-high energies ({\Large $\bullet$}), cosmogenic ($\blacksquare$), source (\rotatebox[origin=c]{45}{$\blacksquare$}), and cosmogenic + source ($\rotatebox[origin=c]{90}{\HexaSteel}$).  Measurement capabilities are expressed in terms of $f_\sigma \equiv \sigma/\sigma_{\rm std}$ for the cross section, and $f_\Phi \equiv \Phi_0/\Phi_{0, {\rm std}}$ for the flux normalization at a neutrino energy of $10^8$~GeV.  The real values are $f_\sigma = 1$ and $f_\Phi = 1$.
  Results are obtained using the baseline choices for the detector resolution: shower energy resolution of $\sigma_\epsilon = 0.1$, with $\epsilon \equiv \log_{10}(E_{\rm sh}^{\rm rec}/E_{\rm sh})$, and angular resolution of $\sigma_{\theta_{z}} = 2^\circ$; see Section~\ref{section:radio_detection_rates} for details.  The shower rates shown are all-sky, \ie, summed over all values of reconstructed direction, $-1 \leq \cos \theta_z^{\rm rec} \leq 1$, and binned in a single bin of reconstructed shower energy, $10^7 \leq E_{\rm sh}^{\rm rec}/{\rm GeV} \leq 10^{10}$.  However, all-sky rates are only referential, and in the statistical analysis used to infer the cross section and flux normalization, we use instead binned shower rates; see Section~\ref{section:cs_extracting_stat_proc} for details.  For the inferred cross section and flux normalization, we show the best-fit values and 68\% credible intervals of their one-dimensional marginalized posteriors.  The statistical procedure returns results for $\log_{10} f_\sigma$ and $\log_{10} f_\Phi$; we show the corresponding results in terms of $f_\sigma$ and $f_\Phi$ to facilitate their interpretation.}
  \centering
  \renewcommand{\arraystretch}{1.4}
  \setlength{\tabcolsep}{2.pt}
  \begin{tabular}{cccccccc}
   \multirow{2}{*}{\#} & 
   \multirow{2}{*}{Type} &
   \multirow{2}{*}{UHE $\nu$ flux model} &
   \multirow{2}{*}{\shortstack[c]{All-sky shower\\rate (10~yr)}} &
   \multicolumn{2}{c}{Inferred cross section} &
   \multicolumn{2}{c}{Inferred flux normalization} \\
   &
   &
   &
   &
   $\log_{10} f_\sigma$ &
   $f_\sigma$ &
   $\log_{10} f_\Phi$ &
   $f_\Phi$ \\
   \hline
   1 &
   {\Large $\bullet$} &
   IceCube HESE (7.5~yr) extrapolated~\cite{IceCube:2020wum} &
   0.73 &
   $-0.39^{+1.12}_{-1.01}$ &
   $0.41^{+4.96}_{-0.37}$&
   $-1.16^{+0.99}_{-1.04}$ &
   $0.07^{+0.61}_{-0.07}$ \\
   2 &
   {\Large $\bullet$} &
   IceCube $\nu_\mu$ (9.5~yr) extrapolated~\cite{IceCube:2021uhz} &
   26.90 &
   $-0.07^{+0.54}_{-0.33}$ &
   $0.85^{+2.10}_{-0.45}$ &
   $+0.06^{+0.32}_{-0.53}$ &
   $+1.15^{+1.25}_{-0.81}$ \\
   3 &
   $\blacksquare$ &
   Heinze {\it et al.}, fit to Auger UHECRs~\cite{Heinze:2019jou} &
   0.71 &
   $-1.75^{+2.23}_{-0.20}$ &
   $0.02^{+3.70}_{-0.01}$ &
   $-1.11^{+1.35}_{-1.07}$ &
   $0.08^{+1.66}_{-0.07}$ \\
   4 &
   $\blacksquare$ &
   Bergman \& van Vliet, fit to TA UHECRs~\cite{Anker:2020lre} &
   332.34 &
   $-0.01^{+0.16}_{-0.13}$ &
   $0.98^{+0.44}_{-0.25}$ &
   $+0.01^{+0.12}_{-0.15}$ &
   $1.02^{+0.33}_{-0.30}$ \\
   5 &
   $\blacksquare$ &
   Rodrigues {\it et al.}, all AGN~\cite{Rodrigues:2020pli} &
   0.89 &
   $-1.45^{+2.14}_{-0.16}$ &
   $0.04^{+4.86}_{-0.01}$ &
   $-1.15^{+1.19}_{-0.90}$ &
   $0.07^{+1.02}_{-0.06}$ \\
   6 &
   \rotatebox[origin=c]{45}{$\blacksquare$} &
   Rodrigues {\it et al.}, all AGN~\cite{Rodrigues:2020pli} &
   107.16 &
   $-0.01^{+0.18}_{-0.15}$ &
   $0.98^{+0.50}_{-0.29}$ &
   $+0.01^{+0.15}_{-0.18}$ &
   $1.02^{+0.42}_{-0.35}$  \\
   7 &
   $\blacksquare$ &
   Rodrigues {\it et al.}, HL BL Lacs~\cite{Rodrigues:2020pli} &
   24.24 &
   $-0.09^{+0.75}_{-0.37}$ &
   $0.81^{+3.76}_{-0.47}$ &
   $+0.09^{+0.36}_{-0.74}$ &
   $1.23^{+1.59}_{-1.01}$ \\
   8 &
   \rotatebox[origin=c]{90}{\HexaSteel} &
   Fang \& Murase, cosmic-ray reservoirs~\cite{Fang:2017zjf} &
   57.41 &
   $-0.04^{+0.36}_{-0.26}$ &
   $0.91^{+1.18}_{-0.41}$ &
   $+0.03^{+0.25}_{-0.33}$ &
   $1.07^{+0.83}_{-0.57}$ \\
   9 &
   \rotatebox[origin=c]{45}{$\blacksquare$} &
   Fang {\it et al.}, newborn pulsars~\cite{Fang:2013vla} &
   125.38 &
   $-0.01^{+0.19}_{-0.16}$ &
   $0.98^{+0.54}_{-0.30}$ &
   $+0.01^{+0.17}_{-0.18}$  &
   $1.02^{+0.49}_{-0.35}$ \\
   10 &
   \rotatebox[origin=c]{45}{$\blacksquare$} &
   Padovani {\it et al.}, BL Lacs~\cite{Padovani:2015mba} &
   57.85 &
   $-0.01^{+0.24}_{-0.21}$ &
   $0.98^{+0.72}_{-0.37}$ &
   $+0.02^{+0.20}_{-0.25}$  &
   $1.05^{+0.61}_{-0.46}$ \\
   11 &
   \rotatebox[origin=c]{90}{\HexaSteel} &
   Muzio {\it et al.}, max. extra $p$ component~\cite{Muzio:2019leu} &
   56.55 &
   $-0.04^{+0.39}_{-0.28}$ &
   $0.91^{+1.33}_{-0.43}$ &
   $+0.03^{+0.27}_{-0.38}$ &
   $1.07^{+0.92}_{-0.62}$ \\
   12 &
   \rotatebox[origin=c]{90}{\HexaSteel} &
   Muzio {\it et al.},fit to Auger $\&$ IceCube~\cite{Muzio:2021zud} &
   17.12 &
   $-0.04^{+0.48}_{-0.36}$ &
   $0.91^{+1.84}_{-0.51}$ &
   $+0.05^{+0.32}_{-0.49}$ &
   $1.12^{+1.22}_{-0.76}$ \\
   \end{tabular}
 \end{ruledtabular}
\end{table*}
\endgroup

Figure~\ref{fig:exposure} shows the expected improvement in the measurements over time, for selected medium-flux and high-flux benchmark models.  In the most optimistic scenario, for flux model 4 (``Bergman \& van Vliet, fit to TA UHECRs"), the 2.5-year uncertainty in the inferred value of the cross section, $f_\sigma$, of $75$~\%, is already comparable to the baseline 10-year uncertainty, of $35$~\%.  The 20-year uncertainty, of $22$~\%, is smaller than the BGR18 theory uncertainty shown in \figu{panorama}.  Results for model 2 (``IceCube $\nu_\mu$ flux (9.5~yr), extrapolated to UHE") are comparable.  
At short exposure times, increased exposure leads to drastic improvements in the measurement uncertainty because the measurement is dominated by statistical uncertainties.  At long exposure times, improvements slow down, as the measurement becomes increasingly dominated by systematic uncertainties, \ie, the energy and angular detector resolution and, to a lesser degree, the background of atmospheric muons.

\begin{figure*}[t]
 \centering
 \includegraphics[width=\textwidth]{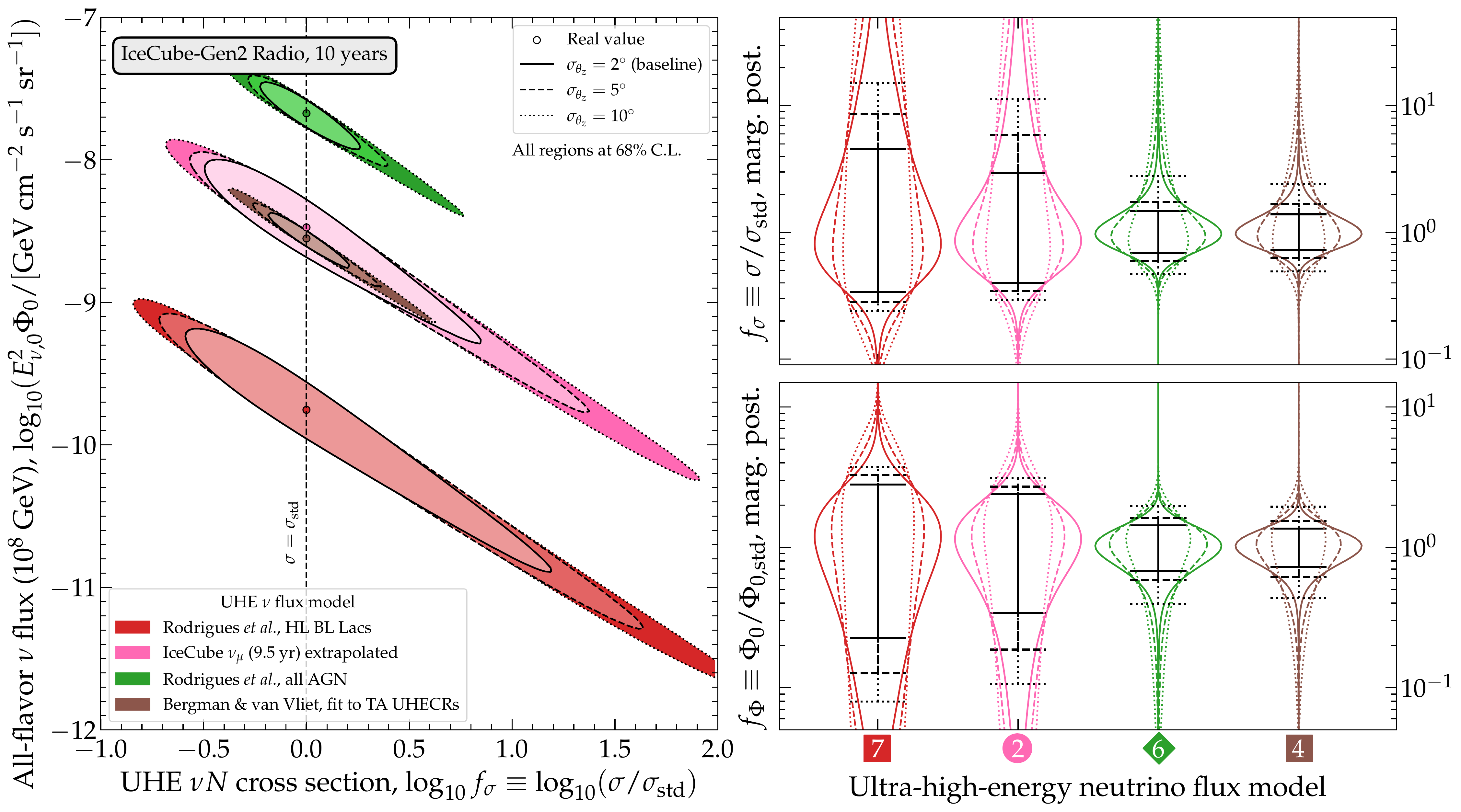}
 \caption{\label{fig:angular_resolution} Impact of the detector angular resolution, $\sigma_{\theta_z}$, on the measurement of the $\nu N$ DIS cross section and flux normalization, for the benchmark UHE neutrino flux models 2, 4, 6, and 7; see \figu{fluxes}.  See Section~\ref{section:cs_extracting_binning} for details about the choice of binning.}
\end{figure*}

Figure~\ref{fig:angular_resolution} shows that angular resolution, $\sigma_{\theta_z}$, is key to making precise measurements of $f_\sigma$ and $f_\Phi$, especially in medium-flux models where the degeneracy between them is more resilient on account of a lower shower rate.  A good angular resolution allows the detector to distinguish between similar arrival directions around the horizon, where the joint sensitivity to $f_\sigma$ and $f_\Phi$ comes from; see Sections~\ref{section:radio_detection_rates} and \ref{section:cs_extracting}.  In \figu{angular_resolution}, we repeat our analysis, but for different choices of values of $\sigma_{\theta_z}$.  For each choice, we set the size of the angular bins used for horizontal and near-horizontal showers, with $\theta_{z}^{\rm rec} \in [80^\circ, 100^\circ]$, equal to $\sigma_{\theta_z}$.  Figure~\ref{fig:angular_resolution} shows, \eg, for flux model 4 (``Bergman \& van Vliet, fit to TA UHECRs"), that using a resolution of $\sigma_{\theta_z} = 5^\circ$ yields $f_\sigma = 0.94_{-0.31}^{+0.73}$ and $f_\Phi = 1.06_{-0.44}^{+0.49}$, and using $\sigma_{\theta_z} = 10^\circ$ yields $f_\sigma = 0.88_{-0.38}^{+1.54}$ and $f_\Phi = 1.14_{-0.64}^{+1.28}$, an increase of $50\%$ and $178\%$ in the uncertainty of $f_\sigma$ and of $41\%$ and $191\%$ in the uncertainty of $f_\Phi$, respectively, compared to the baseline results with $\sigma_{\theta_z} = 2^\circ$ from Table~\ref{tab:sensitivities}.  Results for other flux models change similarly. 


\section{Future directions}
\label{section:limitations}

When generating our forecasts above, we have used state-of-the-art predictions and dedicated simulations to describe the incoming UHE neutrino flux, its propagation through Earth, and its detection in the radio component of IceCube-Gen2, and to jointly measure the cross section and flux normalization.  Yet, there are potential improvements to be implemented in future, revised versions of our calculation.  None of them represent a fundamental limitation of our present analysis.  Below, we present possible future improvements, listed roughly in ascending order of implementation challenge, with the hope of sparking progress in various directions:

\smallskip
{\it Raising the maximum neutrino energy in in-Earth propagation.---}  To compute the propagation of UHE neutrinos through the Earth, we have used {\sc NuPropEarth}~\cite{Garcia:2020jwr, NuPropEarth}, based on the state-of-the-art BGR18 $\nu N$ DIS scattering cross section~\cite{Bertone:2018dse}; see Section~\ref{section:cs_basics}  However, the version of {\sc NuPropEarth} that we used was unable to propagate neutrinos beyond energies of $10^{10}$~GeV.  Fortunately, for the majority of the benchmark flux models 1--12, the flux beyond $10^{10}$~GeV contributes marginally to the shower rate; see \figu{fluxes}.  Only for models 4 and 11 we expect that contribution to be slightly more sizeable.  We recommend raising the maximum allowed neutrino energy in future versions of {\sc NuPropEarth}.

\smallskip
{\it Including the LPM effect in the relation between neutrino and shower energies.---}  The LPM effect is included in the {\sc NuRadioMC} simulations of $\nu_e$-induced CC showers and radio propagation that we use to generate the detector effective volume; see Section~\ref{section:radio_detection_rates}.  However, the energy-dependent relevance of the LPM is not accounted for in the relation between neutrino energy, $E_\nu$, and shower energy, $E_{\rm sh}$, \equ{energy_nu}, because a parametrization of it was not available at the time of writing.  As a result, for $\nu_e$-initiated CC showers, we have assumed that $E_{\rm sh} = E_\nu$ at all energies, whereas in reality the LPM should make $E_{\rm sh} < E_\nu$ at the highest energies.  This may be improved in future work by building a suitable parametrization from dedicated simulations.

\smallskip
{\it Including the contribution of secondary leptons in the effective volumes for $\nu_\mu$- and $\nu_\tau$-initiated CC showers.---}  When generating the effective volumes for $\nu_\mu$- and $\nu_\tau$-initiated CC showers, we have ignored the contribution of the radio emission coming from showers induced by the secondary charged leptons, \ie, the final-state muons and tauons in the CC interactions, due to the additional computational expense involved in simulating them and in parametrizing the relation between neutrino to shower energy in the presence of secondary interactions; see Section~\ref{section:radio_detection_rates}.  Reference~\cite{Garcia-Fernandez:2020dhb} showed that they may enhance the shower rate by up to 25\%, which makes our results above conservative.  Their contribution may be included in future work via dedicated simulations.

\smallskip
{\it Improving the modeling of the angular resolution.---}  To obtain our results, we assumed the same angular resolution for all detected showers; see Section~\ref{section:results}.  However, in reality, the angular resolution of a detected shower depends on its event quality, and on which detector component measured it, \ie, only the shallow one, only the deep one, or multiple components in coincidence.  With a better modelling of the expected experimental uncertainties, our analysis can be redone using different resolution functions depending on the event class.  Further, we have only considered the resolution in zenith angle, not in azimuth angle; the latter might be relevant when analyzing real shower events, with definite celestial coordinates. 

\smallskip
{\it Using an unbinned likelihood analysis.---}  In our analysis above, we binned the expected IceCube-Gen2 showers in reconstructed shower energy and direction (see Section~\ref{section:cs_extracting_binning}), and we used a binned likelihood function, \equ{likelihood}, to infer the values of the $\nu N$ cross section and the neutrino flux normalization.  We chose the bin sizes to be no smaller than the detector energy and angular resolution.  Yet, when inferring the cross section and flux normalization from actual future data detected by IceCube-Gen2, with each shower carrying its own energy and direction resolution, an unbinned likelihood would be preferable, in the style of IceCube TeV--PeV cross-section measurements in \Refs~\cite{Bustamante:2017xuy, IceCube:2020rnc}.  Such unbinned prescription may also be used in forecasts of the cross-section measurement, with modifications of our above procedure.

\smallskip
{\it Including the background of cosmic ray-induced showers.---}  Section~\ref{section:background} stated that the two main non-neutrino backgrounds in the radio array of IceCube-Gen2 are showers induced by atmospheric muons, which we account for in our analysis, and air-shower cores propagating into the ice, which we do not account for, since their rate is still largely unknown.  Early estimates of this background put its rate anywhere between a handful and tens of showers per year, before the application of any cuts or of a surface veto that could mitigate them.  If these estimates are accurate, and if vetoes do not mitigate the rate significantly, then they might render cross-section measurements in even medium-flux models unfeasible.  Ongoing work should clarify the relevance of this background.

\smallskip
{\it Including nuclear effects in the cross section.---}  To obtain our results, we have used the BGR18 cross section built using free-nucleon PDFs.  However, the nucleons with which neutrinos interact are bound in heavy nuclei; this modifies the PDFs and, therefore, the $\nu N$ DIS cross section.  Reference~\cite{Garcia:2020jwr} showed that using nuclear-corrected PDFs in in-Earth propagation could modify the fraction of neutrinos that reach the detector by more than 50\%, especially at high energies.  This is comparable to or greater than the measurement precision that we have forecast; see, \eg, \figu{panorama}.  This should motivate future forecasts of UHE cross-section measurement to devote more attention to the effect of nuclear corrections.

\smallskip
{\it Using flavor identification.---}  Figure~\ref{fig:flavor_ratios} shows that the flavor ratios at the detector carry information about the $\nu N$ cross section.  For instance, comparing the tau-flavor ratio {\it vs.}~the electron-flavor or muon-flavor ratios could give us information about the effect of the cross section on $\nu_\tau$ regeneration.  More generally, flavor ratios have the potential to explore a variety of new neutrino physics; see, \eg, \Refs~\cite{Arguelles:2015dca, Bustamante:2015waa, Ackermann:2019cxh, Arguelles:2019rbn, Song:2020nfh, Arguelles:2022xxa}. Presently, however, flavor identification in neutrino radio-detection is still uncertain, though preliminary results on identifying showers from $\nu_\mu$ and $\nu_\tau$ CC interactions~\cite{Garcia-Fernandez:2020dhb, Glaser:2021hfi} and from $\nu_e$ CC interactions~\cite{Stjarnholm:2021xpj} are promising.  Even if flavor identification is not readily available in IceCube-Gen2, its all-flavor data could be combined with data collected by other planned neutrino telescopes that are predominantly sensitive to UHE $\nu_\tau$~\cite{Abraham:2022jse, Ackermann:2022rqc}, \ie, AugerPrime~\cite{PierreAuger:2016qzd}, BEACON~\cite{Wissel:2020sec}, EUSO-SPB~\cite{Bacholle:2017dye}, GRAND~\cite{GRAND:2018iaj}, POEMMA~\cite{POEMMA:2020ykm}, PUEO~\cite{PUEO:2020bnn}, SKA~\cite{James:2017pvr}, TAMBO~\cite{Romero-Wolf:2020pzh}, TAROGE-M~\cite{Nam:2020hng}, TAx4~\cite{Kido:2020isy}, and Trinity~\cite{Otte:2018uxj, Brown:2021lef}, in order to compute the tau-flavor ratio.

\smallskip
{\it Looking for energy-dependent new physics in the cross section.---}  In our analysis above, we have looked for energy-independent modifications of the Standard Model $\nu N$ DIS cross section.  However, if there are beyond-the-Standard-Model contributions to the cross section, they may instead be energy-dependent and possibly even localized only within a narrow energy window.  In particular, UHE neutrinos could allow us to probe new $\nu N$ resonant interactions via new heavy mediators with masses up to, roughly, $(m_q/{\rm GeV})^{0.5} \times 100$~TeV, where $m_q$ is the mass of the quark involved in the interaction.  Reference~\cite{Huang:2021mki} explored the effect of new energy-dependent contributions to the $\nu N$ cross section, and of resonant contributions to the neutrino-electron cross section, in the context of GRAND, POEMMA and Trinity.  For IceCube-Gen2, attempting to explore energy-dependent modifications of the $\nu N$ cross section within the same sophisticated end-to-end framework that we have developed above quickly escalates the complexity of the tasks involved, and is best left for future work.  

\smallskip
{\it Jointly measuring the $\nu N$ cross section, flux normalization, and spectral shape.---}  In our analysis above, we forecast the joint measurement of the $\nu N$ cross section and flux normalization, but kept the shape of the neutrino energy spectrum fixed to that of one of the benchmark flux models 1--12 taken from the literature; see Section~\ref{section:uhe_neutrinos}.  While this procedure provides us already with useful information about the capability to measure the cross section, ultimately we would also like the shape of the neutrino energy spectrum to be determined from data, rather than assumed.  To do this, future studies could exploit the fact that the sensitivity to the cross section comes almost exclusively from showers around the horizon; see \figu{up_vs_down}.  This suggests that we may use the large number of downgoing, unattenuated showers to measure the shape of the neutrino energy spectrum.


\section{Summary and outlook}
\label{section:summary}

Measuring the neutrino-nucleon ($\nu N$) cross section at ultra-high neutrino energies, above 100~PeV, offers novel insight into the deep structure of protons and neutrons, and sensitivity to potentially transformative new physics.  Yet, it hinges on using ultra-high-energy (UHE) neutrinos of cosmic origin, long sought, but so far undiscovered.  Fortunately, upcoming neutrino telescopes, currently under planning and testing, will have a realistic chance of discovering them in the next 10--20 years, even if their flux is low, as is expected.  Accordingly, we have put forward the first detailed forecasts of the capabilities of IceCube-Gen2~\cite{IceCube-Gen2:2020qha}, one of the leading upcoming neutrino telescopes, to measure the UHE $\nu N$ cross section.  We have endeavored to make our forecasts complete, robust, and realistic.  Our results are encouraging.

The sensitivity to the UHE $\nu N$ cross section stems from the strong directional dependence of the UHE neutrino flux that reaches the detector after propagating through Earth. Due to significant $\nu N$ interactions underground, UHE neutrinos only reach the detector from above---where they undergo nearly no interactions before reaching the detector---and from directions around the horizon---where significant interactions imprint the $\nu N$ cross section on the angular dependence of the flux.  It is from these latter Earth-skimming neutrinos that the sensitivity to the UHE $\nu N$ cross section comes from.

In our forecasts, we have used state-of-the-art ingredients at every stage.  For the $\nu N$ cross section, we adopted as a baseline the recent BGR18 deep-inelastic-scattering (DIS) calculation~\cite{Bertone:2018dse}.  For the fluxes of UHE neutrinos, we explored a wide variety of predictions~\cite{Fang:2013vla, Padovani:2015mba, Fang:2017zjf, Heinze:2019jou, Muzio:2019leu, Rodrigues:2020pli, Anker:2020lre, IceCube:2020wum,  Muzio:2021zud, IceCube:2021uhz}, including extrapolations of the known IceCube TeV--PeV fluxes to ultra-high energies, cosmogenic neutrinos, neutrinos produced inside cosmic-ray sources, and self-consistent models of joint production of the latter two.  We treat each neutrino species separately, $\nu_e$, $\bar{\nu}_e$, $\nu_\mu$, $\bar{\nu}_\mu$, $\nu_\tau$, and $\bar{\nu}_\tau$.  For the neutrino propagation through Earth, we compute $\nu N$ DIS plus sub-dominant interactions, via {\sc NuPropEarth}~\cite{Garcia:2020jwr}.  For  detection, we focus on the radio-detection of neutrino-initiated showers.  We use the same simulation tools as the IceCube-Gen2 Collaboration, {\sc NuRadioMC}~\cite{Glaser:2019cws} and {\sc NuRadioReco}~\cite{Glaser:2019rxw}, to simulate neutrino interactions in the detector, ensuing particle showers, emission of coherent radio signals---Askaryan radiation---its propagation in the Antarctic ice, and its detection in the radio stations.  We account for the detector resolution in shower energy and direction, and for the background of showers from atmospheric muons.

To properly represent the significant uncertainty in the predicted neutrino flux, in our forecasts we have jointly extracted the cross section and the neutrino flux normalization.  To do this, we used a Bayesian statistical approach that accounts for random fluctuations in the number of expected detected showers.  We report results in terms of mean sensitivity, averaged over many random realizations of the predicted shower distributions.

We find that it may be possible to measure the UHE $\nu N$ cross section to within 50\% of the BGR18 prediction within 10~years of exposure of IceCube-Gen2, as long as at least a few tens of neutrino-induced showers are detected in this period; see \figu{panorama}.  In the most optimistic case, we expect that comparable precision could be achieved within 5~years; see \figu{exposure}.  By far, the largest systematic uncertainty in our forecasts comes from the UHE neutrino flux.  Regardless, the level of precision that may be achieved is enough to test the standard prediction of the $\nu N$ DIS cross section~\cite{Bertone:2018dse}, to look for non-linear effects in the nucleon structure~\cite{Lipatov:1976zz, Kuraev:1976ge, Kuraev:1977fs, Balitsky:1978ic}, color-glass condensates~\cite{Gelis:2010nm}, and  sphalerons~\cite{Ellis:2016dgb}, and to identify deviations introduced by a host of new-physics models~\cite{Alvarez-Muniz:2001efi, Connolly:2011vc, Chen:2013dza, Marfatia:2015hva, Mack:2019bps, Huang:2021mki}.

Our forecasts are comprehensive, but we have identified areas where further work is needed.  Most pressing are estimating the effect of the background of showers induced by cosmic rays, improved modelling of reconstruction uncertainties based on individual event quality, and extending our analysis to let both the neutrino flux normalization {\it and} the shape of the neutrino energy spectrum be inferred from data.  Thanks to the flexibility of our calculation framework, these improvements can be easily incorporated.  Work in these directions is ongoing.

We provide our forecasts in the hope of informing the design choices of IceCube-Gen2 and other upcoming neutrino telescopes.  Ultimately, we hope that detailed forecasts like ours provide guidance to tap into their full potential for fundamental-physics research.


\section*{Acknowledgements}

We thank Douglas Bergman, Ke Fang, Alfonso Garc\'ia, Daniel Garc\'ia, Steffen Hallmann, Kohta Murase,  Anna Nelles, Foteini Oikonomou, and Juan Rojo for valuable discussion and input and, especially, Jonas Heinze, Kumiko Kotera, Marco Muzio, Paolo Padovani, Xavier Rodrigues, and Arjen van Vliet for providing or helping to generate the detailed neutrino fluxes used in this work.  MB and VBV are supported by the {\sc Villum Fonden} under project no.~29388.  This work was made possible by Institut Pascal at Universit\'e Paris-Saclay during the Paris-Saclay Astroparticle Symposium 2021, with the support of the P2IO Laboratory of Excellence (programme ``Investissements d’avenir" ANR-11-IDEX-0003-01 Paris-Saclay and ANR-10-LABX-0038), the P2I research departments of the Paris-Saclay University, as well as IJCLab, CEA, IPhT, APPEC, the IN2P3 master project UCMN, and EuCAPT.  This work used resources provided by the High Performance Computing Center at the University of Copenhagen.  The computations and data handling were enabled by resources provided by the Swedish National Infrastructure for Computing (SNIC) at UPPMAX partially funded by the Swedish Research Council through grant agreement no. 2018-05973.


\newpage
\onecolumngrid
\appendix


\section{Posterior distributions for all benchmark UHE neutrino flux models}
\label{section:appendix_posteriors}

\renewcommand{\thefigure}{A\arabic{figure}}
\setcounter{figure}{0}

In the main text we used twelve benchmark UHE neutrino flux models to make our forecasts of cross-section measurements; see Section~\ref{section:uhe_neutrinos}.  For all of them, we showed expected shower rates in IceCube-Gen2 in Table~\ref{tab:event_rates_total} and one-dimensional marginalized posteriors of $f_\sigma$ and $f_\Phi$ in Table~\ref{tab:sensitivities}, but we showed two-dimensional posteriors of $f_\sigma$ and $f_\Phi$ only for four of them, models 2, 4, 6, and 7, in \figu{sensitivity}. 

\begin{figure*}[b]
 \centering
 \includegraphics[width=0.99\textwidth]{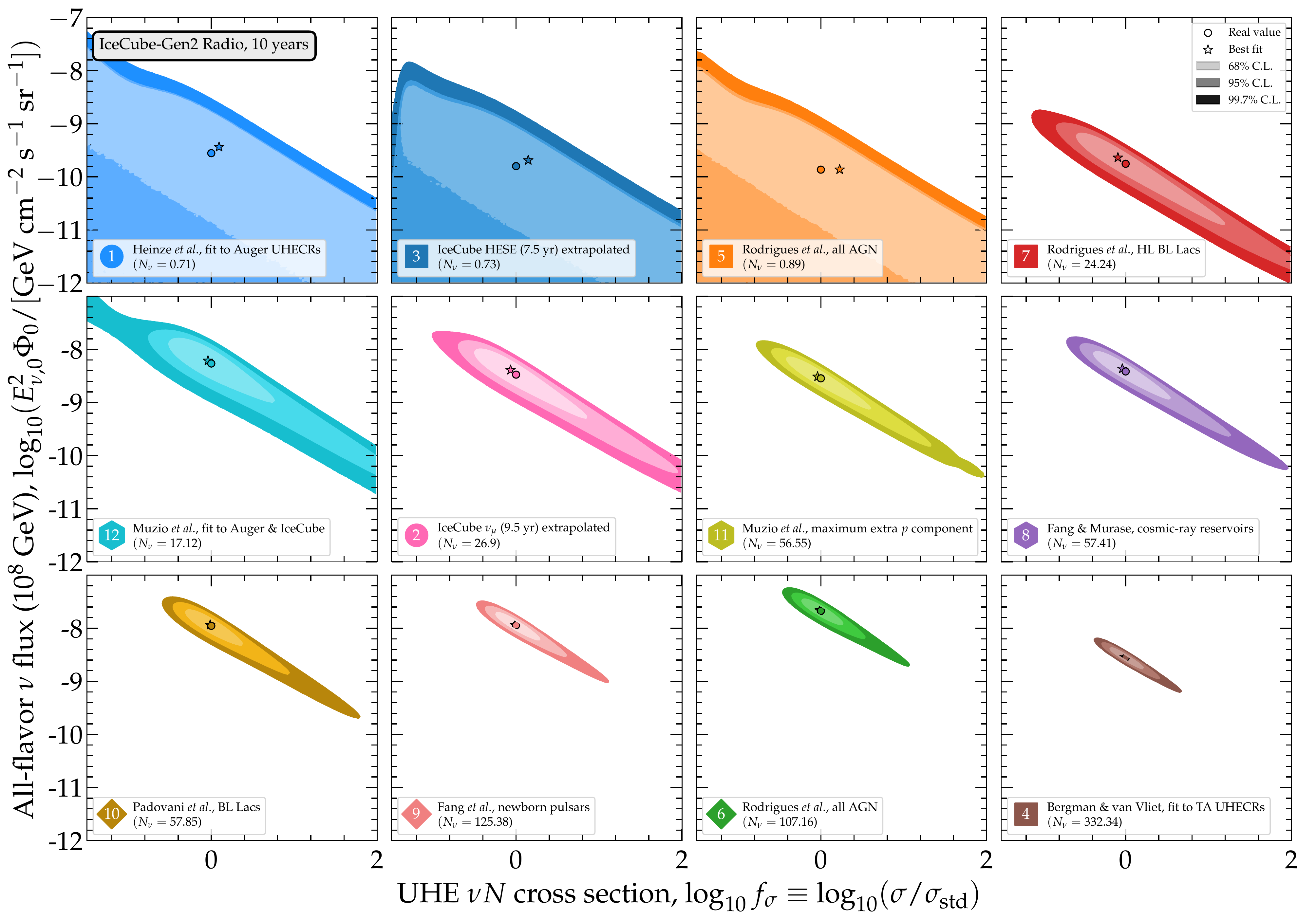}
 \caption{\label{fig:all_2dposteriors}Same as \figu{sensitivity} in the main text, but for all the benchmark UHE neutrino flux models adopted in our analysis; see Section~\ref{section:uhe_neutrinos}.}
\end{figure*}

Figure~\ref{fig:all_2dposteriors} shows the two-dimensional posteriors of $f_\sigma$ and $f_\Phi$ for all benchmark flux models, for the same baseline detector parameters as in \figu{sensitivity}.  From top to bottom and left to right, results are arranged in order of increasing sensitivity, from the most pessimistic to the most optimistic scenarios.

\begin{figure*}[t]
 \centering
 \includegraphics[width=\textwidth]{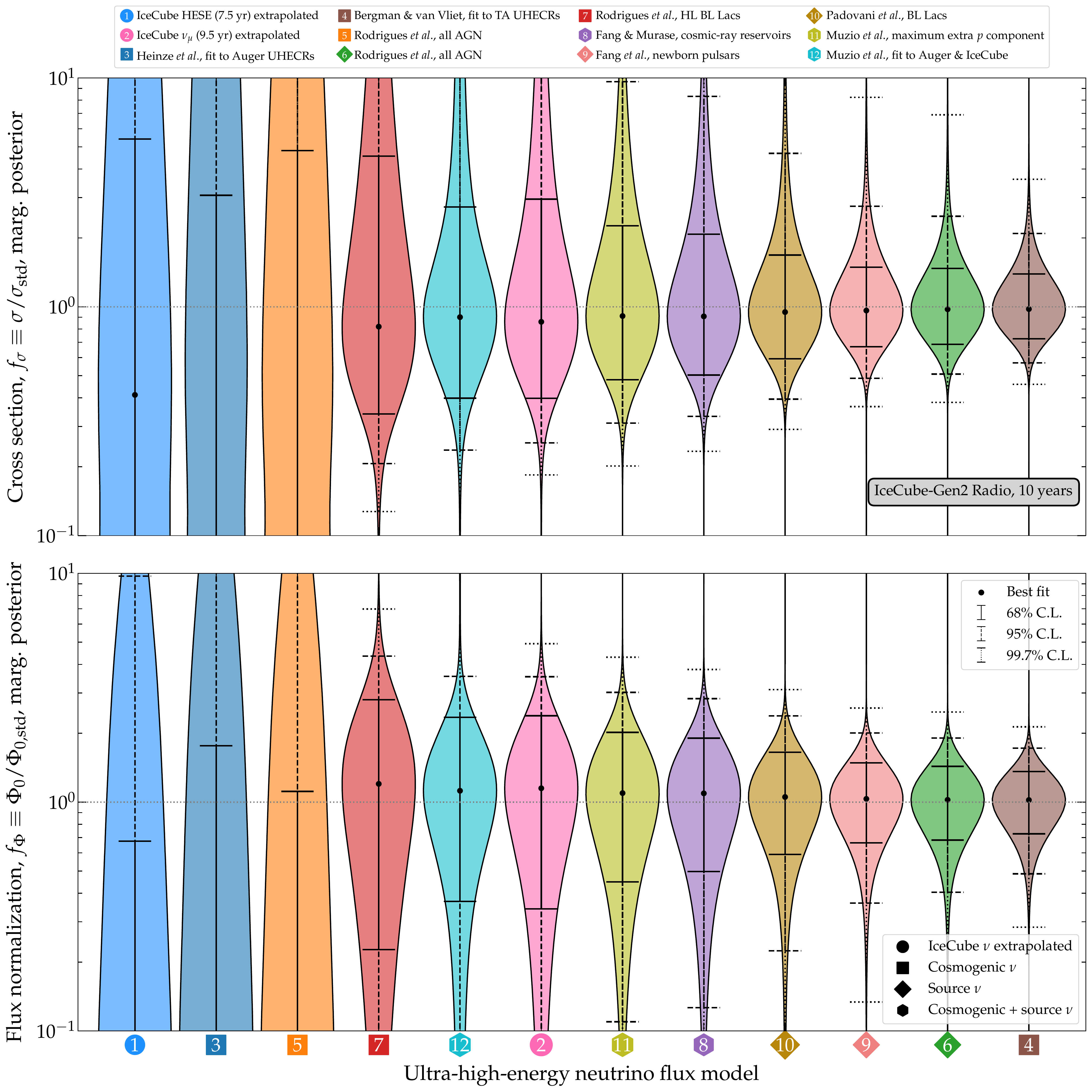}
 \caption{\label{fig:violin}One-dimensional marginalized posteriors of the UHE $\nu N$ cross section, $f_\sigma$, and the UHE neutrino flux normalization, $f_\Phi$, for all the benchmark UHE neutrino flux models adopted in our analysis.  See \figu{all_2dposteriors} for the corresponding two-dimensional posteriors and  Table~\ref{tab:sensitivities} for numerical values.}
\end{figure*}

Figure~\ref{fig:violin} shows the corresponding one-dimensional marginalized posteriors of $f_\sigma$ and $f_\Phi$.  The 68\% C.L.~ranges in this figure match those shown in Table~\ref{tab:sensitivities}.

\pagebreak
\twocolumngrid


%


\end{document}